{}
{}

\documentclass[12pt,a4paper]{article}

\newif\ifpublic\publicfalse
\newif\ifniklas\niklastrue
\newif\ifniklas\niklastrue

\usepackage{subfigure}
\usepackage{color}
\usepackage{tikz}
\usetikzlibrary{decorations.markings}
\usetikzlibrary{shapes,arrows,automata,positioning}
\usepackage{tikz-3dplot}

\usepackage{ifpdf}
\ifniklas\ifpdf\publictrue\else\publicfalse
\PassOptionsToPackage{hypertex}{hyperref}
\PassOptionsToPackage{draft}{graphicx}
\fi\fi

\usepackage{amsmath,amssymb}
\ifpublic\else\usepackage{showkeys}\fi
\usepackage[bookmarks=true,hyperfigures=true]{hyperref}
\usepackage{graphicx}
\usepackage[nosort]{cite}
\usepackage[bulletsep]{collref}

\usepackage[most]{tcolorbox}
\definecolor{block-gray}{gray}{0.85}
\newtcolorbox{exc}{colback=block-gray,
boxrule=0pt,boxsep=0pt,breakable}

\ifniklas

\def\showkeysrefformat#1{{\normalfont\tiny\ttfamily#1}}
\makeatletter
\def\SK@@ref#1>#2\SK@{%
 {\@inlabelfalse\leavevmode\vbox to\z@{%
 \vss\SK@refcolor\rlap{\vrule\raise .75em%
  \hbox{\showkeysrefformat{#2}}}}}}
\makeatother
\fi

\usepackage[a4paper,text={160mm,247mm},centering]{geometry}

\allowdisplaybreaks[3]

\numberwithin{equation}{section}

\usepackage[font=small,labelfont=bf,width=0.85\textwidth]{caption}

\expandafter\def\expandafter\bfseries\expandafter{\bfseries\ifmmode\else\boldmath\fi}
\expandafter\def\expandafter\mdseries\expandafter{\mdseries\ifmmode\else\unboldmath\fi}
\expandafter\def\expandafter\normalfont\expandafter{\normalfont\ifmmode\else\unboldmath\fi}

\RequirePackage{verbatim}

\makeatletter
\newwrite\bibinl@out
\newenvironment{bibtex}[1][\jobname]{%
  \immediate\openout\bibinl@out #1.bib
  \immediate\write\bibinl@out{\@percentchar generated from `\jobname' starting line \the\inputlineno^^J}%
  \def\verbatim@processline{\immediate\write\bibinl@out{\the\verbatim@line}}%
  \@bsphack\let\do\@makeother\dospecials\catcode`\^^M\active\verbatim@start
}%
{\immediate\closeout\bibinl@out\@esphack}
\makeatother

\usepackage{fixmath}

\newcommand{\sfrac}[2]{{\textstyle\frac{#1}{#2}}}
\newcommand{\half}{\sfrac{1}{2}}
\newcommand{\ihalf}{\sfrac{i}{2}}

\newcommand{\alg}[1]{\mathfrak{#1}}

\RequirePackage{amsmath}
\makeatletter
\newcommand{\brk@ord}{\bBigg@{0}}
\newcommand{\brk@ordl}{\mathopen\brk@ord}
\newcommand{\brk@ordr}{\mathclose\brk@ord}
\newcommand{\brk@ordm}{\mathrel\brk@ord}
\newcommand{\brk@var}{\brk@ord}
\newcommand{\brk@varl}{\left}
\newcommand{\brk@varr}{\right}
\newcommand{\brk@varm}{\mathrel\brk@var}
\newcommand{\brk@altname}[3]{\expandafter\def\csname#2\expandafter\@gobble\string#1\endcsname{#1[#3]}}
\newcommand{\brk@usearg}[3]{%
  \def\brk@star{*}\def\brk@blank{}\def\brk@arg{#1}%
  \ifx\brk@arg\brk@blank\def\brk@arg{brk@ord}\fi%
  \ifx\brk@arg\brk@star\def\brk@arg{brk@var}\fi%
  \csname\brk@arg #2\endcsname#3}

\newcommand{\DeclareMathBrackets}[3]{
  \newcommand{#1}[2][]{\brk@usearg{##1}{l}{#2}##2\brk@usearg{##1}{r}{#3}}
  \brk@altname{#1}{big}{big}\brk@altname{#1}{lr}{*}}
\newcommand{\DeclareMathBiBrackets}[4]{
  \newcommand{#1}[3][]{\brk@usearg{##1}{l}{#2}##2#3##3\brk@usearg{##1}{r}{#4}}
  \brk@altname{#1}{big}{big}\brk@altname{#1}{lr}{*}}
\newcommand{\DeclareMathBiMBracketsStar}[4]{
  \newcommand{#1}[3][]{\brk@usearg{##1}{l}{#2}##2\brk@usearg{##1}{m}{#3}##3\brk@usearg{##1}{r}{#4}}
  \brk@altname{#1}{bi}{big}}
\newcommand{\DeclareMathBiBracketsStar}[4]{
  \newcommand{#1}[3][]{\brk@usearg{##1}{l}{#2}##2\brk@usearg{##1}{}{#3}##3\brk@usearg{##1}{r}{#4}}
  \brk@altname{#1}{big}{big}}

\makeatother

\DeclareMathBrackets{\brk}{(}{)}
\DeclareMathBrackets{\sbrk}{[}{]}
\DeclareMathBrackets{\set}{\{}{\}}
\DeclareMathBrackets{\abs}{|}{|}
\DeclareMathBrackets{\eval}{.}{|}
\DeclareMathBrackets{\spn}{\langle}{\rangle}
\DeclareMathBiBrackets{\comm}{[}{,}{]}
\DeclareMathBiBrackets{\acomm}{\{}{,}{\}}
\DeclareMathBiBrackets{\gcomm}{[}{,}{\}}

\DeclareMathOperator{\tr}{tr}

\def\[{\begin{equation}}
\def\]{\end{equation}}

\providecommand{\href}[2]{#2}

\makeatletter
\def\mr@ignsp#1 {\ifx\:#1\@empty\else #1\expandafter\mr@ignsp\fi}%
\newcommand{\multiref}[1]{\begingroup
\xdef\mr@no@sparg{\expandafter\mr@ignsp#1 \: }%
\def\mr@comma{}%
\@for\mr@refs:=\mr@no@sparg\do{\mr@comma\def\mr@comma{,}\ref{\mr@refs}}%
\endgroup}
\renewcommand{\eqref}[1]{(\multiref{#1})}
\makeatother

\makeatletter
\newcommand{\namedref}[2]{\hyperref[#2]{#1~\ref*{#2}}}
\newcommand{\secref}{\@ifstar{\namedref{Section}}{\namedref{Sec.}}}
\newcommand{\appref}{\@ifstar{\namedref{Appendix}}{\namedref{App.}}}
\newcommand{\tabref}{\@ifstar{\namedref{Table}}{\namedref{Tab.}}}
\newcommand{\figref}{\@ifstar{\namedref{Figure}}{\namedref{Fig.}}}
\makeatother

\providecommand{\hypersetup}[1]{}
\providecommand{\texorpdfstring}[2]{#1}

\hypersetup{plainpages=false}
\hypersetup{pdfpagemode=UseNone}
\hypersetup{bookmarksnumbered=true}
\hypersetup{pdfstartview=FitH}
\hypersetup{colorlinks=false}
\hypersetup{citebordercolor={.5 1 .5}}
\hypersetup{urlbordercolor={.5 1 1}}
\hypersetup{linkbordercolor={1 .7 .7}}

\def\ads{{\rm AdS}_5\times {\rm S}^5}
\def\lax{\mathcal{L}}

\newcommand{\de}{\operatorname{d}\!}
\newcommand{\e}{\operatorname{e}}
\newcommand{\eqndot}{\, . }
\newcommand{\eqncom}{\, , }

\newcommand{\MPS}{\mathrm{MPS}}
\newcommand{\Neel}{\mathrm{N\acute{e}el}}

\newcommand{\YM}{{\mathrm{\scriptscriptstyle YM}}}
\newcommand{\renZ}{\mathcal{Z}}

\DeclareMathOperator{\cder}{D}
\newcommand{\scal}{\phi}
\newcommand{\scalc}{\scal^{\text{cl}}}
\newcommand{\scalq}{\tilde{\scal}}
\newcommand{\ferm}{\psi}
\newcommand{\aferm}{\bar{\ferm}}

\newcommand{\cO}{\mathcal{O}}

\newcommand{\gammaE}{\gamma_{\text{E}}}
\newcommand{\peps}{\varepsilon}

\renewcommand{\digamma}{\Psi}
\newcommand{\ham}{\mathcal{H}}

\newcommand{\permop}{\mathbb{P}}

\makeatletter
\let\@keywords\@empty
\let\@subject\@empty
\providecommand{\keywords}[1]{\gdef\@keywords{#1}}
\providecommand{\subject}[1]{\gdef\@subject{#1}}
\def\thetitle{\@title}
\def\theauthor{\@author}
\def\thesubject{\@subject}
\def\thedate{\@date}
\def\thekeywords{\@keywords}
\makeatother
\AtBeginDocument{
\hypersetup{pdftitle={\thetitle}}%
\hypersetup{pdfauthor={\theauthor}}%
\hypersetup{pdfsubject={\thesubject}}%
\hypersetup{pdfkeywords={\thekeywords}}%
}

\title{One-point functions in AdS/dCFT}
\author{Marius de Leeuw}

\begin{document}

\tikzset{middlearrow/.style={
        decoration={markings,
            mark= at position 0.5 with {\arrow{#1}} ,
        },
        postaction={decorate}
    }
}

\tikzstyle{block} = [draw, fill=blue!20, rectangle, 
    minimum height=3em, minimum width=6em]
\pdfbookmark[1]{Title Page}{title}
\thispagestyle{empty}


\vspace*{2cm}
\begin{center}%
\begingroup\Large\bfseries\thetitle\par\endgroup
\vspace{1cm}

\begingroup\scshape\theauthor\par\endgroup
\vspace{5mm}%

\begingroup\itshape
School of Mathematics\\
Trinity College Dublin\\
Dublin, Ireland
\par\endgroup
\vspace{5mm}

\begingroup\ttfamily
mdeleeuw@maths.tcd.ie
\par\endgroup

\vfill

\textbf{Abstract}\vspace{5mm}

\begin{minipage}{12.7cm}
In this review we discuss recent advances in the computation of one-point functions in defect conformal field theories with holographic duals. 
We briefly review the appearance of integrable spin chains in $\mathcal{N}=4$ super Yang--Mills theory and reformulate the problem of computing 
one-point functions to determining overlaps between Bethe states and a Matrix Product State. We will then demonstrate how these overlaps
can be computed by determinant formulas. This work is based on lectures given at the Young Researchers Integrability School and Workshop 2018.
To appear in a special issue of J. Phys. A.

\end{minipage}

\vspace*{4cm}

\end{center}

\newpage

\tableofcontents

\section{Introduction}
\label{sec:intro}

Many exciting and interesting physical phenomena occur when non-trivial boundaries or interfaces are present in a system. Some important and well-known examples are the AdS/CFT correspondence and the quantum Hall effect. In the AdS/CFT correspondence \cite{Maldacena:1997re} the boundary of the anti-de Sitter space plays a pivotal part as the space on which the dual conformal field theory lives. For the quantum Hall effect the non-trivial edge effects play an important role. Unfortunately, while non-trivial boundary conditions can make the physics more interesting, they also make it hard to compute quantities like correlations functions in the corresponding models. 

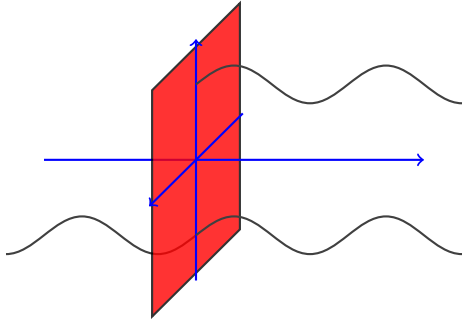
\begin{figure}[h]
\begin{center}
\scalebox{1}{
\begin{tikzpicture}
	[	axis/.style={->,blue,thick},
		axisline/.style={blue,thick},
		cube/.style={opacity=.8, thick,fill=red}]
	
	\draw[axisline] (-2,0,0) -- (0,0,0) node[anchor=west]{};	
        
	\draw[darkgray,thick] (0,-1,0) sin (-.5,-1.25,0)  (-.5,-1.25,0) cos (-1,-1,0)  (-1,-1,0) sin(-1.5,-.75,0)  (-1.5,-.75,0) cos(-2,-1,0) (-2,-1,0) sin (-2.5,-1.25,0)  ;
			
	\draw[cube] (0,-1.5,-1.5) -- (0,1.5,-1.5) -- (0,1.5,1.5) -- (0,-1.5,1.5) -- cycle;
		
	\draw[axis] (0,0,0) -- (3,0,0) ;
	\draw[axis] (0,-1.6,0) -- (0,1.6,0);
	\draw[axis] (0,0,-1.6) -- (0,0,1.6);
	
	\draw[darkgray, thick] (0,1,0) sin (.5,1.25,0)  (.5,1.25,0) cos (1,1,0)  (1,1,0) sin(1.5,.75,0)  (1.5,.75,0) cos(2,1,0) (2,1,0) sin (2.5,1.25,0)  (2.5,1.25,0) cos  (3,1,0) (3,1,0)sin(3.5,.75,0);
	\draw[darkgray, thick] (0,-1,0) sin (.5,-.75,0)  (.5,-.75,0) cos (1,-1,0)  (1,-1,0) sin(1.5,-1.25,0)  (1.5,-1.25,0) cos(2,-1,0) (2,-1,0) sin (2.5,-.75,0)  (2.5,-.75,0) cos  (3,-1,0) (3,-1,0)sin(3.5,-1.25,0);
		
\end{tikzpicture}
}
\caption{Schematic representation of a system with a codimension 1 defect. The defect acts as an interface through which some of the physical modes can propagate. The interface can also support non-trivial defect fields, which could couple to the
physical modes in the bulk.}
\end{center}
\end{figure}

In this review, we will focus on certain quantum field theories with a codimension one defect, see Figure 1. In such field theories, the defect acts as an interface through which some of the physical modes propagate, while others can not. The presence of an interface obviously impacts the correlation functions between fields. Of course, even without complicated boundary conditions, correlation functions in quantum field are very challenging to compute. Beyond one-loop level, computations of correlation functions tend to quickly become unmanageable and actual calculations to all loop orders are basically impossible. Nevertheless, there are some special field theories, such as $\mathcal{N}=4$ super Yang--Mills theory, where all-loop correlation functions are possible to derive. The feature that makes this possible is the presence of integrable structures. 

The notion of integrability originates from classical mechanics, where Liouville's theorem states that if a system has sufficiently many commuting, conserved quantities, then it is solvable by integration by quadratures. This means that the solution to the equations of motion can be constructed by solving a finite set of algebraic equations and performing a finite number of integrals. The notion of integrability later found its way into quantum mechanics and quantum field theory. While quantum integrable models do not really have a quantum analogue of Liouville's theorem, an exact solution can usually still be constructed by a using techniques such as the Bethe Ansatz.

For this reason, the discovery of integrable structures in the planar limit in some field theories from the AdS/CFT correspondence was such an important development. Over 15 years ago integrable spin chains were found in the computation of two-point functions  $\mathcal{N}=4$ super Yang--Mills theory \cite{Minahan:2002ve}. A lot has happened since and thanks to integrability we are now able to compute many quantities to any loop order and we are getting closer to finding an exact solution of $\mathcal{N}=4$ super Yang--Mills theory in the planar limit \cite{Beisert:2010jr}. 

Coming back to the problem of boundary conditions, it is clear that inserting a non-trivial boundary, or defect, will make the physics more interesting, but the correlation functions also much harder to compute. However, introducing a defect in a way which would preserve some measure of integrability could potentially allow us to still find exact solutions for the defect theory. Recently some new models of this type arose from holography. Certain insertions of probe branes on the string theory side of the AdS/CFT correspondence lead to a dual conformal field theory with a co-dimension one defect \cite{DeWolfe:2001pq,Karch:2000gx} (see also \cite{Semenoff:2018ffq} for a recent review). Several brane configurations can be considered, but two set-ups appear to be integrable \cite{deLeeuw:2018mkd}. These correspond to a D5-D3 and a D7-D3 probe brane configuration. In the D3-D5 defect CFT, closed formulas for one-point functions could indeed be derived by using methods from integrability as was first observed in for the SU(2) sector in \cite{deLeeuw:2015hxa,Buhl-Mortensen:2015gfd} and later extended to the general scalar sector \cite{deLeeuw:2018mkd}.

This closed formula was found by mapping the problem of computing one-point functions to the computation of quantum quenches in condensed matter physics. Because of this relation, formulas of the holographic one-point functions were used in quantum quench computations \cite{Mestyan:2017xyk}, while results from the condensed matter community found their way into the defect CFT and AdS/CFT community. This makes the topic of this review particularly dynamic and broad as it brings together topics such as, string theory, conformal field theory, and condensed matter physics.

The aim of these notes is to review in a pedagogical way the Bethe ansatz technqiues necessary for the computation of scalar products of spin-chain states and how those can be employed to compute one-point
functions in the context of AdS/CFT. This article will mainly focus on the integrability approach to the computation of one-point functions and it will include numerous recent developments.

This review is organized as follows. First we will discuss the effect of inserting a codimension one defect into a conformal field theory. We will indeed see that the insertion of the defect has a large impact on the form of correlation functions. In particular, a richer class of correlation functions is allowed. Second we will discuss how integrable spin chains arise in the computation of correlation functions in theories such as $\mathcal{N}=4$ SYM. We also review various defect set-ups in $\mathcal{N}=4$ SYM theory. After this we exploit the integrability to find exact solutions for the eigenvalues and eigenstates of the integrable spin chain. We will briefly review both the coordinate and algebraic Bethe Ansatz. In the next section we compute one-point functions and discuss their properties. At the end we give a quick overview of doing quantum computations. We end with some discussions and outlook.

\section{Correlation functions and defects}

In this section we will briefly discuss the form of correlation functions in conformal field theories and the effects of partially breaking conformal symmetry by the insertion of a codimension one defect.

\subsection{Correlation functions in Conformal Field theories}

The presence of conformal symmetry in a field theory is very powerful and restrictive. For example, looking at the one-point function of a scalar operator $\mathcal{O}$, we find from invariance under translations that they need to be constant. Scaling invariance then dictates that they vanish, \textit{i.e.}
\begin{align}
\langle \mathcal{O} \rangle = 0.
\end{align}
More generally, because of conformal symmetry, all correlation functions are fixed in terms of the conformal data $(\Delta,\lambda)$. The $\Delta$'s are the conformal dimensions of operators and the $\lambda$'s are called structure constants and describe three-point functions. 
Specifically, the space-time dependence of two-point functions is completely fixed by the scaling dimensions of the operators
\begin{equation}
\label{twopoint}
  \langle \mathcal{O}_i(x) \bar{\mathcal{O}}_j(y) \rangle = \frac{M_{ij}}{|x-y|^{\Delta_i + \Delta_j}}\eqncom
\end{equation}
where $M_{ij} = 0$ for $\Delta_i \neq \Delta_j$.
Conformal symmetry also fixes the three-point function up to the structure constant $\lambda_{i j k}$, which appears in the operator product expansion (OPE):
\begin{align}\label{eq:GeneralOPE}
\mathcal{O}_i(x) \mathcal{O}_j(y) = \frac{M_{ij}}{|x-y|^{\Delta_{i}+\Delta_{j}}} + \sum_k \frac{\lambda_{i j}{}^{k}} {|x-y|^{\Delta_{i}+\Delta_{j}-\Delta_k} }\,C(x-y,\partial_y)\mathcal{O}_k(y)\eqncom
\end{align}
where the sum over $k$ runs over conformal primary operators and the differential operator $C$ in \eqref{eq:GeneralOPE} accounts for the presence of conformal descendants. The indices on $\lambda$ can be raised and lowered with the matrix $M$. The normalisation of $C$ is such that 
$C(x-y,\partial_y)=1+O(x-y)$.
The scaling dimensions $\Delta_i$ and the structure constants $\lambda_{i j k }$ then determine all higher-point functions via repeated use of the OPE \eqref{eq:GeneralOPE}. 

\subsection{Defect CFTs}

A particularly interesting way to break conformal symmetry is by inserting a defect or interface in the theory. This leads to non-trivial boundary conditions, which break part of the conformal algebra. The class of defects that are the main topic of this work are codimension 1 defects in four-dimensional theories from AdS/CFT. Without loss of generality, we put our defect at the $x_3=0$ plane and consider correlation functions of operators on the right hand side, see Figure 2.

\begin{figure}[h]
\begin{center}
\scalebox{1}{
\begin{tikzpicture}
	[	axis/.style={->,blue,thick},
		axisline/.style={blue,thick},
		cube/.style={opacity=.8, thick,fill=red}]
	\draw[cube] (0,-1.5,-1.5) -- (0,1.5,-1.5) -- (0,1.5,1.5) -- (0,-1.5,1.5) -- cycle;
	
	\draw[axisline] (-2,0,0) -- (0,0,0) ;
        \draw[thick,fill] (2,0,1) circle (.05) node[right]{$\mathcal{O}_j(y)$};
        \draw[thick,fill] (2,0,-1) circle (.05) node[right]{$\mathcal{O}_i(x)$};
        \draw[thick,dotted, <->] (1.9,0,-1) -- (0.1,0,-1) node[pos=.4, above]{$x_3$};
        \draw[thick,dotted, <->] (1.9,0,1) -- (0.1,0,1) node[pos=.4, below]{$y_3$};

	\draw[axis] (0,0,0) -- (3,0,0) node[anchor=west]{$x_3$};
	\draw[axis] (0,-1.6,0) -- (0,1.6,0) node[anchor=south]{$x_0$};
	\draw[axis] (0,0,-1.6) -- (0,0,1.6) node[anchor=east]{$x_{1,2}$};	
	
		
\end{tikzpicture}
}
\caption{Representation of operator insertions in the bulk in a theory with a codimension one interface at $x_3=0$.}
\end{center}
\end{figure}
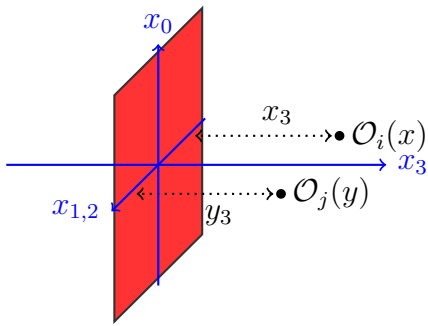

\paragraph{Correlation functions}
The correlation functions in a conformal field theory with a codimension one defect are less restricted than for a usual CFT. This is due to the fact that such a defect reduces the conformal symmetry to the conformal symmetry in one dimension lower. 

For instance, in contrast to a proper CFT, one-point functions of operators $\cO_i$ can be non-vanishing. One-point functions are now fixed up to a constant $a_i$ \cite{Cardy:1984bb}
\begin{equation}
\label{eq: form of the one-point function}
 \langle \mathcal{O}_i(x)\rangle=\frac{a_i}{x_3^{\Delta_{i}}}\eqncom
\end{equation}
where $x_3$ is the distance from the operator to the defect, see Figure 2.
Furthermore, two-point functions in a dCFT can be non-vanishing for operators of unequal scaling dimensions and are fixed to be of the form
\begin{align}\label{eq:TwoPoint}
\langle \mathcal{O}_i(x) \mathcal{O}_j(y) \rangle = \frac{f_{ij}(\xi)}{x_3^{\Delta_i}y_3^{\Delta_j}}\eqncom
\end{align}
where $f(\xi)$ is a function of the conformal ratio $\xi=\frac{|x -y|^2}{4 x_3 y_3}$. 

The spectrum of conformal dimensions for bulk operators is unchanged. This is because $\Delta$ is determined by the two-point function when the operators are taken close together, \textit{i.e.} the distance between the two operators is a lot smaller than the distance to the defect and the effect of the defect can be neglected. Moreover, if we fix the normalization for the two-point function such that, far away from the defect, the two-point function is unit-normalized $\lim_{\xi\rightarrow 0} \xi^{-\frac{\Delta_i+\Delta_j}{2}} f_{ij}  = \delta_{ij}$ then $a_{i}$ is uniquely defined. We find that $a_i$ is a part of the boundary conformal data that determines all correlation functions in the defect CFT.

\section{Spin chains from field theory}

In the remainder of this work, we will restrict to $\mathcal{N}=4$ SYM theory with gauge group $U(N)$. This conformal field theory naturally arises in holographic set-ups. There are many reviews and lecture notes that discuss various aspects of this field theory, see \textit{e.g.} \cite{DHoker:2002nbb, Beisert:2010jr, deLeeuw:2017cop}. In these lectures, the main focus will be on the appearance of integrable spin chains and in this section we will briefly discuss the basic features of $\mathcal{N}=4$ SYM theory relevant to this. In particular, we discuss how integrable spin chains describe the spectrum of conformal dimensions $\Delta$. The interested reader can find more details on the various computations in the references listed above.

\subsection{$\mathcal{N}=4$ super Yang--Mills theory}

\paragraph{Field content and Action}
The field content of $\mathcal{N}=4$ SYM theory consists of a four-dimensional gauge field $A_\mu$, six real scalars $\phi_i$ and four four-dimensional Majorana fermions $\psi$. The action of $\mathcal{N}=4$ SYM theory is given by
\begin{multline}
 \label{SYMaction}
  S_{{\mathcal N}=4}=\frac{2}{g_\YM^2}\int \de^4x\,\tr\biggl[ -\frac{1}{4}F_{\mu\nu}F^{\mu\nu}-\frac{1}{2}\cder_\mu\scal_i\cder^\mu\scal_i\\+\frac{i}{2}\aferm\Gamma^\mu\cder_\mu\ferm +\frac{1}{2}\aferm\Gamma^i\comm{\scal_i}{\ferm}+\frac{1}{4}\comm{\scal_i}{\scal_j}\comm{\scal_i}{\scal_j}\biggr]\eqncom
\end{multline}
where the field strength $F_{\mu\nu}$ and the covariant derivatives $\cder_\mu$ are defined in the usual way 
\begin{equation}
 \begin{aligned}
  & \hspace{.5cm}F_{\mu\nu}=\partial_\mu A_\nu-\partial_\nu A_\mu-i\comm{A_\mu}{A_\nu}\eqncom\\
  \cder_\mu\scal_i&=\partial_\mu\scal_i-i\comm{A_\mu}{\scal_i}\eqncom \hspace{0.5cm}
  \cder_\mu\psi=\partial_\mu\psi-i\comm{A_\mu}{\psi}\eqndot
 \end{aligned}
\end{equation}
This action can be obtained by reducing $\mathcal{N}=1$ SYM theory in ten dimensions down to four dimensions. The $\Gamma$ denotes the ten-dimensional gamma matrices which describe the coupling of the ten-dimensional fermion to the bosons. 

\paragraph{Operators}
All fields transform in the adjoint representation of the gauge group $U(N)$. We can build gauge-invariant local composite operators by taking traces of products of fields that transform covariantly under the gauge group, \textit{e.g.} $\mathcal{O} = \tr (\phi_1 \phi_2 F^2 \ldots)$. Moreover, we can take products of such single-trace operators to obtain multi-trace operators, such as $\tr (\phi_1\phi_2) \tr (\phi_3\phi_4)$.

We will work in the 't Hooft (or planar) limit, where $g_\YM\to 0$, $N\to\infty$ while the 't Hooft coupling $\lambda= g_\YM^2 N$ is kept fixed~\cite{tHooft:1973alw}. In this limit, only planar Feynman diagrams contribute to correlation functions. This means that interactions that lead to splitting and joining of traces are suppressed. Thus, it is sufficient to look at operators with a fixed number of traces, such as single-trace operators.

We will mainly restrict to a particular subset of operators comprised of two scalar operators. More precisely, we consider the two complex scalar fields $X$ and $Y$, defined by
\begin{align}
  &X = \phi_1 + i\phi_4, &&Y = \phi_2 + i\phi_5\eqncom
\end{align}
and consider single trace operators build out of these two fields. This subsector is called the SU(2) subsector and it is a closed sector to all loop orders.

\paragraph{Spectral problem}

The spectral problem is the problem of finding the spectrum of conformal dimensions $\Delta$. The quantum correction $\Delta-\Delta^{(0)}$ is called the anomalous dimension. For concreteness, let us consider single-trace operators with classical conformal dimension $\Delta^{(0)}=L$ in the SU(2) sector. These single trace operators form a vector space spanned by operators of the form
\begin{align}
\cO = \Psi^{S} \, \tr [\phi_{s_1} \ldots \phi_{s_L}]\eqncom
\label{eq:O-from-Psi}
\end{align}
where $s_i = \,\uparrow,\downarrow$ and $\phi_\uparrow=X,\phi_\downarrow = Y$. Due to cyclicity of the trace, the coefficient $\Psi^{S}$ is invariant under cyclic permutations $ \Psi^{\{s_1,s_2,\ldots,s_L\}} = \Psi^{\{s_L,s_1,s_2,\ldots,s_{L-1}\}} $. Clearly, $\Psi$ can be seen as a vector in $\bigotimes_L \mathbb{C}^2 = \mathbb{C}^{2^L}$ since each index $s_i$ takes two values. 

One can then explicitly compute the two-point function of two such operators to one-loop order in dimensional regularization and find \cite{deLeeuw:2017cop}
\begin{multline}
  \langle\cO_1(x)\bar{\cO}_2(y)\rangle_\peps = L\, 4^L g^{2L}\Bigg[\frac{\pi^\peps\mu^{2\peps}\Gamma(1-\peps)}{[(x-y)^2]^{1-\peps}}\Bigg]^{L} \\
  \times \langle \Psi_2 | 1
            -g^2\left(\frac{1}{\peps} + 1 + \gammaE +\log(\pi|x-y|^2)\right) \ham
            | \Psi_1 \rangle  +g^2O(\peps) + O(g^4)\eqndot
  \label{eq:SU2-two-point-spin-chain}
\end{multline}
Here, the inner product $\langle\Psi_2 | \Psi_1 \rangle = (\Psi^S_2)^\dag\Psi^S_1$ is the standard inner product on $\mathbb{C}^{2^L}$ and we have defined the coupling constant $g^2= \lambda/16\pi^2$. Moreover, $\ham$ is given by
\begin{align}\label{eq:HfromD}
  \ham = 2 \sum_{i=1}^L (1 - \mathbb{P}_{i,i+1})\eqncom 
\end{align}
where $\permop_{i,i+1}$ now denotes the operator which permutes two neighboring spins,
\begin{equation}
  (\permop_{i,i+1} \Psi)^{\{s_1,\ldots,s_L\}} = \Psi^{\{s_1,\ldots,s_{i-1},s_{i+1},s_{i},s_{i+2},\ldots,s_L\}}\eqndot
\end{equation}
In other words, we see that the conformal states are simply the eigenvectors of $\ham$ and the anomalous dimension of $\cO$ is then the corresponding eigenvalue. This maps the problem of finding the anomalous dimensions to the spectral problem of the so-called Heisenberg spin chain. 

\paragraph{Normalized operators} It is convenient to normalize our field theory operators such that the two-point functions are unit-normalized, \textit{i.e.} $M_{ij} = \delta_{ij}$ in \eqref{twopoint}. Let $|\mathbf{u}\rangle$ be an eigenvector of $\ham$ with corresponding eigenvalue $\Delta^{(1)}$. Then from \eqref{eq:SU2-two-point-spin-chain} we see that the explicit normalized operator is given by
\begin{equation}\label{eq:OviaU}
  \mathcal{O} \equiv  \left(\frac{1}{2g}\right)^{L}\frac{\mathcal{Z}}{\sqrt{L}}
    \frac{\tr\prod_{l=1}^L\Big(\langle\uparrow_l \!\! |\otimes X+\langle\downarrow_l \!\! |\otimes Y\Big) | \mathbf{u} \rangle }{\sqrt{ \langle \mathbf{u} | \mathbf{u} \rangle}}\eqndot
\end{equation}
The normalisation factor $\renZ$ can be derived from \eqref{eq:SU2-two-point-spin-chain}
\begin{equation}
  \renZ = 1 + g^2\frac{\Delta^{(1)}}{2}\left(\frac{1}{\peps} + 1 + \gammaE + \log\pi\right) + O(g^4) \eqndot
    \label{eq:Z-our-scheme}
\end{equation}
This follows by imposing that the only one-loop correction to the two-point function is $\propto \log (x-y)^2$.

\paragraph{The Heisenberg spin chain}

Let us now discuss the properties of the spin chain model that arose in \eqref{eq:SU2-two-point-spin-chain} in a bit more detail. The Heisenberg spin chain of length, consists of $L$ sites, with on each site a spin-$\half$ particle. The particle at site $i$ can have spin up or down, so it generates a local two-dimensional Hilbert space $V_i = \mathbb{C}^2$. The total Hilbert space $H$ of the spin chain is simply
\begin{align}
H = \bigotimes_{i=1}^L V_i = \mathbb{C}^{2^L}.
\end{align}
Let us also introduce the usual spin operators $S^{x,y,z}_i$. They act locally on site $i$ and hence they satisfy commutation relations 
\begin{align}
[S^a_i,S^b_j] = \delta_{ij}\epsilon^{abc}S^c_i.
\end{align}
It is also convenient to consider the raising and lowering operators $S^\pm = S^x \pm i S^y$, such that
\begin{align}
&S^+|\!\uparrow\rangle =0, &&S^-|\!\uparrow\rangle =|\!\downarrow\rangle , &&S^z|\!\uparrow\rangle =\half|\!\uparrow\rangle\nonumber\\
&S^+|\!\downarrow\rangle =|\!\uparrow\rangle, &&S^-|\!\downarrow\rangle =0 , &&S^z|\!\downarrow\rangle =-\half|\!\downarrow\rangle.
\end{align}
We can realize the spin operators in the usual way via the Pauli matrices $ \vec{S} = \frac{1}{2}(\sigma_1,\sigma_2,\sigma_3)$. It is easy to check that the permutation operator can also be expressed in terms of Pauli matrices as
$P = \half(1+ \sigma^a\otimes \sigma^a)$.

Thus we can express the spin chain Hamiltonian \eqref{eq:HfromD} in terms of a nearest neighbor spin-spin interaction. More precisely, we write the Hamiltonian in terms of the Hamiltonian density $\mathcal{H}_{i,i+1}$
\begin{align}\label{eq:Ham}
& \mathcal{H} =  \sum_{i=1}^L \mathcal{H}_{i,i+1}, &&
\mathcal{H}_{i,i+1} = 1 - 4 \, \vec{S}_{i} \cdot \vec{S}_{i+1}.
\end{align}
Depending on the overall sign of the Hamiltonian, the spins in the chain want to either align or anti-align. Hence this is a rudimentary model of (ferro)magnetism. We will restrict to periodic boundary conditions by identifying $\vec{S}_{L+1} = \vec{S}_{1}$. There are more general boundary conditions that are compatible with integrability, described by so-called reflection matrices $K$ \cite{Sklyanin}.

\paragraph{Symmetries} Let us also look at the symmetries of the spin chain. Consider the operator corresponding to the total spin in the $z$ direction
\begin{align}
S^z = \sum_i S^z_i\eqndot
\end{align}
It measures the difference between the total number of up and down spins.

Since $[S^z,\mathcal{H}]=0$, we can restrict to states where the total number of up and down spins is fixed. Moreover, we also know that the sum of up and down spins should sum to $L$. This means that we can restrict to subsets of a fixed number of spins up (and down) and that we can label eigenstates by two numbers $L$, the length of the spin chain, and $M$ the number of spins down.

Similarly, it is easy to check that all the total spin operators $\vec{S} = \sum_i \vec{S}_i$ commute with the Hamiltonian. The spin operators form an SU(2) algebra and consequently this spin chain has SU(2) as part of its symmetry algebra. For this reason, the Heisenberg spin chain is sometimes referred to as the SU(2) spin chain. This means that the eigenstates of the Hamiltonian will arrange themselves into SU(2) multiplets.

\paragraph{More general sectors} In this section we restricted our discussion to the SU(2) subsector. When one considers the complete spectrum anomalous dimensions, more general integrable spin chains appear. For instance, when restricting to all scalar fields at one-loop level, the spectrum is described by an integrable SO(6) spin chain \cite{Minahan:2002ve}. At higher loop levels, the integrable structures become more intricate as the interaction range of the Hamiltonian increases.

\subsection{\texorpdfstring{$\mathcal{N}=4$}{N=4} SYM theory with a defect}

Via the AdS/CFT correspondence, $\mathcal{N}=4$ SYM theory is dual to type IIB superstring theory on $\ads$. From a string theory perspective there are three string theory set-ups which are dual to a co-dimension one defect versions of $\mathcal{N}=4$ SYM theory. These set-ups correspond to the insertion of a D5 or D7 probe brane in the $\ads$ space. These branes will intersect the usual stacks of D3 branes that underly the AdS/CFT conjecture. The number of D3 branes on both sides of the probe brane can differ, resulting in a new parameter $k$, which corresponds to a non-zero background flux. How exactly these branes intersect then determines the exact form of the defect field theory and how it depends on the parameter $k$. Some quantities have been computed on the string theory side of the duality by a supergravity computation in the so-called double scaling limit
\begin{align}
&k\rightarrow\infty,
&&\lambda\rightarrow\infty,
&& \frac{\lambda}{k^2} = \mathrm{fixed}.
\end{align}
In order to compare any computations we will do in the field theory with string theory predictions, we see that we need to understand the large $k$ limit.  
In what follows we will only work on the field theory side of the duality, so let us list the three defect field theory set-ups and describe their properties.

\paragraph{General properties}

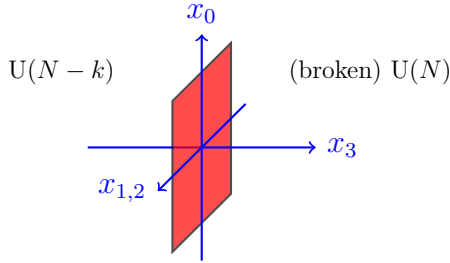
\begin{figure}[h]
\begin{center}
\scalebox{1}{
\begin{tikzpicture}
	[	axis/.style={->,blue,thick},
		axisline/.style={blue,thick},
		cube/.style={opacity=.7, thick,fill=red}]

	\draw[axisline] (-1.5,0,0) -- (0,0,0) node[anchor=west]{};	
		
	\draw[cube] (0,-1,-1) -- (0,1,-1) -- (0,1,1) -- (0,-1,1) -- cycle;

	\draw[axis] (0,0,0) -- (1.5,0,0) node[anchor=west]{$x_3$};
	\draw[axis] (0,-1.5,0) -- (0,1.5,0) node[anchor=south]{$x_0$};
	\draw[axis] (0,0,-1.5) -- (0,0,1.5) node[anchor=east]{$x_{1,2}$};	
		
	\node[anchor=east] at (-1,1,0) {\scalebox{0.8}{$\mathrm{U}(N-k)$}};
	\node[anchor=west] at (1,1,0) {\scalebox{0.8}{(broken) $\mathrm{U}(N)$}};
	
\end{tikzpicture}
}
\caption{The co-dimension 1 defect.}\label{fig:dCFT}
\end{center}
\end{figure}

In all defect versions of $\mathcal{N}=4$ SYM theory, a codimension-one defect/interface is inserted at $x_3=0$ and divides space into two regions, $x_3>0$ and $x_3<0$. On both sides of the defect we consider $\mathcal{N}=4$ SYM theory but with different gauge groups. The gauge group for $x_3<0$ is  $U(N-k)$, while the gauge group for $x_3>0$ is $U(N)$, see Figure \ref{fig:dCFT}. The $U(N)$  symmetry for $x_3>0$ is broken by some of the scalar fields acquiring a non-zero vacuum expectation value so that the gauge symmetry there also effectively is $U(N-k)$. In addition to the usual action of $\mathcal{N}=4 $ SYM theory, the system also has a three-dimensional action involving fields that are confined to the defect. In this review, however, since we will focus on correlation functions of bulk operators up to one-loop, the defect action does not play a role. 

The $U(N)$ symmetry for $x_3>0$ is broken by assigning a vacuum expectation value for the scalar fields $\phi_i$. The explicit form of the vev of the scalar fields depends on the model.
\begin{itemize}
\item The D3-D5 defect is parameterized by SU(2) representations \cite{Constable:1999ac}. More precisely,
for $x_3>0$
\begin{align}\label{eq:vev}
&\phi _i^{\rm cl} = -
 \frac{(t_i)_{k\times k} \oplus 0_{(N-k)\times (N-k)}}{x_3}
&\phi ^{\rm cl}_{4,5,6}=0\eqncom&& i=1,2,3\eqncom
\end{align}
where the $k\times k$ matrices $t_{1,2,3}$ form a $k$-dimensional unitary, irreducible representation of SU(2)
\begin{equation}\label{eq:su2relations}
 \left[t_i,t_j\right]=i\varepsilon _{ijk}t_k\eqndot
\end{equation}
For $x_3<0$, all classical fields are vanishing.

\item The D3-D7 $\alg{su}(2)\oplus\alg{su}(2)$ defect solution is parameterized by two SU(2) representations \cite{Kristjansen:2012tn}. More precisely,
for $x_3>0$
\begin{align}
&\phi _{1,2,3}^{\rm cl} = -
 \frac{(t_{1,2,3})_{k_1\times k_1} \oplus 0}{x_3}
&\phi ^{\rm cl}_{4,5,6}=-
 \frac{(t_{1,2,3})_{k_2\times k_2} \oplus 0}{x_3}\eqncom \eqncom
\end{align}
where the matrices $t_{1,2,3}$ again form a unitary, irreducible representation of SU(2).
\item For the D3-D7 $\alg{s0}(5)$ set-up, the classical fields take the values~\cite{Constable:2001ag}
\begin{eqnarray} 
\phi_i ^{\text{cl}}= \frac{G_i \oplus 0_{N-k}}{\sqrt{8}\,x_3}, \quad i = 1,\ldots,5, \qquad \phi_6^{\text{cl}} = 0, \hspace{0.5cm} x_3>0,  \label{Solution1}
\end{eqnarray}
whereas they vanish for $x_3<0$. Here,
the $G_i$ are matrices whose commutators generate a $k$-dimensional irreducible representation of $\mathfrak{so}(5)$. Such matrices can be constructed starting from the four-dimensional gamma matrices~\cite{Castelino:1997rv,deLeeuw:2016ofj}.
\end{itemize}
In the remainder of this review, we will focus on the D3-D5 set-up, as it is the most simple and most developed. We will comment on the other set-ups in Section \ref{sec:Disc}.

\subsection{The D3-D5 defect}

\paragraph{Representation of \texorpdfstring{SU(2)}{SU(2)}} In order to describe the D3-D5 vacuum expectation values, we need the explicit form of the $k$-dimensional irreducible representation. Let $E^i{}_j$  be the standard $k\times k$ matrix unities $E^i{}_j$ that are zero everywhere except for a 1 at position $(i,j)$. These matrices satisfy the relation $E^i{}_{j}E^k{}_{l} = \delta^k{}_j E^i{}_l$.

Then we define the usual spin raising and lowering matrices
\begin{align}\label{eq:tgenerators}
& t_+= \sum_{i=1}^{k-1} \sqrt{i(k-i)}E^i{}_{i+1}\eqncom
&& t_- =\sum_{i=1}^{k-1}  \sqrt{i(k-i)} E^{i+1}{}_{i}\eqndot
\end{align}
The $k$-dimensional SU(2) representation is then given by
\begin{align}\label{eq:defrep}
&t_1 = \frac{t_+ +t_-}{2}\eqncom
&&t_2 = \frac{t_+ -t_-}{2i}\eqncom
&& t_3 =\sum_{i=1}^k  \frac{1}{2}(k-2i+1)E^{i}{}_{i}\eqndot
\end{align}
It is easy to check that these matrices satisfy the commutation relations \eqref{eq:su2relations}. 
For the special case $k=2$, the representation matrices are multiples of the Pauli matrices: $t_i|_{k=2}=\frac{1}{2}\sigma_i$.

\paragraph{Spin chains} 
We are interested in computing one-point functions of conformal operators in the SU(2) sector at tree-level. As we saw in the previous section, these conformal operators correspond to the eigenvectors of the Heisenberg Hamiltonian \eqref{eq:HfromD}. 

At tree level, the one-point function of an operator is simply obtained by inserting the classical solution \eqref{eq:vev}. Clearly, only operators consisting solely of scalar operators can have a non-zero one-point function. This means that we restrict to single-trace operators of the form
\begin{align}
\mathcal{O} = \Psi^{i_1\ldots i_L} \tr( \scal_{i_1} \ldots \scal_{i_L})\eqndot
\end{align}
Inserting the vev $\phi _i^{\rm cl} = -t_i/x_3$ from \eqref{eq:vev} into such an operator $\cO$ then gives us at tree level
\begin{align}\label{eq:vevO}
\langle \mathcal{O} \rangle^{cl} = (-1)^L\Psi^{i_1\ldots i_L}\frac{ \tr( t_{i_1} \ldots t_{i_L})}{x_3^L}\eqndot
\end{align}
For any explicitly given operator, the above expression can straightforwardly be evaluated, but this is hardly a constructive approach. Note that the one-point function depends on the defect parameter $k$ through the matrices $t_i$, which form a $k$-dimensional representation of SU(2). We will keep the dependence on $k$ implicit.

The first step to a more systematic approach of computing one-point functions is to reformulate the computation of a one-point function in the spin chain language. This is done by noting  that \eqref{eq:vevO} can be written as an inner product between the state $|\mathbf{u}\rangle$ corresponding to our operator $\mathcal{O}$ via \eqref{eq:OviaU} and a so-called matrix product state (MPS) \cite{deLeeuw:2015hxa}:
\begin{align}\label{eq:MPS}
&& |\MPS\rangle_{k,L}= \tr \prod_{n=1}^L  \Big[t_1 \otimes |\!\uparrow\rangle_n  +  t_2 \otimes|\!\downarrow\rangle_n\Big]:=
\sum_{\vec{i}} \tr[ t_{i_1}\ldots t_{i_L}] \,|e_{i_1}\ldots e_{i_L}\rangle\eqncom
\end{align}
where $e_1 = \uparrow, e_2 =\downarrow$
The subscript $n$ stands for the usual embedding in the $L$-fold tensor product, while the matrices $t_i$ and the trace are in colour space.  The MPS depends on the length $L$ of the spin chain and the dimension of the representation $k$. In order to keep the notation light, we will from now on omit the subscripts $k,L$. In particular, in the spin chain picture, the information from the defect is encoded in the MPS through the matrices $t$. These matrices encode the breaking of the gauge symmetry due to the defect by their algebra relations and the flux parameter $k$ which describes their dimensions.

\noindent
 
Using the explicit relation between the Bethe states and the unit normalized field-theory operators  \eqref{eq:OviaU},
the problem of computing a one-point function then reduces to computing the following quantity
\begin{align}\label{eq:CK}
&\langle\mathcal{O}\rangle_k^{cl} = (-1)^L\left(\frac{1}{2g}\right)^{L}\frac{1}{\sqrt{L}} \frac{C_k}{x_3^L}\eqncom
&&C_k = \frac{\langle\MPS|\mathbf{u} \rangle}{\sqrt{\langle\mathbf{u} |\mathbf{u} \rangle}}\eqndot
\end{align}
In other words, we have mapped the problem of computing a one-point function to computing an overlap on the Heisenberg spin chain.

Finally, note that there is the important subtlety that $C_k$ is only defined up to a phase. In our identification of the field-theory operator and spin chain state $|\mathbf{u}\rangle$, we can always insert an additional phase factor. This obviously leaves the two-point function invariant, but it will affect the overlap with the MPS. In order to fix this ambiguity, we will always choose the overall phase such that $C_k$ is real and positive.

\section{Solving the Heisenberg spin chain}

Now that we have successfully mapped the problem of computing one- and two-point functions in the SU(2) sector to problems on the Heisenberg spin chain, we can make use of the powerful techniques to compute those correlation functions. The Heisenberg spin chain was first solved in 1931 H. Bethe \cite{Bethe1931}. Bethe introduced an Ansatz for the eigenvectors of the Heisenberg Hamiltonian \eqref{eq:Ham}, which allowed him to find the eigenvectors and eigenvalues of the Hamiltonian. This way of solving a spin chain model is usually referred to as the coordinate Bethe Ansatz. In the 80s the Leningrad school developed an alternative method of solving the spin chains which is rooted in algebra. This method goes under the name of the algebraic Bethe Ansatz or the quantum inverse scattering method \cite{Faddeev:1996iy}. 

Both approaches have their merits. The coordinate Ansatz is very explicit in terms of eigenvectors, which makes it ideally suited for the computation of one-point functions. On the other hand, a lot of useful properties and formulas are more readily derived in terms of the algebraic Bethe Ansatz. In this section we will briefly review both approaches. Both versions of the Bethe Ansatz have various reviews (\textit{e.g.} \cite{Levkovich-Maslyuk:2016kfv} ) devoted to them and we refer the reader to those for more details. 

\subsection{Coordinate Bethe Ansatz}\label{sec:CBA}

The main idea behind Bethe's Ansatz is to consider spins down as quasi-particles, called magnons, propagating on a ground state with all spins up. These magnons will move along the spin chain with a certain momentum $p$, which is quantized by the Bethe equations.

\paragraph{Vacuum} The Heisenberg Hamiltonian \eqref{eq:Ham} exhibits SU(2) symmetry. This implies in particular that the total spin in the $z$-direction is preserved. Hence, the state where all spins are pointing in the same direction is necessarily an eigenstate of the Hamiltonian. This is the ferromagnetic vacuum
\begin{align}\label{eq:DefVac}
|0\rangle = |\!\uparrow\uparrow\ldots\uparrow\uparrow\rangle.
\end{align}
It is easy to check that  $\mathcal{H}|0\rangle = 0$.

\paragraph{Magnons} The next step is to consider an eigenstate where some of the spins are flipped. In general, a spin at position $n$ can be flipped down by acting with the local lowering operator $S^-_n$. In this way, we can write
\begin{align}
 S^-_{n_1}\ldots S^-_{n_M}|0\rangle = | \ldots \downarrow_{n_1} \ldots \downarrow_{n_M}\ldots\rangle .
\end{align}
An eigenstate with $M$ flipped spins is then of the form
\begin{align}
|\psi\rangle = \sum_{1\leq n_1<\ldots<n_M\leq N} a(n_1,\ldots,n_M) \,  S^-_{n_1}\ldots S^-_{n_M}|0\rangle,
\end{align}
with some coefficients $a(n_1,\ldots,n_M)$ that are to be determined. Furthermore, the periodic boundary condition of our model is now formulated as
\begin{align}
a(n_2,\ldots,n_M,n_1+L) = a(n_1,\ldots,n_M).
\end{align}
The Bethe Ansatz postulates the form of these coefficients to be
\begin{align}\label{eq:BAgeneral}
a(n_1,\ldots,n_M) = \sum_{\sigma\in S_M} A_\sigma(p_1,\ldots,p_M) e^{ip_{\sigma_i}n_i},
\end{align}
where the $p_i$ are complex numbers and the sum runs over all permutations of length $M$. This is a plane-wave type Ansatz where the $p_i$ play the role of momenta. It is also useful to introduce a phase $\theta$ such that $A_\sigma = e^{\ihalf\theta_\sigma}$.

\paragraph{A single magnon} Let us first consider the case where only one of the spins in the ground state is flipped. Let us see how the Hamiltonian acts on such a term
\begin{align}\label{eqn:Hon1spin}
\mathcal{H}|\ldots\downarrow_n\ldots\rangle =  4|\ldots\downarrow_n\ldots\rangle - 2|\ldots\downarrow_{n-1}\ldots\rangle - 2 |\ldots\downarrow_{n+1}\ldots\rangle .
\end{align}
We see that the flipped spin can hop between sites or remain on the same site.

Let us now consider the flipped spin as a quasi-particle with momentum $p$ following the Bethe Ansatz \eqref{eq:BAgeneral}
\begin{align}\label{eq:BA1magnon}
|p\rangle = \sum_{n=1}^L e^{ipn}S_n^-|0\rangle,
\end{align}
which is basically the discrete version of a plane-wave. 

For the moment let us ignore the boundary conditions and look at one term in \eqref{eq:BA1magnon}. From \eqref{eqn:Hon1spin} we see that the action of the Hamiltonian results in
\begin{align}
\mathcal{H}|p\rangle = \ldots 2 \left[2 e^{ipn} - e^{ip(n+1)} - e^{ip(n-1)}\right] |\ldots\downarrow_n\ldots\rangle  \ldots
\end{align}
Thus, we see that $|p\rangle$ is an eigenstate of the Hamiltonian and its corresponding energy is
\begin{align}
E = 4 \sin^2 \frac{p}{2}.
\end{align}
It is simply a particle with momentum $p$ and energy $E$ moving in the vacuum state. 

Imposing periodic boundary conditions should imply a quantization condition on the momentum. This can for example be seen by acting with the Hamiltonian on the $L$th site. Instead of getting a contribution from $n=L+1$ we get a contribution from $n=1$. We find that \eqref{eq:BA1magnon} is an eigenstate with eigenvalue $E$ provided that $a(n+L)=a(n)$, or
\begin{align}
e^{ipL} = 1.
\end{align}
This is the usual momentum quantization condition for a particle on a circle of length $L$. 

\paragraph{Two magnons.} In the case of two magnons, a new situation arises. If the two magnons are separated by two sites or more, they behave as two single magnons since the action of the Hamiltonian only affects nearest neighbors. However, when the two excitations are on neighboring sites, there is a contact term.

According to the Bethe Ansatz \eqref{eq:BAgeneral}, the wave function takes the form
\begin{align}\label{eq:twoMagnon}
|p_1,p_2\rangle = \sum_{n_1<n_2}\left[A_{12} e^{i(p_1n_1+p_2n_2)}+A_{21} e^{i(p_2n_1+p_1n_2)}\right]S_{n_1}^-S_{n_2}^-|0\rangle.
\end{align}
Of course, the eigenstate can always be normalized in any way we like, so only the ratio $\mathcal{A} = A_{21}/A_{12}$ has physical meaning. 

We can interpret the terms in wave function \eqref{eq:twoMagnon}  as two magnons with momenta $p_1,p_2$ at positions $n_1,n_2$. The wave function is divided into two parts; the first magnon can be either to the left or the right of the second magnon. The two different regions are connected by the constant $\mathcal{A}$ which describes the scattering when the two magnons passing each other.

Let us now look in detail on how the Hamiltonian acts on this state. Firstly, when $n_1>n_2+1$, the wave function clearly behaves like two disjoint copies of the one magnon eigenstate. From this we can immediately read off that, if this state is an eigenstate, then the eigenvalue is necessarily given by
\begin{align}
\mathcal{H}|p_1,p_2\rangle \stackrel{!}{=} (E(p_1) + E(p_2))|p_1,p_2\rangle.
\end{align}
This reinforces the idea that we have quasi-particles on the spin chain whose energies are additive.

Now consider the case when the magnons are neighbors, say at position $n$ and $n+1$.  It is easy to see that the only terms that can be mapped to this term by the Hamiltonian are $|\ldots \uparrow\downarrow\downarrow\uparrow\ldots\rangle, |\ldots \downarrow\uparrow\downarrow\uparrow\ldots\rangle$ and $|\ldots \uparrow\downarrow\uparrow\downarrow\ldots\rangle$. Demanding that \eqref{eq:BAgeneral} is an eigenstate, then fixes $\mathcal{A}$. More precisely, from the explicit form of the Bethe ansatz we derive
\begin{align}
-\frac{1}{2}A_{12}\left[e^{ip_1(n-1) + ip_2(n+1)}-2e^{ip_1 n + ip_2(n+1)}+e^{ip_1 n + ip_2(n+2)}\right] -\frac{g^2}{2} A_{21} \left[p_1\leftrightarrow p_2\right] =\nonumber\\
(E(p_1) + E(p_2))(A_{12}e^{ip_1 n + ip_2(n+1)} + A_{21} e^{ip_2 n + ip_1(n+1)}).
\end{align}
From this it follows that the two magnon wave function is an eigenstate of the Hamiltonian exactly if
\begin{align}
\mathcal{A}(p_1,p_2)\equiv e^{\ihalf(\theta_{21}-\theta_{12})}= \frac{\cot\frac{p_1}{2}-\cot\frac{p_2}{2}-2i}{\cot\frac{p_1}{2}-\cot\frac{p_2}{2}+2i}.
\end{align}

It is easy to see that $\mathcal{A}$ is a pure phase if $p_{1,2}$ are real. We call $\mathcal{A}$ the scattering matrix (S-matrix). The S-matrix satifies the following condition
\begin{align}
\mathcal{A}(p_1,p_2)\mathcal{A}(p_2,p_1)=1.
\end{align}
In our derivation we have again neglected the boundary conditions. Periodicity implies
\begin{align}
&e^{ip_1L} = \mathcal{A}(p_2,p_1),&& e^{ip_2L} = \mathcal{A}(p_1,p_2).
\end{align}
These are called the Bethe equations and they are  the equivalent of the quantization condition of particles on a circle. 

The Bethe equations can be interpreted as follows. Take one of the particles and move it around the circle. When it encounters the other particle, it picks up a scattering phase. Then, when the particle is back at its original position, the wave function is unchanged up to a phase.

\paragraph{Multiple magnons.} The fact that this spin chain is integrable manifests itself when adding more magnons. It turns out that just knowing the dispersion relation and the two-magnon scattering matrix $\mathcal{A}$ is enough to construct all eigenvectors.

For instance, for the case of three magnons, the Bethe Ansatz \eqref{eq:BAgeneral} includes six coefficients $A_{123},A_{132},A_{321},\ldots$.  However, all of these coefficients can be expressed in terms of $\mathcal{A}$ that we encountered in the two-magnon case. More precisely, one finds that $A_\sigma = e^{ \ihalf \sum_{i<j}\theta_{\sigma_i\sigma_j}}$. 

The physical picture behind this decomposition of $A_\sigma$ is that magnons always scatter pairwise. You can interpret for instance $A_{321}/A_{123}$ as the three magnon scattering process $123 \rightarrow 321$. The above decomposition means that this process factorizes into three processes where first magnons 1 and 2 scatter, then magnons 1 and 3 and then magnons 2 and 3. This is the defining property of integrable systems; any scattering process factorizes into two-particle scatterings. 

In general, the eigenvectors of the Hamiltonian take the form
\begin{align}
|\vec{p}\rangle = |p_1,\ldots,p_M\rangle = \sum_{n_1<n_2<\ldots}a(n_1,\ldots,n_M)S^-_{n_1}\ldots S^-_{n_M}|0\rangle.
\end{align}
with the specific choice
\begin{align}\label{eq:genCBA}
a(n_1,\ldots,n_M) = \sum_{\sigma\in S_M} e^{ip_{\sigma_i}n_i + \ihalf \sum_{i<j}\theta_{\sigma_i\sigma_j}}.
\end{align}
Thus, all the relevant data of the scattering and spectrum is fully encoded in the dispersion relation and the two-body scattering. 

\paragraph{Bethe equations.} Now that we know the full eigenvectors, let us write down the corresponding Bethe equations that arise from the periodicity conditions
\begin{align}
e^{ip_iL} = \prod_{j\neq i} \mathcal{A}(p_j,p_i).
\end{align}
The spectrum can then be found by solving the Bethe equations and summing the energies of the different magnons. 

\subsection{Monodromy and R-matrix}

In this section we discuss a different approach to study the Heisenberg spin chain. This method goes under the name of algebraic Bethe Ansatz or quantum inverse scattering method \cite{Faddeev:1996iy}. The main idea is to reformulate the eigenvalue problem in a more algebraic setting. 

\paragraph{Lax operator.} The basic tool of the algebraic Bethe Ansatz approach is the so-called Lax operator $\lax$. Consider a chain with $L$ sites and corresponding Hilbert space $H= \bigotimes_i V_i$. The spaces $V_i\sim \mathbb{C}^2$ are called quantum spaces. To this chain we add an additional, auxiliary space $V_a$, which in this case will again simply be $\mathbb{C}^2$. This  can be interpreted as adding a test particle to the chain. The Lax operator is an operator $\lax_{i,a}:V_i\otimes V_a \rightarrow V_i\otimes V_a$ and for the Heisenberg model it is of the form
\begin{align}
\lax_{n,a}(u) = u\, 1\otimes1+ i S^a_n\otimes\sigma^a,
\end{align}
where $\sigma^\alpha$ are the usual Pauli matrices
\begin{align}
&\sigma^x = \begin{pmatrix} 0 & 1 \\ 1 & 0 \end{pmatrix},
&&\sigma^y = \begin{pmatrix} 0 & -i \\ i & 0 \end{pmatrix},
&&\sigma^z = \begin{pmatrix} 1 & 0 \\ 0 & -1 \end{pmatrix}.
\end{align}
They are related to the spin operators as $S^a = \half\sigma^a$. We can view the Lax operator $\lax$ as a matrix on the auxiliary space with entries operators acting in the quantum space
\begin{align}\label{eq:LviaS}
\lax_{n,a} = \begin{pmatrix}
u + i S^z_n & i S^-_n \\
i S^+_n & u - i S^z_n \\
\end{pmatrix}.
\end{align}
The next ingredient we will need are the commutation relations between the entries of the Lax operator. Let us introduce the permutation operator
\begin{align}\label{defP}
P = \half(1+\sigma^a\otimes\sigma^a) = 
\begin{pmatrix}
1 & 0 & 0 & 0 \\
0 & 0 & 1 & 0 \\
0 & 1 & 0 & 0 \\
0 & 0 & 0 & 1 
\end{pmatrix}.
\end{align}
The operator $P$ is called the permutation operator because it acts like a permutation on tensors
\begin{align}
P (x\otimes y) = y\otimes x.
\end{align}
It is then straightforward to check that the commutation relations of the entries of the Lax matrix can be written in the following compact form
\begin{align}\label{eqn:FCR}
R_{ab}(u_1-u_2)\lax_{i,a}(u_1)\lax_{i,b}(u_2)=
\lax_{i,b}(u_2)\lax_{i,a}(u_1)R_{ab}(u_1-u_2),
\end{align}
where the operator $R$ is called the quantum R-matrix and it has the following form
\begin{align}\label{eq:RXXX}
R_{ab}(\lambda) = \lambda 1 + i P_{ab}.
\end{align}
The commutation relations \eqref{eqn:FCR} are called the fundamental commutation relations. 

The fundamental commutation relations imply that the R-matrix needs to satisfy a consistency condition called the quantum Yang-Baxter equation. By using the fundamental commutation relations, we can rewrite the product of three Lax operators in two different ways. More precisely,
\begin{align}
\lax_1\lax_2\lax_3 = R_{12}^{-1}\lax_2\lax_1\lax_3R_{12} = 
R_{12}^{-1}R_{13}^{-1}\lax_2\lax_3\lax_1R_{13} R_{12} =
R_{12}^{-1}R_{13}^{-1}R_{23}^{-1}\lax_3\lax_2\lax_1R_{23}R_{13} R_{12},
\end{align}
while we can also write
\begin{align}
\lax_1\lax_2\lax_3 = R_{23}^{-1}\lax_1\lax_3\lax_2R_{23} = 
R_{23}^{-1}R_{13}^{-1}\lax_3\lax_1\lax_2R_{13} R_{23} =
R_{23}^{-1}R_{13}^{-1}R_{12}^{-1}\lax_3\lax_2\lax_1R_{12}R_{13} R_{23}.
\end{align}
For both of these rewritings to coincide, one has to impose that the R-matrix satisfies the equation
\begin{align}\label{eq:YBEgen}
R_{12}R_{13} R_{23} = R_{23}R_{13} R_{12}.
\end{align} 
So, in order for the fundamental commutation relations to form a consistent set of commutation relations, the R-matrix needs to satisfy the Yang-Baxter equation \eqref{eq:YBEgen}.

\paragraph{Monodromy and transfer matrix.} From the Lax operator, we define the monodromy matrix $T$
\begin{align}\label{eq:TviaL}
T_a = \lax_{L,a}(u)\ldots \lax_{1,a}(u).
\end{align}
It again can be seen as a two by two matrix acting on the auxiliary space whose entries are operators that act on the physical Hilbert space
\begin{align}
T(u) = \begin{pmatrix} A(u) & B(u) \\ C(u) & D(u) \end{pmatrix}.
\end{align}
First of all, we need the commutation relations between the different entries of the monodromy matrix. This can be found by making repeated use of the fundamental commutation relations for the Lax operator. We find
\begin{align}\label{eq:FCR}
R_{ab}(u_1-u_2) T_a(u_1) T_b(u_2) =  T_b(u_2) T_a(u_1)R_{ab}(u_1-u_2).
\end{align}
Let us furthermore introduce the transfer matrix
\begin{align}
t = \mathrm{tr}_a T_a = A+D. 
\end{align}
Rewriting the fundamental commutation relations in the following way
\begin{align}
T_a(u_1) T_b(u_2) =  R_{ab}(u_1-u_2)^{-1}T_b(u_2) T_a(u_1)R_{ab}(u_1-u_2),
\end{align}
and subsequently taking the trace over both auxiliary spaces $a,b$ we find by cyclicity of the trace that
\begin{align}\label{eq:tcomm}
[t(u_1),t(u_2)]=0.
\end{align}
The Lax operator is linear in the spectral parameter $u$. This means that $t$ is a polynomial of degree $L$ in $u$ whose coefficients are operators acting on the spin chain. From \eqref{eq:tcomm} it is then easy to see that all of these operators commute. In this way we have found a family of mutually commuting operators. It turns out that the Hamiltonian is a member of this family. 

In order to see that this is indeed the case, we expand the transfer matrix around the point $u=\ihalf$. From the properties of the permutation matrix (see below \eqref{eq:RXXX}) we find that
\begin{align}\label{eq:TandP}
T_a(\ihalf) = i^L P_{L,a}P_{L-1,a}\ldots P_{1,a} = i^L P_{12}P_{23}\ldots P_{L-1,L} P_{L,a},
\end{align}
Now it is easy to take the trace in the auxiliary space since $\mathrm{tr}_a P_{L,a}=1_L$. Thus we find
\begin{align}\label{eq:MomDef}
t(\ihalf) = i^L P_{12}P_{23}\ldots P_{L-1,L} =: i^L \mathcal{U},
\end{align}
which simply is the shift operator $\mathcal{U}$. By definition the shift operator is identified with the momentum operator as
\begin{align}
\mathcal{U} = e^{ip}.
\end{align}
Next, we look at the next order in our expansion around $u=\ihalf$ and we take the derivative of the transfer matrix. Direct computation gives
\begin{align}\label{eq:HviaTpre}
\left.\frac{dT_a}{du}\right|_{\lambda=\ihalf} = i^{L-1}\sum_n P_{L,a}\ldots \hat{P}_{n,a}\ldots P_{1,a}=
i^{L-1}\sum_n P_{12}P_{23}\ldots P_{n-1,n+1}\ldots P_{L-1,L}P_{L,a},
\end{align}
where $\hat{P}$ means that the corresponding term is absent. Then using the expression for $t(\ihalf)$ we quickly find
\begin{align}\label{eq:HviaT}
\left.\frac{d}{du}\ln t(u)\right|_{u=\ihalf} = -i\sum_{n}P_{n,n+1}.
\end{align}
We immediately see that the logarithmic derivative of the transfer matrix gives the Heisenberg Hamiltonian. 

\subsection{Algebraic Bethe Ansatz}

What remains is finding the spectrum of the Hamiltonian. Since the tranfer matrix defines a set of commuting quantities, we can diagonalize them simultaneously. In particular, we will now derive the eigenvalues and eigenstates of the transfer matrix directly. The Hamiltonian can be constructed from $t$ and so we automatically find the spectrum of the Hamiltonian.

\paragraph{Ground state.} Consider again the reference state $|0\rangle$ from \eqref{eq:DefVac}. It is an eigenstate of the transfer matrix. In particular, from the explicit expression \eqref{eq:LviaS} of the Lax operator we readily derive
\begin{align}
&A|0\rangle = \prod_n(u+iS_n^z)|0\rangle = (u+\ihalf)^L|0\rangle,
&&D|0\rangle = (u-\ihalf)^L|0\rangle,
&&C|0\rangle = 0.
\end{align}
Hence, $|0\rangle$ is an eigenstate with eigenvalue
\begin{align}
t(u)|0\rangle = (A+D)|0\rangle  =  \left[(u+\ihalf)^L + (u-\ihalf)^L\right]|0\rangle.
\end{align}
Now we need to find a way to flip spins in such a way that we generate eigenstates of the Hamiltonian. We will do this by using the operator $B$.

\paragraph{Commutation relations.} 

The key ingredient of the algebraic Bethe Ansatz are the fundamental commutation relations between the different operators from the monodromy matrix \eqref{eq:FCR}. Let us work out some of these explicitly. We have $T_1(u) = T(u) \otimes 1$ and $T_2(v) = 1\otimes T(v)$, hence
\begin{align}
&T_1(u)
=
\begin{pmatrix} 
A(u) & 0 & B(u) & 0 \\
0 & A(u) & 0 & B(u) \\
C(u) & 0 & D(u) & 0 \\
0 & C(u) & 0 & D(u) 
\end{pmatrix},
&T_2(v) 
=
\begin{pmatrix} 
A(v) & B(v) & 0 & 0\\
C(v) & D(v) & 0 & 0 \\
0 & 0 & A(v) & B(v) \\
0 & 0 & C(v) & D(v) 
\end{pmatrix}.
\end{align}
Then, by the explicit form of the R-matrix
\begin{align}\label{eq:Rexpl}
R =\begin{pmatrix}
u-v+i & 0 & 0 & 0 \\
0 & u-v & i & 0 \\
0 & i & u-v & 0 \\
0 & 0 & 0 & u-v+i
\end{pmatrix},
\end{align}
the fundamental commutation relation imply the following relations between the operators $A,B$ and $D$
\begin{align}
&B(u)B(v)  = B(v)B(u),\\
&A(u)B(v)  = \frac{u-v-i}{u-v} B(v)A(u) + \frac{i}{u-v}B(u)A(v),\label{eq:ABcom}\\
&D(u)B(v)  = \frac{u-v+i}{u-v} B(v)D(u) - \frac{i}{u-v}B(u)D(v).
\end{align}
For the eigenstate with $M$ flipped spins, we make the following Ansatz for the wave function
\begin{align}\label{algBstate}
|\mathbf{u}\rangle =  B(u_1)\ldots B(u_M)|0\rangle.
\end{align}
The next step is to compute the action of the transfer matrix on a state of the form \eqref{algBstate} and investigate when this will be an eigenvector of $t$ (and hence of the Hamiltonian). 

The tranfer matrix is simply given by $A+D$ and we already know how $A$ and $D$ act on the vacuum $|0\rangle$. You can then evaluate $A|u_1,\ldots, u_M\rangle $ by repeatedly applying \eqref{eq:ABcom} and moving $A$ through the $B$ operators.
We find
\begin{align}
A(u)|\mathbf{u}\rangle = 
(u+\ihalf)^L\prod_n \frac{u-u_n-i}{u-u_n} |\mathbf{u}\rangle  + 
\sum_n W^A_n|\ldots u_{n-1},u,u_{n+1}\ldots\rangle 
\end{align}
for some coefficients $W^A_n$.
The coefficients $W$ can be determined in the following way. Since the operators $B$ commute with each other, we can write
\begin{align}
|\mathbf{u}\rangle = B(u_n)\prod_{i\neq n}B(u_i)|0\rangle,
\end{align}
for any $n$. It is then easy to see that the only term that can contribute to $W^A_n$ is the one where $A(u)$ and $B(u_n)$ commute but interchange spectral parameters, \textit{i.e.} the second term in \eqref{eq:ABcom}. Then the $A$ operator simply commutes through all the other $B$'s. By this argument we find
\begin{align}
W^A_n = \frac{i(u_n+\ihalf)^L}{u-u_n}\prod_{j\neq n}\frac{u_n-u_j-i}{u_n-u_j}
\end{align}
The discussion with the commutators involving $D$ is completely analogous and we get
\begin{align}
D(u)|\mathbf{u}\rangle = 
(u-\ihalf)^L\prod_n \frac{u-u_n+i}{u-u_n} |\mathbf{u}\rangle  + 
\sum_n W^D_n|\ldots u_{n-1},u,u_{n+1}\ldots\rangle 
\end{align}
where
\begin{align}
W^D_n = -\frac{i(u_n-\ihalf)^L}{u-u_n}\prod_{j\neq n}\frac{u_n-u_j+i}{u_n-u_j}.
\end{align}
Combining these two results, we see that the state $|\mathbf{u}\rangle $ is indeed an eigenstate with eigenvalue $\Lambda$ 
\begin{align}\label{eq:Lambda}
\Lambda = (u+\ihalf)^L\prod_{j=1}^M\frac{u-u_j-i}{u-u_j} + (u-\ihalf)^L\prod_{j=1}^M\frac{u-u_j+i}{u-u_j},
\end{align}
if and only if the term $W^A_n+W^D_n$ vanishes. The requirement that $W^A_n+W^D_n$ vanishes puts restrictions on the parameters $u_i$ in the form of a set of equations
\begin{align}\label{eq:BAE}
\left[\frac{u_n+\ihalf}{u_n-\ihalf}\right]^L = \prod_{j\neq n}\frac{u_n-u_j+i}{u_n-u_j-i},
\end{align}
which are exactly the Bethe equations we found in the coordinate Bethe Ansatz after we identify the rapidity $u$ with the momentum $p$ as $u = \half \cot \sfrac{p}{2}$.

Finally, notice that the Bethe equations can also be read off from the eigenvalue itself. If we require that the eigenvalue $\Lambda$ is an analytic function of $u$, then is should not have any poles. However, at $u=u_j$, $\Lambda$ clearly has an apparent pole. In order to make this pole only superficial we need to impose the Bethe equations. In other words, the Bethe equations can also be found from the analytical properties of the transfer matrix. 

\paragraph{Baxter's TQ relation} For future convenience, let us introduce the Baxter Q-function
\begin{align}\label{eq:defQ}
Q(v) = \prod_{i=1}^M (v-u_i).
\end{align}
This is a polynomial of degree $M$ with zeroes at the positions of the Bethe roots of the eigenvector of the transfer matrix. In particular, we can rewrite the eigenvalue of the transfer matrix as
\begin{align}
\Lambda(v,\mathbf{u})= (v+\ihalf)^L \frac{Q(v-i)}{Q(v)} + (v-\ihalf)^L \frac{Q(v+i)}{Q(v)}.
\end{align}
Or, we find that the transfer matrix satisfies the equation
\begin{align}
\Lambda(v,\mathbf{u}) Q(v)= (v+\ihalf)^L Q(v-i) + (v-\ihalf)^L Q(v+i),
\end{align}
which is called Baxter's TQ relation. The form of $\Lambda$ and the Bethe equations can then be derived by looking at the analytic properties of the above equation. 

\paragraph{Dispersion relation.} The last step is to find the eigenvalues of the Hamiltonian. We have an explicit expression \eqref{eq:Lambda} for the eigenvalues, so let us determine both the energy and momentum. First, the momentum was defined via \eqref{eq:MomDef}
\begin{align}
e^{ip} = i^L t(\ihalf) 
\end{align}
Substituting the eigenvalue into this equation then results in
\begin{align}
&p = \sum_{j}p_j, && p_j =-i \log\frac{u_j-\ihalf}{u_j+\ihalf}\quad \Leftrightarrow\quad u = \frac{1}{2}\cot\frac{p}{2}.
\end{align}
This is defines the relation between momentum and rapidity. 

The energy is then found from the logaritmic derivative \eqref{eq:HviaT} and gives
\begin{align}\label{eq:EABA}
E = \sum_{n=1}^M \frac{1}{u_n^2+\frac{1}{4}},
\end{align}
which, in turn, perfectly agrees with the energy we found in the coordinate Bethe Ansatz. We have successfully reproduced the spectrum of the Hamiltonian that we found with the help of the coordinate Bethe Ansatz. 

\paragraph{Conserved charges} Let us also compute the eigenvalues of the other conserved charges that are generated by the transfer matrix. If we write
\begin{align}\label{eq:charges}
\ln t(u) = \sum_n -i Q_{n+1} (u-\ihalf)^n,
\end{align}
then $Q_1 = p$, $Q_2 = E$ and the other conserved charges $Q_r$ have eigenvalues $q_r$ given by
\begin{align}\label{eq:Qeigenv}
q_r = \sum_n \frac{i}{r-1}\Bigg[\frac{1}{(u_n+\ihalf)^{r-1}} - \frac{1}{(u_n-\ihalf)^{r-1}} \Bigg].
\end{align}
Finally, let us address the explicit form of the operators $Q_r$. By following the derivation of the Hamiltonian \eqref{eq:HviaTpre}, it is easy to work out $Q_3$
\begin{align}
&Q_3 = \sum_i \mathcal{Q}_{i,i+1,i+2}
& \mathcal{Q}_{i,i+1,i+2}= [\mathcal{H}_{i,i+1},\mathcal{H}_{i+1,i+2}].
\end{align}
We see that $Q_3$ is an operator with a next-to-nearest neighbor interaction, \textit{i.e.} it has range 3. In general $Q_r$ will be an operator with range $r$.

\subsection{The symmetry algebra}

From the FCR \eqref{eq:FCR} we can also deduce the properties of Bethe state under the SU(2) symmetry.

\paragraph{The SU(2) algebra}

The algebraic Bethe Ansatz revolved about the observation that at $\ihalf$, the Lax operator reduces to the permutation operator. There is a second special point, namely at $u\rightarrow\infty$. In fact at this point $L(u) \rightarrow u 1$ and expanding the monodromy matrix leads to the following
\begin{align}
T(u) \xrightarrow{u\rightarrow\infty} u^L + u^{L-1} i \sum_n S^a_n\otimes \sigma^a + \ldots
\end{align}
Recall from our definition of the coordinate Bethe Ansatz that we can realize the total spin operators, that form an SU(2) algebra, exactly as $S^a = \sum_n S^a_n$ and hence
\begin{align}
T \rightarrow u^L + u^{L-1} i \, S^a \otimes \sigma^a + \ldots.
\end{align}
In other words, we see the SU(2) symmetry generator appearing in this limit. 

Let us make this more explicit. Let us consider the FCR at the limit where we send one of the rapidities to $\infty$. We get
\begin{align}
\left[(u-v) + i P_{ab}\right]T_a(u) (v^L + i S^a\otimes \sigma^a)_b = 
(v^L + i S^a\otimes \sigma^a)_b T_a(u) \left[(u-v) + i P_{ab}\right]
\end{align}
The leading order terms cancel out and we are left with the subleading terms. They give (we express $P$ via the Pauli matrices)
\begin{align}
[T(u),S^a+\half\sigma^a]\otimes \sigma^a=0.
\end{align}
This describes how the different terms of the monodromy matrix transform under the global symmetry generators
\begin{align}
[S^a,T] = \half[T,\sigma^a].
\end{align}
Thus, for example we find
\begin{align}
[S^z,T]=\half[
\begin{pmatrix} A & B \\ C & D \end{pmatrix},
\begin{pmatrix} 1 & 0 \\ 0 & -1 \end{pmatrix}] = 
\begin{pmatrix} 0 & -B \\ C & 0 \end{pmatrix}.
\end{align}
Working out these relations then leaves us with the following commutation relations
\begin{align}
&[S^z,B] = -B, && [S^+,B] = A-D.
\end{align}
We can now easily check from the fundamental commutation relations that the transfer matrix (and thus the Hamiltonian as well) displays SU(2) symmetry. We have that
\begin{align}
&[S^z,A] = [S^z,D] = 0,
&& [S^+,A]=-[S^+,D] = -C,
&& [S^-,A]=-[S^-,D] = B,
\end{align}
from which it easily follows that $[S^a,t]= [S^a,A+D] = 0$

\paragraph{Highest weight states}

Not only have we shown that the Hamiltonian respects SU(2) symmetry, but since we also know how the spin operators commute with the creation operator $B$.
By these commutation relations we can study the action of the symmetry operators on the eigenstates of the transfer matrix. In particular we will show that they are highest weight states.

First, it is easy to see that the ground state $|0\rangle$ is a highest weight state since
\begin{align}
&S^+ |0\rangle = 0, && S^z |0\rangle = \frac{L}{2}|0\rangle.
\end{align}
The general state $|\mathbf{u}\rangle$, is similarly an eigenstate of $S^z$ and its eigenvalue readily follows from the commutator of $B$ with $S^z$
\begin{align}
S^z |\mathbf{u}\rangle = (\frac{L}{2}-M) |\mathbf{u}\rangle.
\end{align}
Next we need to show that $S^+ |\mathbf{u}\rangle = 0$. We can write
\begin{align}\label{eq:HWS}
S^+ |\mathbf{u}\rangle &= \sum_j B(u_1)\ldots (A(u_j)-D(u_j))\ldots B(u_M) |0\rangle\nonumber\\
& = \sum_j V_j B(u_1)\ldots \hat{B}(u_j) \ldots B(u_M) |0\rangle
\end{align}
Then by similar arguments as in the case for the Algebraic Bethe Ansatz one can show that $V_j=0$ if one imposes the Bethe equations. 

\subsection{Graphical Representation}

In order to discuss one-point functions, it is useful to introduce some graphical notation.

\paragraph{Building blocks} For spin-$\half$,
the Lax operator and the R-matrix are basically the same object via \eqref{defP} and the fact that $S_i = \half\sigma_i$. More precisely,
\begin{align}\label{eqRandL}
R_{ab}(u) = \mathcal{L}_{ab}(u-\ihalf).
\end{align}
The components of the Lax- and  R-matrix can be pictorially represented as follows
\begin{center}
\begin{tikzpicture}
\draw(-1.5,0) node[left ] {$R^{a_1 s_1}_{a_2 s_2}(u-v) ~= ~~~$} ;
\draw (-1,0) node[left ] {$u, a_1$} -- (1,0) node[right] {$a_2$};
\draw (0,-1) node[below ] {$v,s_1$} -- (0,1) node[above ] {$s_2$};
\end{tikzpicture}
\end{center}
where $a_{1,2},s_{1,2} =\, \uparrow,\downarrow$. The basic properties of the R-matrix can then also be shown pictorially. For instance, the unitarity property $R_{12}(u)R_{21}(-u)=1$ reduces to
\begin{center}
\begin{tikzpicture}
\draw (-2,-0.5) -- (-1,0.5);
\draw (-2,0.5) -- (-1,-0.5);
\draw (-3,-0.5) node[left]{$u$} -- (-2,0.5);
\draw (-3,0.5) node[left]{$v$} -- (-2,-0.5);
\draw (0,0) node {$=$} ;
\draw (1,0.5)node[left]{$v$} -- (3,0.5);
\draw (1,-0.5) node[left]{$u$}-- (3,-0.5);
\end{tikzpicture}
\end{center}
where we have suppressed al the indices on the external legs. Indices on internal legs are always assumed to be summed over.
Analogously, the Yang-Baxter equation can then simply be drawn as
\begin{center}
\begin{tikzpicture}
\draw (-2.5,-1) node[left ] {$1$} -- (-1,1) node[right] {};
\draw (-2.5,1) node[left ] {$2$} -- (-1,-1) node[above ] {};
\draw (-2,-1) node[below ] {$3$} -- (-2,1) node[above ] {};
\draw (0,0) node {$=$} ;
\draw (2.5,-1) node[right] {} -- (1,1) node[left] {$2$};
\draw (2.5,1) node[right ] {} -- (1,-1) node[left ] {$1$};
\draw (2,-1) node[below ] {$3$} -- (2,1) node[above ] {};
\end{tikzpicture}
\end{center}
The fundamental commutation relations \eqref{eq:FCR} take the same form.

We can then also represent the building blocks of the Algebraic Bethe Ansatz. The monodromy matrix then can be represented as a product of R-matrices
\begin{center}
\begin{tikzpicture}
\draw (-.7,0) node {$T_a(u) ~=~~ u$} ;
\draw (3.5,-1)node[below]{$\ihalf$}   -- (3.5,1) ;
\draw (3,-1)node[below]{$\ihalf$}   -- (3,1) ;
\draw[dotted] (2.5,0.5)  -- (2,0.5) ;
\draw (1,-1)node[below]{$\ihalf$}  -- (1,1) ;
\draw (1.5,-1)node[below]{$\ihalf$}   -- (1.5,1) ;
\draw (.5,0)  -- (4,0) ;
\end{tikzpicture}
\end{center}
The factors of $\ihalf$ correspond to the shift in the identification between the Lax matrix and the R-matrix \eqref{eqRandL}. The components of the monodromy matrix are the obtained by specifying the indices on the auxiliary legs. For instance, 
\begin{center}
\begin{tikzpicture}
\draw (-.7,0) node {$A_a(u) ~=~~ u\uparrow~$} ;
\draw (3.5,-1)node[below]{$$}   -- (3.5,1) ;
\draw (3,-1)node[below]{$$}   -- (3,1) ;
\draw[dotted] (2.5,0.5)  -- (2,0.5) ;
\draw (1,-1)node[below]{$$}  -- (1,1) ;
\draw (1.5,-1)node[below]{$$}   -- (1.5,1) ;
\draw (.5,0)  -- (4,0)node[right]{$\uparrow\, .$} ;
\end{tikzpicture}
\end{center}
The transfer matrix is defined as the trace of the monodromy matrix, so it is obtained by closing the external line corresponding to the auxiliary space
\begin{center}
\begin{tikzpicture}
\draw (0.2,-0.1) node[above left] {$t_a(u) ~=~u\,$} ;
\draw(1,1) [domain=90:270] plot ({0.5+ 0.25*cos(\x)}, {0.25+0.25*sin(\x)});
\draw(1,1) [domain=-90:90] plot ({4 + 0.25*cos(\x)}, {0.25+0.25*sin(\x)});
\draw (3.5,-1)  -- (3.5,1) ;
\draw (3,-1)  -- (3,1) ;
\draw[dotted] (2.5,0.25)  -- (2,0.25) ;
\draw[dashed] (.5,0.5)  -- (4,0.5) ;
\draw (1,-1)  -- (1,1) ;
\draw (1.5,-1)  -- (1.5,1) ;
\draw (.5,0)  -- (4,0)node[right]{$ .$} ;
\end{tikzpicture}
\end{center}
\noindent All algebraic relations that have been derived in the previous sections can then be elegantly represented by pictures. For instance, the fundamental commutation relations take a simple pictorial form that resembles train tracks
\begin{center}
\begin{tikzpicture}
\draw (-5.5,-1) node[left ] {$u$} -- (-4,1) ;
\draw (-5.5,1) node[left ] {$v$} -- (-4,-1) ;
\draw (-4,1)  -- (-.5,1) ;
\draw (-4,-1)  -- (-.5,-1) ;
\draw (-3.5,-1.5)  -- (-3.5,1.5) ;
\draw (-3,-1.5)  -- (-3,1.5) ;
\draw[dotted] (-2.5,0)  -- (-1.5,0) ;
\draw (-1,-1.5)  -- (-1,1.5) ;
\draw (0,0) node {$=$} ;
\draw (5.5,-1) node[left ] {$u$} -- (4,1) ;
\draw (5.5,1) node[left ] {$v$} -- (4,-1) ;
\draw (4,1)  -- (0.5,1) ;
\draw (4,-1)  -- (0.5,-1) ;
\draw (3.5,-1.5)  -- (3.5,1.5) ;
\draw (3,-1.5)  -- (3,1.5) ;
\draw[dotted] (2.5,0)  -- (1.5,0) ;
\draw (1,-1.5)  -- (1,1.5) ;
\end{tikzpicture}
\end{center}

\paragraph{Algebraic Bethe Ansatz}

The vacuum $|0\rangle$ is the state with all spins pointing up, so
\begin{center}
\begin{tikzpicture}
\draw(-1/2,0) node[left]{$|0\rangle ~ =$};
\draw[dotted] (1.5,0) -- (2.5,0) ;
\draw (0,-1/2) node[below ] {$\uparrow$} -- (0,1/2);
\draw (1,-1/2) node[below ] {$\uparrow$} -- (1,1/2);
\draw (3,-1/2) node[below ] {$\uparrow$} -- (3,1/2);
\draw (4,-1/2) node[below ] {$\uparrow$} -- (4,1/2);
\end{tikzpicture}
\end{center}
The action of $A,D$ on the vacuum can then directly be read of from the fact that there is only one configuration of internal indices that contribute
\begin{center}
\begin{tikzpicture}
\draw (-6.5,-1/2) node[below ] {$\uparrow$} -- (-6.5,1/2) node[above ] {$\uparrow$};
\draw (-5.5,-1/2) node[below ] {$\uparrow$} -- (-5.5,1/2) node[above ] {$\uparrow$};
\draw[dotted] (1.5,0) -- (2.5,0) ;
\draw (-3.5,-1/2) node[below ] {$\uparrow$} -- (-3.5,1/2) node[above ] {$\uparrow$};
\draw (-2.5,-1/2) node[below ] {$\uparrow$} -- (-2.5,1/2) node[above ] {$\uparrow$};
\draw(-1/2,0) node[left]{$=$};
\draw (-7,0) node[left ] {$u, \uparrow$} -- (-5.5,0) ;
\draw[dashed] (-6,0) -- (-3.5,0) ;
\draw (-3.5,0)-- (-2,0) node[right] {$\uparrow$};
\draw (0,-1/2) node[below ] {$\uparrow$} -- (0,1/2);
\draw (1,-1/2) node[below ] {$\uparrow$} -- (1,1/2);
\draw (3,-1/2) node[below ] {$\uparrow$} -- (3,1/2);
\draw (4,-1/2) node[below ] {$\uparrow$} -- (4,1/2);
\draw (4.5,0) node[right ] {$\times ~ (u+\ihalf)^L$};
\end{tikzpicture}
\end{center}

\begin{center}
\begin{tikzpicture}
\draw (-6.5,-1/2) node[below ] {$\uparrow$} -- (-6.5,1/2) node[above ] {$\uparrow$};
\draw (-5.5,-1/2) node[below ] {$\uparrow$} -- (-5.5,1/2) node[above ] {$\uparrow$};
\draw[dotted] (1.5,0) -- (2.5,0) ;
\draw (-3.5,-1/2) node[below ] {$\uparrow$} -- (-3.5,1/2) node[above ] {$\uparrow$};
\draw (-2.5,-1/2) node[below ] {$\uparrow$} -- (-2.5,1/2) node[above ] {$\uparrow$};
\draw(-1/2,0) node[left]{$=$};
\draw (-7,0) node[left ] {$u, \downarrow$} -- (-5.5,0) ;
\draw[dashed] (-6,0) -- (-3.5,0) ;
\draw (-3.5,0)-- (-2,0) node[right] {$\downarrow$};
\draw (0,-1/2) node[below ] {$\uparrow$} -- (0,1/2);
\draw (1,-1/2) node[below ] {$\uparrow$} -- (1,1/2);
\draw (3,-1/2) node[below ] {$\uparrow$} -- (3,1/2);
\draw (4,-1/2) node[below ] {$\uparrow$} -- (4,1/2);
\draw (4.5,0) node[right ] {$\times ~ (u-\ihalf)^L$};
\end{tikzpicture}
\end{center}
Bethe states are created by acting with $B$ on the vacuum state, so they can be represented as follows
\begin{center}
\begin{tikzpicture}
\draw (-6.5,-1/2) node[below ] {$\uparrow$} -- (-6.5,1/2) ;
\draw (-5.5,-1/2) node[below ] {$\uparrow$} -- (-5.5,1/2) ;
\draw (-3.5,-1/2) node[below ] {$\uparrow$} -- (-3.5,1/2) ;
\draw (-2.5,-1/2) node[below ] {$\uparrow$} -- (-2.5,1/2) ;
\draw(-1/2,0) node[left]{$= |u\rangle$};
\draw (-7,0) node[left ] {$u, \downarrow$} -- (-5.5,0) ;
\draw[dashed] (-6,0) -- (-3.5,0) ;
\draw (-3.5,0)-- (-2,0) node[right] {$\uparrow$};
\end{tikzpicture}
\end{center}

\subsection{Correlation functions and norms}

So far the focus has been on the spectrum, \textit{i.e.} the eigenvalues of the Hamiltonian. However, in order to compute more general correlation functions, such as the one-point functions in the defect CFT, we need to compute norms and inner products of Bethe states as well.

\paragraph{Algebraic vs. Coordinate}

Bethe vectors obtained via the coordinate Bethe Ansatz or the algebraic Bethe Ansatz will differ by an overall normalization. The proportionality factor between the coordinate and the algebraic Bethe Ansatz vectors is given by \cite{Izergin1989,Escobedo:2010xs}
\begin{align}\label{eq:HybridAlg}
|\vec{u}\rangle^{algebraic} = \Big[ \prod_{j}\frac{i(u_j-\ihalf)^L}{u_j+\ihalf} \prod_{i<j} \frac{u_i-u_j+ i}{u_i-u_j}\Big] |\vec{u}\rangle^{coordinate} .
\end{align}
Note that this factor depends explicitly on the normalization of the Lax matrix \eqref{eq:LviaS}. By definition, we have that the action of B on a state $|\psi\rangle$ is
\begin{align}\label{eq:actionB}
B(v)|\psi\rangle &= \langle\uparrow\! | \lax_{L,a}(v) \ldots \lax_{1,a}(v) |\psi\rangle\otimes |\!\downarrow\rangle, \nonumber\\
&= \sum_{m=1}^L \Big[ \prod_{m=1}^{n-1}(v + i S^z_m) \Big] S^-_n \Big[ \prod_{m=n+1}^{L}(v - i S^z_m)  \Big] |\psi\rangle\otimes |\!\downarrow\rangle.
\end{align}
Formula \eqref{eq:HybridAlg} can be proven by noticing that in terms of the rapidities $u_i$, the coefficients of the Bethe states are rational functions. This means that they are fully fixed by the zeroes, poles and asymptotic behavior. 

It is easy to see that in the coordinate Bethe Ansatz, the wave function vanishes when two Bethe roots coincide. Similarly from \eqref{eq:actionB} it is easily checked that $B(v)B(v+i)|0\rangle =0$, which indicates a zero at $v_i-v_j+i$. Combining these two terms gives rise to the factor $\prod_{i<j} \frac{u_i-u_j+ i}{u_i-u_j}$ in \eqref{eq:HybridAlg}. The rest of the prefactor \eqref{eq:HybridAlg} is straightforwardly derived along similar lines.

\paragraph{Slavnov's determinant}

Slavnov derived a formula for the inner product between an on-shell and an off-shell Bethe vector \cite{Slavnov1989}
\begin{align}
S_M(\mathbf{u},\vec{v}) : = \langle0|\prod_{i=1}^M C(v_i)\prod_{j=1}^M B(u_j)|0\rangle,
\end{align}
where $\mathbf{u}$ are a solution of the Bethe equations.
Clearly the number of flipped spins in both vectors need to be the same for their inner product to be non-zero. The overlap $S_M$ can be expressed as
\begin{align}\label{eq:Slavnov}
S_M(\mathbf{u},\vec{v})  = \frac{\det W}{\det V} \prod_{i=1}^M (u_i+\ihalf)^L,
\end{align}
where
\begin{align}
&W_{ij} := \frac{d}{d u_i} \Lambda(v_j,\mathbf{u}),
&& V_{ij} : = \frac{1}{v_i-u_j}.
\end{align}
and $\Lambda$ is the eigenvalue of the transfer matrix \eqref{eq:Lambda}. 

\paragraph{Norm of Bethe states}

A special case of the Slavnov determinant formula when $v\rightarrow u$. In this case, it computes the norm of the Bethe state $|\mathbf{u}\rangle$, which was first postulated by Gaudin \cite{Gaudin:1976sv}, see also \cite{Korepin:1982gg}. Both matrices $V,W$ in \eqref{eq:Slavnov} are divergent in this case, but this is easily resolved by setting $v_i = u_i+\epsilon$ and sending $\epsilon\rightarrow0$. This gives
\begin{align}\label{eq:Norm1}
&\langle \mathbf{u}|\mathbf{u}\rangle = \det \tilde{G} \, \prod_{i=1}^M (u_i+\ihalf)^L,
&& \tilde{G}_{ij} = \mathrm{Res}_{v_j=u_i} \bigg[\frac{d}{d u_i} \Lambda(v_j,\mathbf{u})\bigg].
\end{align}
As we saw in our discussion of the algebraic Bethe Ansatz, that eigenvalue $\Lambda$ at the point $v=u_i$ is related to the Bethe equations. It is indeed possible to rewrite \eqref{eq:Norm1} in terms of the Bethe equations. More precisely, for the coordinate Bethe Ansatz we find
\begin{align}\label{eq:NormCBAE}
{}^{coord}\langle \mathbf{u} | \mathbf{u}\rangle^{coord} = Q(\ihalf)Q(-\ihalf) \det G_{ij},
\end{align}
where
\begin{align}
&G_{ij} = \partial_{u_i} \Phi_j, 
&& \Phi_j = -i\log\! \bigg[ \left(\frac{u_j+\ihalf}{u_j-\ihalf}\right)^L \prod_{n\neq j}\frac{u_n-u_j+i}{u_n-u_j-i}\bigg].
\end{align}
The norm of the algebraic Bethe Ansatz then follows from \eqref{eq:HybridAlg}. 

\subsection{Integrable quenches}\label{sec:quench}

A quantum quench describes the evolution of a quantum system being in a particular eigenstate $|\psi_0\rangle$ of a Hamiltonian when the latter is instantaneously changed to a different Hamiltonian \cite{quench}. The initial eigenstate is generically not an eigenstate of the new Hamiltonian and the system will then have a non-trivial time evolution. A simple example would be the system in an eigenstate of the XXX spin chain in which suddenly the nearest neighbor interaction strength changes in the $z$ direction, making it into a so-called XXZ spin chain. 

In order to study the time evolution of the quenched system, one needs to compute the overlaps between the initial state and the eigenstates of the Hamiltonian $\langle \psi_0 | \lambda \rangle$. The computation of these overlaps, however, is a complicated problem. First, there is a large range of different initial states to consider. Second, overlaps between states are hard to compute in general. Only in special cases an exact formula, such as the Slavnov formula \eqref{eq:Slavnov}, is known. 

In view of these difficulties, it would be good to classify a family of integrable quenches for which, potentially, a closed formula can be derived. Recently a proposal has been put forward for a definition of an integrable quench \cite{Piroli:2017sei}.

\paragraph{Definition of integrable quench}

Consider an integrable spin chain with transfer matrix $t(u)$. An initial state $|\psi_0\rangle$ is integrable if it satisfies \cite{Piroli:2017sei}
\begin{align}\label{eq:IntDef}
\sigma t(v) \sigma |\psi_0\rangle = t(v) |\psi_0\rangle \eqncom
\end{align}
where $\sigma$ is the parity operator
\begin{align}
\sigma: v_1\otimes\ldots\otimes v_L \mapsto v_L\otimes\ldots\otimes v_1.
\end{align}
The parity operator clearly reverses rapidities, \textit{i.e.} $u\rightarrow-u$. This implies that the odd conserved charges \eqref{eq:charges} need to annihilate the initial state
\begin{align}
Q_{2n+1} |\psi_0\rangle = 0.
\end{align}
We will later see that this has implications for the overlap formulas.

\paragraph{Motivation}

The idea behind the integrability condition \eqref{eq:IntDef} is the fact that correlation functions involving some initial state can alternatively be seen as correlation functions on an open spin chain where the initial state corresponds to a non-trivial boundary condition. For integrable initial states, these boundary states seem to correspond to integrable reflection matrices. Integrability condition \eqref{eq:IntDef} is motivated by extending results from two-dimensional integrable quantum field theories in the presence of boundaries \cite{Ghoshal:1993tm}.

For a Euclidean field theory, there are two equivalent ways to introduce a Hamiltonian picture, namely the time direction can be taken along the boundary or perpendicular to the boundary, see Figure 4. In the first case, the boundary represents a non-trivial boundary conditions on the fields, while in the second case the boundary can rather be interpreted as an initial state $|B\rangle$. The question if the boundary preserves some measure of integrability can then be reformulated in terms of a condition on the state $|B\rangle$ \cite{Ghoshal:1993tm}, which is reminiscent of \eqref{eq:IntDef} \cite{Piroli:2017sei}. 

\begin{figure}[h]
\begin{center}
\begin{tikzpicture}
\filldraw[draw= lightgray,fill=lightgray] (-8.5,-2) rectangle (-5.5,2);
\filldraw[draw=black,fill=black] (-8.9,-2) rectangle (-5.2,-1.9);
\draw[->](-7,-1.5)  -- (-7,1.5) node[pos=.6, right ] {$t$} ;
\draw[<->](-8,-1.5) -- (-6,-1.5) node[pos=.6, below ] {$x$} ;
\filldraw[draw= lightgray,fill=lightgray] (-3,-1.9) rectangle (1,1.9);
\filldraw[draw=black,fill=black] (-3,-2) rectangle (-3.2,2);
\draw[<->](-2.5,-1.8)  -- (-2.5,1.5) node[pos=.6, right ] {$t$} ;
\draw[->](-2.5,-.5)  -- (0.5,-.5) node[pos=.6, below ] {$x$} ;
\end{tikzpicture}
\end{center}
\caption{Two equivalent ways of describing a two-dimensional Euclidean field theory with a boundary. In the left picture, the time coordinate is chosen perpendicular to the boundary so that it has the interpretation of a non-trivial initial state. In the right picture, time runs parallel to the boundary, which then correspond to non-trivial boundary conditions on the fields.}
\end{figure}
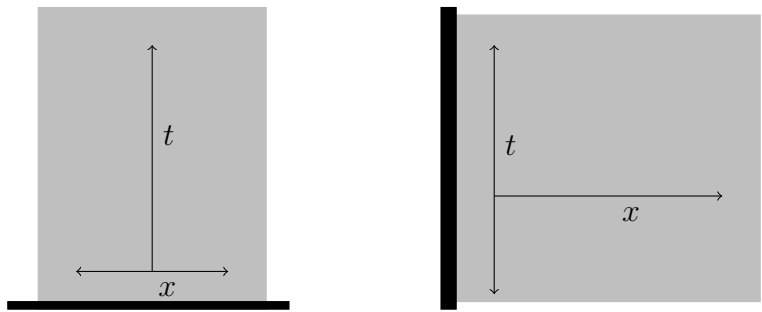

For the case of spin chains, let us look at an example of the overlap between a three magnon state on a length six spin chain and an initial state $|\psi_0\rangle$. This corresponds to the partition function corresponding to the following diagram
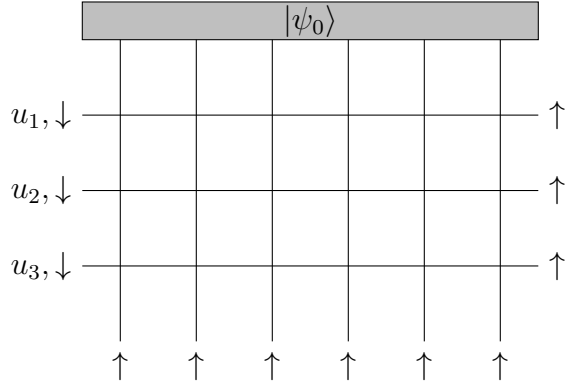
\begin{center}
\begin{tikzpicture}
\filldraw[draw=black,fill=lightgray] (-7,2) rectangle (-1,2.5)node[pos=.5] {$|\psi_0\rangle$};
\draw (-6.5,-2) node[below ] {$\uparrow$} -- (-6.5,2) ;
\draw (-5.5,-2) node[below ] {$\uparrow$} -- (-5.5,2) ;
\draw (-4.5,-2) node[below ] {$\uparrow$} -- (-4.5,2) ;
\draw (-3.5,-2) node[below ] {$\uparrow$} -- (-3.5,2) ;
\draw (-2.5,-2) node[below ] {$\uparrow$} -- (-2.5,2) ;
\draw (-1.5,-2) node[below ] {$\uparrow$} -- (-1.5,2) ;
\draw (-7,1) node[left ] {$u_1, \downarrow$} -- (-1,1) node[right] {$\uparrow$};
\draw (-7,0) node[left ] {$u_2, \downarrow$} -- (-1,0) node[right] {$\uparrow$};
\draw (-7,-1) node[left ] {$u_3, \downarrow$} -- (-1,-1) node[right] {$\uparrow$};
\end{tikzpicture}
\end{center}
We can then rotate the picture and exchange physical and auxiliary spaces. The partition function is clearly still the same. However, now the state $|\psi_0\rangle$ can be seen as a boundary condition on the spin chain and the overlap corresponds to the inner product between two domain walls (states with all spins up or down). 
\begin{center}
\begin{tikzpicture}[rotate=-90]
\filldraw[draw=black,fill=lightgray] (-7,2) rectangle (-1,2.5)node[pos=.5] {$|\psi_0\rangle$};
\draw (-6.5,-2) node[left ] {$\uparrow$} -- (-6.5,2) ;
\draw (-5.5,-2) node[left ] {$\uparrow$} -- (-5.5,2) ;
\draw (-4.5,-2) node[left ] {$\uparrow$} -- (-4.5,2) ;
\draw (-3.5,-2) node[left ] {$\uparrow$} -- (-3.5,2) ;
\draw (-2.5,-2) node[left ] {$\uparrow$} -- (-2.5,2) ;
\draw (-1.5,-2) node[left ] {$\uparrow$} -- (-1.5,2) ;
\draw (-7,1) node[above ] {$u_1, \downarrow$} -- (-1,1) node[below] {$\uparrow$};
\draw (-7,0) node[above ] {$u_2, \downarrow$} -- (-1,0) node[below] {$\uparrow$};
\draw (-7,-1) node[above ] {$u_3, \downarrow$} -- (-1,-1) node[below] {$\uparrow$};
\end{tikzpicture}
\end{center}
In \cite{Piroli:2017sei,Piroli:2018ksf,Piroli:2018don,Pozsgay:2018dzs,PPV} it was shown for wide range of spin chains that integrable initial states factorize into a product of two site states $|\psi_0\rangle \sim \bigotimes K_{ab}  |ab\rangle$. The matrix $K_{ab}$ is related to an integrable reflection matrix. In particular, integrable matrix product states can be shown to correspond to solutions of a twisted boundary Yang--Baxter equation \cite{PPV}. This can be worked out for various classes of matrix product states and spin chains including the ones that are relevant for holographic defect CFTs.

Thus, the overlap between a Bethe state and an integrable initial state schematically takes the form
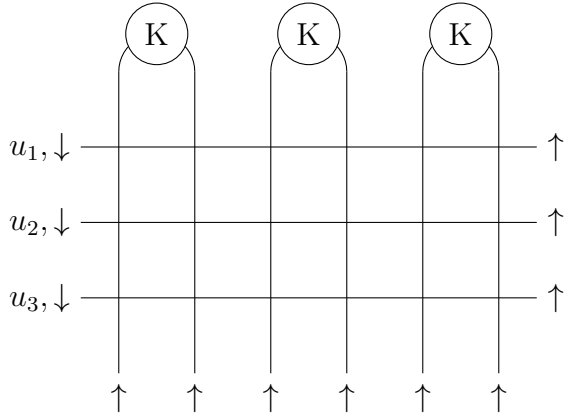
\begin{center}
\begin{tikzpicture}
\draw(1,1) [domain=0:180] plot ({-6+ 0.5*cos(\x)}, {2+0.5*sin(\x)});
\draw(1,1) [domain=0:180] plot ({-4+ 0.5*cos(\x)}, {2+0.5*sin(\x)});
\draw(1,1) [domain=0:180] plot ({-2+ 0.5*cos(\x)}, {2+0.5*sin(\x)});
\draw(-6,2.5) node[draw,shape=circle,fill=white]{K};
\draw(-4,2.5) node[draw,shape=circle,fill=white]{K};
\draw(-2,2.5) node[draw,shape=circle,fill=white]{K};
\draw (-6.5,-2) node[below ] {$\uparrow$} -- (-6.5,2) ;
\draw (-5.5,-2) node[below ] {$\uparrow$} -- (-5.5,2) ;
\draw (-4.5,-2) node[below ] {$\uparrow$} -- (-4.5,2) ;
\draw (-3.5,-2) node[below ] {$\uparrow$} -- (-3.5,2) ;
\draw (-2.5,-2) node[below ] {$\uparrow$} -- (-2.5,2) ;
\draw (-1.5,-2) node[below ] {$\uparrow$} -- (-1.5,2) ;
\draw (-7,1) node[left ] {$u_1, \downarrow$} -- (-1,1) node[right] {$\uparrow$};
\draw (-7,0) node[left ] {$u_2, \downarrow$} -- (-1,0) node[right] {$\uparrow$};
\draw (-7,-1) node[left ] {$u_3, \downarrow$} -- (-1,-1) node[right] {$\uparrow$};
\end{tikzpicture}
\end{center}
And indeed, for such configurations, when the K-matrix is reflecting, a determinant formula is known due to Tsuchiya \cite{Tsuchiya}. In fact all initial states for which an exact overlap formula is known seem to satisfy \eqref{eq:IntDef} \cite{Piroli:2017sei}. For instance, for the XXZ spin chain it was argued that a factorized formula for the overlap between a Bethe state and a MPS can only exist when the Y-system relations are satisfied (\textit{i.e.} when the system is integrable) \cite{PozsgayXXZ}. Unfortunately, the results of \cite{Tsuchiya} do not directly apply to the one-point functions in defect CFTs. Only in the SU(2) sector and for $k=2$ a generalization of this approach was carried out \cite{Foda:2015nfk}. Particularly for higher rank spin chains, finding determinant formulas for overlap functions is a notoriously hard problem.

\section{D3-D5 one-point functions}

Now that we have developed a full toolbox with integrability techniques, let us return to the D3-D5 defect version of $\mathcal{N}=4$ SYM theory and compute the one-point functions. We need to compute the quantity $C_k$ \eqref{eq:CK} where $|\mathbf{u}\rangle$ is some given eigenstate of the Heisenberg spin chain.

\subsection{Integrability}\label{sec:D3D5int}

Let us first prove \eqref{eq:IntDef} for the D3-D5 MPS \eqref{eq:MPS} that corresponds to the defect \cite{deLeeuw:2018mkd}. First, it is easy to see that the MPS is parity even
\begin{align}
\sigma|\MPS\rangle &= \sum_{\vec{i}} \tr[ t_{i_L}\ldots t_{i_1}] \,|e_{i_1}\ldots e_{i_L}\rangle \nonumber\\
&= \sum_{\vec{i}} \tr[ t^T_{i_1}\ldots t^T_{i_L}] \,|e_{i_1}\ldots e_{i_L}\rangle \nonumber \\
&= \sum_{\vec{i}} (-1)^{\#t_2} \tr[ t_{i_1}\ldots t_{i_L}] \,|e_{i_1}\ldots e_{i_L}\rangle \nonumber\\
&= |\MPS\rangle \eqndot
\end{align}
The last step uses the fact that there must be an even number of $t_2$'s, which is proven in section \ref{sec:DefectGen}. 

Next, we consider the action of the transfer matrix on the MPS. From the decomposition of the transfer matrix $t$ in terms of Lax operators \eqref{eq:TviaL}, we see that we can write
\begin{align}\label{eq:TMPS}
t(v) |\MPS\rangle &= 
\tr_{a,b} \lax_{b,n}(v) \prod_n \big[(t_1)_a \otimes |\!\uparrow\rangle_n + (t_2)_a \otimes |\!\downarrow\rangle_n\big] \nonumber\\ 
&=\tr_{a,b} \prod_n \big[\tau_1 \otimes |\!\uparrow\rangle_n + \tau_2 \otimes |\!\downarrow\rangle_n\big]\eqncom\nonumber\\
&=\sum_{\vec{i}} \tr[ \tau_{i_1}\ldots \tau_{i_L}] \,|e_{i_1}\ldots e_{i_L}\rangle\eqncom
\end{align}
where
\begin{align}
&\tau_1 = \begin{pmatrix}
(v+\ihalf) t_1 & 0 \\
it_2 & (v-\ihalf) t_1 
\end{pmatrix},
&&\tau_2 = \begin{pmatrix}
(v-\ihalf) t_2 & it_1 \\
0 & (v+\ihalf) t_2 
\end{pmatrix}\eqndot
\end{align}
The $\tau$-matrices have the property that there is a similarity transformation $U$ such that $\tau_1^T = U\tau_1 U^{-1}$ and $\tau_2^T = -U\tau_2 U^{-1}$.
\noindent Thus, combining everything, we find
\begin{align}
\sigma t(v) \sigma |\MPS\rangle &= \sum_{\vec{i}} (-1)^{\#t_2}\tr[ \tau_{i_L}\ldots \tau_{i_1}] \,|e_{i_1}\ldots e_{i_L}\rangle \nonumber\\
&= \sum_{\vec{i}} \tr[ \tau^T_{i_1}\ldots \tau^T_{i_L}] \,|e_{i_1}\ldots e_{i_L}\rangle \nonumber\\
&= \sum_{\vec{i}} (-1)^{2\#t_2}\tr[ \tau_{i_1}\ldots \tau_{i_L}] \,|e_{i_1}\ldots e_{i_L}\rangle \nonumber\\
&=  t(v) |\MPS\rangle\eqndot
\end{align}
Note that this proof is independent of $k$ (apart from the explicit form of the similarity transformation) and holds for any representation. This shows that the defect preserves integrability and is a clear sign that a closed formula for $C_k$ exists.

\subsection{Generalities}\label{sec:DefectGen}

There are some general observations that can be made for $C_k$ that restrict the operators that have a non-vanishing one-point function.

\paragraph{Restricting L, M} It is easy to show that $C_k$ is only non-vanishing if both $L$ and $M$ are even, where $M$ is the number of 
excitations or equivalently the number of Bethe roots. Namely, the Lie algebra of SU(2) admits an isomorphism where two of the $t$'s are mapped to $-t$. This isomorphism is realized by a similarity transformation 
\begin{align}
t_i \rightarrow U t_i U^{-1}.
\end{align}
For example, consider the case when $(t_1,t_2,t_3)\rightarrow(-t_1,-t_2,t_3)$, which is generated by the similarity transformation $U = \sum_{i} (-1)^i E^i_i$. This immediately implies that 
\begin{align}
(-1)^L\langle\MPS|\mathbf{u} \rangle &=  \langle \mathbf{u}| \Big( \tr \prod_{n=1}^L  \Big[-t_1 \otimes |\!\uparrow\rangle_n  -  t_2 \otimes|\!\downarrow\rangle_n\Big] \Big)\nonumber\\
&= \sum_{\vec{i}} \tr[Ut_{i_1}\ldots t_{i_L}U^{-1}]  \langle \mathbf{u}|e_{i_1}\ldots e_{i_L}\rangle\nonumber\\
&= \langle\MPS|\mathbf{u} \rangle \eqndot
\end{align}
This means that $L$ has to be even. From the similarity transformation that sends $(t_1,t_2,t_3)\rightarrow(t_1,-t_2,-t_3)$ you find that $M$ has to be even in order for $C_k$ to be non-trivial.

\paragraph{Restricting $\mathbf{u}$} Apart from these restrictions on the quantum numbers, for a non-zero overlap with the MPS we also find some restrictions on the Bethe roots. 
The integrability criterion \eqref{eq:IntDef} clearly implies all odd charges vanish. From \eqref{eq:Qeigenv} we then see that this implies that
\begin{align}
0 = \langle\MPS| Q_{2n+1}|\mathbf{u} \rangle = \langle\MPS|\mathbf{u} \rangle \sum_i \frac{i}{2n}\Big[\frac{1}{(u_i+\ihalf)^{2n}} - \frac{1}{(u_i-\ihalf)^{2n}} \Big] 
\end{align}
This gives an additional restriction on the rapidities. The only way for this to be satisfied is if $\mathbf{u} = -\mathbf{u}$, \textit{i.e.} all rapidities come in pairs $\{u_1,\ldots, u_{\frac{M}{2}},-u_1,\ldots,-u_{\frac{M}{2}}\}$. 

\paragraph{The vacuum} The first state to consider is the ferromagnetic vacuum \eqref{eq:DefVac}, which corresponds to the operator $\tr X^L$.
Its one-point function is given by
\begin{align}\label{eq:CBPS}
C_k = \frac{\langle\MPS|0\rangle}{\sqrt{\langle0|0\rangle}} = \tr( t_1^L) = 2^{-L}\sum_{i=1}^k (k-2i+1)^L =  2\frac{B_{L+1}(\frac{k+1}{2})}{L+1}\eqncom
\end{align}
where $B_{L+1}$ is the Bernoulli polynomial with index $L+1$. We see that the one-point function is a polynomial of degree $L+1$ in $k$ .

\subsection{One-point functions for \texorpdfstring{$k=2$}{k=2}}

\paragraph{Formula} The simplest case that we can consider for $M>0$ is the case $k=2$. This actually turns out to be a fundamental building block for the general $k$ case. For $k=2$, the $t$-matrices are simply related to the Pauli matrices: $t_i=\frac{1}{2}\sigma_i$. They satisfy the following relations:
\begin{align}
&t_i^2 = \sfrac{1}{4}\eqncom
&& t_i t_j = -t_j t_i  \quad\mathrm{for}\quad i\neq j.
\end{align}
This means that the inner product of the MPS with a Bethe state dramatically simplifies. In particular, any trace factor can be easily evaluated: 
\begin{align}\label{eq:MPSk2}
\tr ( t_1^{n_1-1}t_2 t_1^{n_2-n_1-1} t_2\ldots ) = (-1)^{n_1+n_2+\ldots} \tr( t_1^{L-M}t_2^{M}) = 2^{1-L}(-1)^{n_1+n_2+\ldots} \eqncom
\end{align}
where we used that $L,M$ are even. Let us consider the inner product of a Bethe state from the coordinate Bethe Ansatz \eqref{eq:genCBA} with the MPS. We see that it takes the following form:
\begin{align}\label{eq:MPSk2}
\langle\MPS|\mathbf{u}\rangle = 2^{1-L}\sum_{\sigma \in S_{M}} \sum_{n_i} \e^{i
\sum_m (p_{\sigma_m} n_m + \frac{1}{2}\sum_{j<m} \theta_{\sigma_j\sigma_m} )} (-1)^{n_1 + \ldots +n_M}\eqndot
\end{align}

\noindent To describe the overlap for a general number of excitations $M$, we introduce the following function
\begin{align}
K_{ij} : = \frac{1}{2} \left[ \frac{1 + 4 u_i^2} {1 + (u_i + u_j)^2} + \frac{1 + 4 u_i^2} {1 + (u_i - u_j)^2}  \right]\eqncom
\end{align}
and the following $\frac{M}{2}\times \frac{M}{2}$ matrix
\begin{align}
F_{ij} : = \bigg(L - \sum_{n=1}^{M/2} K_{in}\bigg) \delta_{ij} + K_{ij}\eqndot
\end{align}
The overlap is then given by
\begin{align}
\langle\MPS|\mathbf{u}\rangle_{k=2} = 2^{1-L}
\det F \sqrt{\frac{Q(\ihalf)}{Q(0)}}\eqncom
\end{align}
where  $Q(u) = \prod_{i=1}^M (u-u_i)$ is the Baxter polynomial, defined in \eqref{eq:defQ}. This formula of determinant type was first found by explicitly working out \eqref{eq:MPSk2}  for $M=0, 2,4,6$ by performing a sequence of nested geometric sums and then trying to rewrite the results as a determinant \cite{deLeeuw:2015hxa}. The resulting formula can then be checked numerically against states with more excitations. However, a direct proof of this formula is also possible by making contact with the condensed matter literature, as we will discuss in the next section.

In order to finally obtain the one-point function $C_2$, we need to divide by the norm of the Bethe state \eqref{eq:NormCBAE}. For states with paired rapidities $|\mathbf{u}\rangle = |-\mathbf{u}\rangle $, the norm formula factorises. Let us order the roots as $\{u_1,\ldots, u_{\frac{M}{2}}, -u_1,\ldots, -u_{\frac{M}{2}}\}$ and introduce the following $\frac{M}{2}\times\frac{M}{2}$ dimensional matrices $G_{\pm}$:
\begin{align}
\label{eq: definition of G pm}
&G_{\pm} = \partial_{u_m} \Phi_n \pm \partial_{u_{m+\frac{M}{2}}} \Phi_n\eqncom
\end{align}
then $\det G = \det G_+ \det G_-$. In terms of these matrices, the one-point function for $k=2$ can finally be written as
\begin{align}\label{eq:SU2quotient}
C_2 = 2^{1-L}\sqrt{ \frac{Q(\frac{i}{2})}{Q(0)}}\sqrt{ \frac{\det  G_+}{\det  G_-}} \eqndot
\end{align}
This means in particular that $\langle\MPS|\mathbf{u}\rangle_{k=2} \sim \det G_+$. 

\paragraph{N\'eel state}

Relation \eqref{eq:SU2quotient}  can be rigorously proven by using results from the condensed matter literature. The key idea is to map the MPS for $k=2$ to a known state. More precisely, it turns out that the MPS is equivalent equivalent to the so-called N\'eel state:
\begin{align}
|\Neel\rangle = |\! \uparrow\downarrow\uparrow\downarrow\ldots\rangle + |\! \downarrow\uparrow\downarrow\uparrow\ldots\rangle\eqndot
\end{align}
The N\'eel state is a state at half-filling, \textit{i.e.}\ it has $M=L/2$. It can be shown \cite{deLeeuw:2015hxa} that 
\begin{align}
2^L \Big(\frac{i}{2}\Big)^M|\MPS \rangle \Big|_{M=L/2}= |\Neel\rangle  + S^- |\ldots\rangle\eqndot
\end{align}
More precisely, let $|\MPS_M \rangle$ be the component of the MPS that have $M$ spins down, then
\begin{align}
|\Neel\rangle =\Big(\frac{i}{2}\Big)^{\frac{L}{2}} \sum_{s=0}^{L/2} i^s\frac{(S^-)^s}{s!}  |\MPS_{\frac{L}{2}-s} \rangle\eqndot
\end{align}
This can be shown by direct computation. From \eqref{eq:MPSk2} we see that the coefficient in front of a component of the MPS with downarrows at positions $n_1\ldots n_M$ has coefficient $2^{1-L}(-1)^{n_1+\ldots +n_M}$. The action of $S^-$ simply flips a spin with coefficient 1. Now, let us look at the coefficient at half filling and let's consider a configuration with flipped spins at positions $n_1\ldots n_\frac{L}{2}$. Each flipped spin comes either with a factor of $(-1)^{n}$ or with a factor of $1^{n}=1$. Let two spins be adjacent, say $n_1=n_2-1$. Then the terms proportional to $(-1)^{n_1+n_2}=-1$ and $(1)^{n_1+n_2}$ and the terms proportional to $(-1)^{n_1}$ and $(-1)^{n_2} = -(-1)^{n_1} $ will cancel pairwise. For the state at half-filling this means that all flipped spins must be separated by one site, which exactly gives the N\'eel state.

We already saw that one of the remarkable properties of the Bethe ansatz is that the Bethe states are highest-weight states, \textit{cf.} \eqref{eq:HWS}. This means that $S^+ |\mathbf{u}\rangle =0$ and thus for any Bethe state with $M=L/2$ the overlap of the MPS is the same as the overlap of the Bethe state with the N\'eel state, i.e.\ $2^{L-M} i^M\langle \mathbf{u}|\MPS \rangle= \langle \mathbf{u}|\Neel\rangle$. This is a problem that has been studied and solved in the condensed-matter literature \cite{Brockmann1, Brockmann2, Pozsgay}.

This interesting relationship can be extended to general excitation numbers. Let $M=L/2-2m$, then
\begin{align}
2^L \Big(\frac{i}{2}\Big)^M(2m)!|\MPS \rangle \Big|_{M=L/2-2m}= (S^+)^{2m}|\Neel\rangle  + S^- |\ldots\rangle\eqndot
\end{align}
The state $(S^+)^{2m}|\Neel\rangle$ is called the $(2m)$-raised N\'eel state \cite{Brockmann}. This means that the sought-after one-point functions can be rewritten in terms of a condensed-matter problem and the results from the condensed-matter literature \cite{Brockmann1, Brockmann2, Pozsgay} then provide proofs of the formulas that we just presented above.

Alternatively, for the SU(2) sector and $k=2$ one can use similar arguments to ones presented in section \ref{sec:quench} by considering partition functions of a spin chain with open boundary conditions. Indeed, the explicit reflection matrix corresponding to the MPS can be found and the corresponding determinant formula can be derived in that case \cite{Foda:2015nfk}.

\subsection{General \texorpdfstring{$k$}{k}}

The one-point function for general $k$ can be derived from the case $k=2$ in a recursive way. This is due to the fact that there is a recursive relation between matrix product states with different values of $k$: 
\begin{align}\label{eq:recurMPS}
|\mathrm{MPS}\rangle_{k+2} = \frac{t({\textstyle\frac{ik}{2}})}{(k-1)^L} \, |\mathrm{MPS}\rangle_k - \left(\frac{k+1}{k-1}\right)^L |\mathrm{MPS}\rangle_{k-2}\eqncom
\end{align}
where $k\geq2$ and $|\mathrm{MPS}\rangle_0 = |\mathrm{MPS}\rangle_1=0$. 

The idea behind the proof of formula~(\ref{eq:recurMPS}) is similar to the one we used to prove that the defect is integrable. We consider the local action of the Lax operator on the matrix product state. Indeed, setting $v = \frac{ik}{2}$ in \eqref{eq:TMPS} yields
\begin{align}\nonumber
\lax_{ia}({\textstyle\frac{ik}{2}}) \left[ \left\langle \uparrow_i \right| \otimes t_1^{(k)} +\left\langle \downarrow_i \right|\otimes t_2^{(k)}  \right]  =
\Big[\frac{i(k-1)}{2}\Big]^L\left(\left\langle \uparrow_i \right|\otimes \tau_1^{(k)} +\left\langle \downarrow_i \right|\otimes \tau_2^{(k)}\right)
\end{align}
where the matrices $\tau_{1,2}^{(k)}$ are given by
\begin{align}
&\tau_1^{(k)} = 
\begin{pmatrix}
\frac{k+1}{k-1} t^{(k)}_1 & 0 \\
\frac{2}{k-1}t^{(k)}_2 & t^{(k)}_1
\end{pmatrix}\eqncom
&&\tau_2^{(k)} = 
\begin{pmatrix}
t_2^{(k)} & \frac{2}{k-1}t_1^{(k)} \\
0 & \frac{k+1}{k-1}t_2^{(k)}
\end{pmatrix}\eqndot
\end{align}
The important observation is now that there exists a similarity transformation $A$ such that
\begin{align}
A \tau_i^{(k)} A^{-1} = \begin{pmatrix}
t_i^{(k+2)} & 0 \\
\star_i & \frac{k+1}{k-1}t_i^{(k-2)}
\end{pmatrix}\eqncom
\end{align}
where $\star_i$ stands for some irrelevant non-trivial entries~\cite{Buhl-Mortensen:2015gfd}.
This relation immediately proves the recursion relation~(\ref{eq:recurMPS}) by writing the transfer matrix as a product of Lax operators and using the form of the $\tau$ matrices to relate it to $ |\mathrm{MPS}\rangle_{k\pm2}$.

The Bethe states  $|\mathbf{u}\rangle$ are eigenvectors of the transfer matrix with eigenvalues $\Lambda(v|\mathbf{u})$. The recursion relation \eqref{eq:recurMPS} then fixes all overlap functions $C_k$ for even $k$ in terms of $C_2$ and $C_0\equiv0$ by the following recursion relation:
\begin{align}
C_{k+2}  = \Lambda\left(\tfrac{ik}{2}\middle|\{u_i\}\right) C_{k} - \left(\frac{k+1}{k-1}\right)^L C_{k-2}\eqndot
\end{align}
This then implies the following explicit form for the one-point function for $k>2$:
\begin{align}\label{eq:SU2genK}
C_k =i^L T_{k-1}(0)\sqrt{\frac{Q(\frac{i}{2})Q(0)}{Q^2(\frac{ik}{2})} }\sqrt{\frac{\det  G_+}{\det  G_-}} \eqncom
\end{align}
where 
\begin{align}\label{eq:Transfer}
T_n(u) =
\!\! \sum_{a=-\frac{n}{2}}^{\frac{n}{2}}\!\! (u+ia)^L\frac{Q(u+\frac{n+1}{2}i)Q(u-\frac{n+1}{2}i)}{Q(u+(a-\frac{1}{2})i)Q(u+(a+\frac{1}{2})i)}\eqndot
\end{align}
The function $T_n(u)$ can be identified as the transfer matrix of the Heisenberg spin chain where the auxiliary space is the $(n+1)$-dimensional representation.

Since the recursion relation \eqref{eq:recurMPS} goes in steps of two, the $k=2$ result extends to all even $k$. Of course, equation \eqref{eq:SU2genK} is well-defined for any $k$ and from numerical examples it is easily seen that it also works for odd $k$. By using \eqref{eq:recurMPS}, we see that for a proof of \eqref{eq:SU2genK} for odd $k$ we only need a proof for $k=3$. This is still an open question. However, there seems to be a remarkable relation between $C_2$ and $C_3$. From \eqref{eq:SU2genK}, we find
\begin{align}
C_3 = 2^L \frac{Q(0)}{Q(\frac{i}{2})}C_2\eqndot
\end{align}
This suggests that $C_3$ and $C_2$ are related by Q-operators \cite{Bazhanov:2010ts} rather than a transfer matrix, which has been checked for states with length up to 8 \cite{Buhl-Mortensen:2015gfd}. Actually, this results is a special case of the overlap formula for matrix product states considered in \cite{Piroli:2018ksf}, (see also \cite{PozsgayXXZ}), where a conjecture for this formula was put forward and motivated by considering the thermodynamic limit. A complete proof for finite $L,M$ is still missing however. 

\subsection{Dependence on $k$}

Let us study how the one-point functions depend on the defect parameter $k$. This is particularly important to understand the large $k$ behavior, which is relevant for string theory. To start with, we consider the BPS state \eqref{eq:CBPS}. In this case, the one-point function is given by a Bernoulli polynomial of degree $L+1$. For non-protected operators from the SU(2) sector the $k$ dependence is unclear. From the closed formula \eqref{eq:SU2genK}, we see that, in general, $C_k$ will depend rationally on $k$. However, we will show that the dependence becomes polynomial on solutions of the Bethe equations. 

In order to show this, let us partially fraction the product of Baxter polynomials in the denominator of the transfer matrix \eqref{eq:Transfer}
\begin{align}
 \frac{j^L}{Q((j-\half)i)Q((j+\half)i)} = 
-\sum_{i=1}^{M/2} \frac{1}{Q^\prime(u_i)} \Biggl( &\frac{u_i+\frac{i}{2}}{Q(u_i+i)} \! \left[\frac{j^{L-1}}{j-i(u_i+\frac{i}{2})}+\frac{j^{L-1}}{j+i(u_i+\frac{i}{2})}\right] 
\\&  -
  \frac{u-\frac{i}{2}}{Q(u_i-i)}  \left[\frac{j^{L-1}}{j-i(u_i-\frac{i}{2})}+\frac{j^{L-1}}{j+i(u_i-\frac{i}{2})}\right] \Biggr) \nonumber\eqncom
\end{align}
where we used the fact that the rapidities are paired and we denote $Q^\prime(v) = \frac{d}{dv}Q(v) $.
Each term can be further worked out using the identity
\begin{align}
\sum_{j=\frac{1-k}{2}}^{\frac{k-1}{2}} \frac{j^{L-1}}{j-a} = a^{L-1}\left[\Psi(\tfrac{1-k}{2}-a)-\Psi(\tfrac{1+k}{2}-a) \right]-2 \sum_{m=1}^{L/2} a^{L-2m}\frac{B_{2m-1}}{2m-1}\eqncom
\end{align}
where $\Psi$ is the digamma function and $B_{2m-1} = B_{2m-1}(\frac{k+1}{2})$ is the Bernoulli polynomial with index $2m-1$. Using the fact that $L$ is even, we also have that
$
\sum_j \frac{j^{L-1}}{j-a} = \sum_j \frac{j^{L-1}}{j+a}\eqndot
$
This implies
\begin{align}
\sum_{j=\frac{1-k}{2}}^{\frac{k-1}{2}} \frac{j^L}{Q((j-\half)i)Q((j+\half)i)} &= 
\sum_i\frac{4i^L}{Q^\prime(u_i)}  \bigg\{
\frac{i}{2}\frac{(u_i+\frac{i}{2})^{L}}{Q(u_i+i)}\frac{k}{u_i^2+ \frac{k^2}{4}}  \\\nonumber
&+\sum_{m=1}^{L/2} \left[\frac{(u_i+\frac{i}{2})^{L-2m+1}}{Q(u_i+i)}+\frac{(u_i-\frac{i}{2})^{L-2m+1}}{Q(u_i-i)} \right] 
\frac{B_{2m-1}}{(2m-1)i^{2m}} \\
&- \frac{i}{2}\left[\frac{(u_i+\frac{i}{2})^{L}}{Q(u_i+i)}+\frac{(u_i-\frac{i}{2})^{L}}{Q(u_i-i)} \right] \!\!\! \left[\Psi(-\tfrac{k}{2}-iu_i )-\Psi(\tfrac{k}{2}-iu_i) \right]\bigg\}\eqndot\nonumber
\end{align}
Now let us compare the left- and right-hand side of the above equation. In particular, we see that in the limit $u_1\rightarrow\infty$ the left-hand side scales like $u_1^{-4}$. In order for the above equation to hold, this means that the right-hand side must display the same behaviour. This means that the sum in the second line can only run up to $\frac{L}{2}-M+1$. One can indeed check that the coefficients in front of the Bernoulli polynomials with higher indices vanish. Then, upon using the Bethe equations \eqref{eq:BAE}, we arrive at
\begin{align}
& \sum_{j=\frac{1-k}{2}}^{\frac{k-1}{2}} \frac{j^L}{Q((j-\half)i)Q((j+\half)i)}  =\\ 
&\sum_i\frac{4i^L}{Q^\prime(u_i)}  \frac{(u_i+\frac{i}{2})^{L}}{Q(u_i+i)}\Biggl\{ \frac{i}{2}\frac{k}{u_i^2+ \frac{k^2}{4}}   
+\sum_{m=1}^{\frac{L}{2}-M+1} \left[\frac{1}{(u_i+\frac{i}{2})^{2m-1}}- \frac{1}{(u_i-\frac{i}{2})^{2m-1}} \right] 
\frac{B_{2m-1}}{(2m-1)i^{2m}}\Biggr\}\eqndot \nonumber
\end{align}
We then recognize the conserved charges $q_r$ \eqref{eq:Qeigenv}, so
\begin{align}
\sum_{j=\frac{1-k}{2}}^{\frac{k-1}{2}} \frac{j^L}{Q((j-\half)i)Q((j+\half)i)} =  \sum_i\frac{4i^{L+1}}{Q^\prime(u_i)}  \frac{(u_i+\frac{i}{2})^{L}}{Q(u_i+i)}\left[\frac{k/2}{u_i^2+ \frac{k^2}{4}}  -
\sum_{m=1}^{\frac{L}{2}-M+1} \frac{q_{2m}(u_i)}{i^{2m}}  B_{2m-1}\right].
\end{align}
We can now insert this into \eqref{eq:SU2genK} to obtain
\begin{align}
C_k=2 C_2
(2i)^L\sum_i\frac{Q(0)}{Q^\prime(u_i)}  \frac{(u_i+\frac{i}{2})^{L}}{Q(u_i+i)}\Biggl[ \frac{\frac{ik}{2}\, Q(\frac{ik}{2})}{u_i^2+ \frac{k^2}{4}}  -
Q(\tfrac{ik}{2})\sum_{m=1}^{\frac{L}{2}-M+1} \frac{q_{2m}(u_i)}{i^{2m-1}} B_{2m-1}\Biggr].
\end{align}
Notice that since $Q(\frac{ik}{2}) = \prod_i [u_i^2 +\frac{k^2}{4} ]$, we find that the one-point function is a polynomial of degree $L-M+1$. It is no longer given by a single Bernoulli polynomial, but rather by a sum of them.

\subsection{Descendants}

Descendant states can be obtained from the highest weight Bethe eigenstates by sending some of the rapidities to infinity. This process is best described using the coordinate Bethe ansatz, 
 \begin{align}
 \lim_{u_k\rightarrow \infty} |\{u_j\}\rangle =  S^{-}|\{u_j\}_{j\neq k}\rangle \eqndot
 \end{align}
For descendant states with $M$ finite and $N-M$ infinite roots, one has the following expression for the norm~\cite{Escobedo:2010xs}:
\begin{align} \label{eq:NormDesc}
^{{\mbox{\footnotesize coord}}}\langle\{u_j,\infty^{N-M}\} |\{u_j,\infty^{N-M}\}\rangle^{{\mbox{\footnotesize coord}}}
=\frac{(L-2M)!(N-M)!}{(L-M-N)!} \, \,
^{{\mbox{\footnotesize coord}}}\langle\{u_j\} |\{u_j\}\rangle^{{\mbox{\footnotesize coord}}}\eqndot
 \end{align}
For the overlap, we find a similar relation for $k=2$:
\begin{align}\label{eq:MPSdesc}
 \langle \mbox{MPS} |\{u_j,\infty^{N-M}\}\rangle^{{\mbox{\footnotesize coord}}} =
\frac{(N-M)!(\frac{L}{2}-M)!}{(\frac{N-M}{2})!(\frac{L-M-N}{2})!}
 \langle \mbox{MPS} |\{u_j\}\rangle^{{\mbox{\footnotesize coord}}}\eqndot
\end{align}
So far we do not have a proof of this formula, but it has been checked for chains up to $L=18$. In particular, one finds that restricted to a fixed number of flipped spins
\begin{align}
(S^{+})^{N-M}|\mbox{MPS}\rangle_{M} = \frac{(N-M)!(\frac{L}{2}-M)!}{(\frac{N-M}{2})!(\frac{L-M-N}{2})!} |\mbox{MPS}\rangle_{N}  + S^{-}|\cdots\rangle\eqndot
\end{align}
where the second term vanishes upon taking the inner product with a Bethe state since Bethe states are highest weight states. An alternative approach to find a closed formula for descendant states is to lake the limit $u\rightarrow\infty$ in the general overlap function \cite{Brockmann1}. This leads to a slightly modified version of the determinant formula. 

Summarising, from \eqref{eq:NormDesc} and \eqref{eq:MPSdesc} we see that the one-point functions of descendant operators are proportional to those of the corresponding primary operators. The proportionality factor is a simple combinatorical factor depending on $L,M,N$.

\subsection{The SU(3) sector and the nested Bethe Ansatz}

One-point function formulas can be found in other scalar sectors of $\mathcal{N}=4$ SYM theory as well \cite{deLeeuw:2016umh,deLeeuw:2018mkd}. Before discussing the results of the complete scalar sector, we look at an intermediate case; the SU(3) sector. This sector is closed at one-loop and consists of three fields which form an SU(3) spin chain. 

The SU(3) spin chain is reminiscent of the SU(2) spin chain that we discussed in great detail. In fact the Hamiltonian density in both spin chains is given by \eqref{eq:HfromD}
\begin{align}
\mathcal{H}_{ij} = 1- \mathbb{P}_{ij},
\end{align}
However, in the case of the SU(3) spin chain, we have three degrees of freedom at each site, so the local Hilbert space is $\mathbb{C}^3$, with basis elements $|e_{1,2,3}\rangle$. Correspondingly, the matrix product state takes the more general form
\begin{align}\label{MPSreal}
|\mathrm{MPS_k}\rangle  = \sum_{\vec{i}} \mathrm{tr} [t_{i_i}\ldots t_{i_L}] |e_{i_1} \ldots e_{i_L}\rangle.
\end{align}
Also for this more general MPS, one can show that it is integrable by a proof analogous to the one presented in Section \ref{sec:D3D5int}. In order to compute one-point functions we need to find the eigenstates of the SU(3) spin chain. However, in order to deal with the extra degrees of freedom of the spin chain, the Bethe Ansatz needs to be generalized to a so-called nested Bethe Ansatz. 

\paragraph{Nested Bethe Ansatz} For the computation of one-point functions, the coordinate Bethe Ansatz is important, so let us briefly discuss the nested coordinate Bethe Ansatz for SU(3). For a recent review on the Algebraic nested Bethe Ansatz see \textit{e.g.} \cite{Levkovich-Maslyuk:2016kfv}. Similar to the coordinate Bethe Ansatz discussed in section \ref{sec:CBA}, we start by defining a vacuum state
\begin{align}
|0\rangle = |e_1\ldots e_1\rangle.
\end{align}
This is again an eigenstate with eigenvalue 0. 

The next step is to consider excitations on this vacuum. Now we have two choices; we can put $e_2$ or $e_3$ in the vacuum state. In the SU(2) case, eigenstates can be labelled by two parameters, the length $L= \#e_1 + \# e_2$ and the number of excitations $M=\#e_2$. For the SU(3) chain we need three parameters $L= \#e_1 + \# e_2 + \# e_3$, $M= \# e_2 + \# e_3$ and $N= \# e_3$ to specify a sector. Moreover, instead of the spin operators $S^\pm, S^z$ we now have to work with generators $E^i_j$ defined such that $E^i_j|e_k\rangle = \delta^i_k |e_j\rangle$.

The idea of the nested Bethe Ansatz is to write the Bethe vectors in a nested form. More precisely, we introduce an extra set of auxiliary Bethe roots $\{v_i\}_{i=1,\ldots,N}$ and write a generalization of the Bethe vectors \eqref{eq:genCBA}. For SU(2) the Bethe Ansatz is of the form
\begin{align}
|\mathbf{u}\rangle = \sum_{1\leq m_1<\ldots<m_M\leq L} a(m_1,\ldots,m_M) (E^1_2)_{m_1} \ldots (E^1_2)_{m_M}|0\rangle,
\end{align}
where 
\begin{align}
a(m_1,\ldots,m_M)  = \sum_{\sigma\in S_{M}} A_\sigma(u) \prod_{i=1}^M \Big[\frac{u_{\sigma_i} +\ihalf}{u_{\sigma_i} -\ihalf}\Big]^{m_i}.
\end{align}
For SU(3) we make the following Ansatz
\begin{align}\label{eq:NBA}
|\mathbf{u},\mathbf{v}\rangle = &\sum_{m_i}
a(m_1,\ldots,m_M)
\Bigg[ 
\sum_{n_i}
\tilde{a}(n_1,\ldots,n_M)
(E^2_3)_{n_1}\ldots (E^2_3)_{n_N}
\Bigg] (E^1_2)_{m_1}\ldots (E^1_2)_{m_M}|0\rangle.
\end{align}
The term between the square brackets exactly takes the form of a SU(2) type Bethe Ansatz, which is why this approach is called the Nested Bethe Ansatz. It decomposes the wave function into Bethe vectors of models with lower rank. The coefficient $\tilde{a}$ depends on the auxiliary parameters $v$ and is explicitly given by
\begin{align}
\tilde{a}(n_1,\ldots,n_N) = \sum_{\tau\in S_N}  A_\tau(v)\prod_{j=1}^N \frac{1}{v_{\tau_j}-u_{\sigma_{n_j}} +\ihalf} \prod_{k=1}^{n_j}\frac{v_{\tau_j}-u_{\sigma_k} +\ihalf}{v_{\tau_j}-u_{\sigma_k} -\ihalf}.
\end{align}
This is the wave function of an inhomogeneous SU(2) spin chain, where the rapidities $u_i$ play the role of inhomogeneities.

\paragraph{The Bethe equations}
It can be checked that the state \eqref{eq:NBA} is an eigenstate of the SU(3) spin chain Hamiltonian with eigenvalue \eqref{eq:EABA}
\begin{align}
E = \sum_{i=1}^M \frac{1}{u_i^2 + \frac{1}{4}}.
\end{align}
Notice that $E$ does not depend on the auxiliary parameters $v$. Imposing periodic boundary conditions on a state $|\mathbf{u},\mathbf{v} \rangle$ with labels $(L,M,N)$ lead to a set of nested Bethe equations which are given by
\begin{align}
1&= \Big( \frac{u_m - \ihalf}{u_m + \ihalf} \Big)^L 
\prod^M_{n\neq m} \frac{u_m - u_n + i }{u_m - u_n - i} 
\prod^N_{n=1} \frac{u_m - v_n - \ihalf }{u_m - v_n - \ihalf} ,\label{eq:BAEsu3a}\\
1&=\prod^M_{m= 1} \frac{v_n - u_m - \ihalf }{v_n - u_m + \ihalf} \prod^N_{m\neq n} \frac{v_n - v_m + i }{v_n - v_m - i}.\label{eq:BAEsu3b}
\end{align}

\paragraph{Norm}

The formula for the Gaudin norm \eqref{eq:NormCBAE} can be generalized to the SU(3) case. We now need two norm functions corresponding to the two Bethe equations  \eqref{eq:BAEsu3a} and \eqref{eq:BAEsu3b}
\begin{align}\label{eq:PhiSU3}
\phi^v_m &:= -i \log\left[
\Big( \frac{v_m - \ihalf}{v_m + \ihalf} \Big)^L 
\prod^M_{n\neq m} \frac{v_m - v_n + i }{v_m - v_n - i} 
\prod^N_{n=1} \frac{v_m - w_n - \ihalf }{v_m - w_n - \ihalf} 
\right],\\
\phi^w_n &:= -i \log\left[\prod^M_{m= 1} \frac{w_n - v_m - \ihalf }{w_n - v_m + \ihalf} \prod^N_{m\neq n} \frac{w_n - w_m + i }{w_n - w_m - i} \right].
\end{align}
The norm of a Bethe state is then given by \cite{Escobedo:2010xs,Escobedo:2012ama}
\begin{align}
\langle\mathbf{u},\mathbf{v}|\mathbf{u},\mathbf{v} \rangle = \prod_{i=1}^M \Big[ u_i^2 + \sfrac{1}{4}\Big] \det_{(M+N)\times(M+N)} \partial_I \phi_J,
\end{align}
where the generalized indices $I,J = 1,\ldots M, M+1,\ldots M+N$ run over both the momentum carrying and auxiliary Bethe roots.

\paragraph{One-point functions} 
Now that we have computed the eigenstates of the SU(3) spin chain, we can take their overlap with the MPS \eqref{MPSreal}. Let us consider the case $k=2$ now and postpone the general $k$ result to the next section.  A first observation is that since \eqref{MPSreal} corresponds to an integrable MPS, the Bethe roots again have to be paired in the sense that
\begin{align}
\{u_i,v_j\} = \{-u_i,-v_j\}.
\end{align}
Just as in the SU(2) case, it can readily be shown that $M$ always has to be even. However, now there is a new possible solution where $N$ is odd and one of the auxiliary roots $v_i=0$. Nevertheless, due to the pair structure of the Bethe root configurations the determinant of the norm matrix factorizes in the same way as in case for the SU(2) sub-sector~\cite{Brockmann1,deLeeuw:2016umh}
\begin{align}
&\det G = \det G_+ \cdot \det G_-.
\end{align}
Assuming that there is a closed formula for the one-point functions in this case, it should satisfy the property that for states with $N=0$, it should reduce to \eqref{eq:SU2quotient}. This suggest a natural generalization 
\begin{align}\label{eq:C2SU3}
C_2^{SU(3)} = 2^{1-L} \sqrt{\frac{Q_u(\ihalf)Q_v(\ihalf)}{Q_u(0)\bar{Q}_v(0)}} \sqrt{\frac{\det G_+}{\det G_-}},
\end{align}
where $Q_{u,v}$ are the Baxter polynomials depending on $u,v$, respectively and $\bar{Q}$ omits any zero roots. Consequently, \eqref{eq:C2SU3} can be checked numerically against explicit one-point functions and perfect agreement is found for states of spin chains up to length 14. This provides very strong evidence that \eqref{eq:C2SU3} indeed is the determinant formula for the SU(3) spin chain, but a direct proof is still missing. Note that \eqref{eq:C2SU3} is actually one of the few non-trivial overlap functions that that can be written in a closed form for spin chains with higher rank.

\subsection{Full scalar sector}

It is possible to find the one-point function formula for arbitrary values of $k$ for the full scalar sector of $\mathcal{N}=4$ SYM theory \cite{deLeeuw:2016umh,deLeeuw:2018mkd}. The scalar sector has six scalar fields and at one-loop it forms a corresponding SO(6) spin chain \cite{Minahan:2002ve}.
The matrix product state is also integrable for the SO(6) spin chain and takes the more general form
\begin{align}\label{MPSreal}
|\mathrm{MPS}\rangle  = \sum_{\vec{i}} \mathrm{tr} [t_{i_i}\ldots t_{i_L}] |\phi_{i_1} \ldots \phi_{i_L}\rangle.
\end{align}
The Bethe eigenstates that diagonalize the transfer matrix of the SO(6) spin chain are now characterized by three sets of Bethe roots
\begin{align}
\{u_{j}\}_{j=1}^{M}, \hspace{1.0cm}
\{v^{\pm}_{j}\}_{j=1}^{N_\pm}
\end{align}
which satisfy the following nested Bethe equations
\begin{align}\label{eq:SO6BAE}
1&=
\bigg(\frac{u_i-\ihalf}{u_i+\ihalf}\bigg)^L\prod_{j\neq i}^{M}\frac{u_i-u_j+i}{u_i-u_j-i}\prod_{k=1}^{N_+}\frac{u_i-v^+_{k}-\ihalf}{u_i-v^+_k+\ihalf}\prod_{k=1}^{N_-}\frac{u_i-v^-_{k}-\ihalf}{u_i-v^-_k+\ihalf}, \nonumber\\
1&=  \prod_{l\neq i}^{N_+}\frac{v^+_i - v^+_l + i}{v^+_i-v^+_l-i} \prod_{k=1}^{M}\frac{v^+_i -u_k-\ihalf }{v^+_i -u_k+\ihalf}, \\
1&=  \prod_{l\neq i}^{N_-}\frac{v^-_i - v^-_l + i}{v^-_i-v^-_l-i} \prod_{k=1}^{M}\frac{v^-_i -u_k-\ihalf }{v^-_i -u_k+\ihalf}.\nonumber
\end{align}
The quantum numbers $(L,M,N_\pm)$ that characterize a states correspond to the length ($L$), total number of excitations ($M$) and the number of excitations of a specific type $(N_\pm)$. The explicit form of the eigenstates is again of nested type can be found for instance in~\cite{Basso:2017khq}.

Just as for SU(2) and SU(3), we find that all roots must come in pairs $\{u_i,v_i^{\pm}\}=\{-u_i,-v_i^{\pm}\}$. Looking at the results in the SU(2) and SU(3) subsector~\cite{deLeeuw:2015hxa,Buhl-Mortensen:2015gfd}, we see that the overlap formulas are expressible in terms of a few building blocks, namely the
Baxter polynomials and the Gaudin matrix $G$. So, just as for the SU(3) case, we introduce the standard Baxter Q-functions
\begin{align}
&Q_1(x) = \prod_{i=1}^M (x-u_i),
&&Q_\pm(x) = \prod_{j=1}^{N_\pm} (x-v^\pm_j),
\end{align}
as well as the reduced Baxter Q-functions
\begin{align}
\bar{Q}_\pm(x) = \prod_{j=1;v^\pm_j\neq 0}^{N_\pm} (x-v^\pm_j),
\end{align}
where we omit the zero roots in the product.
Furthermore, we now need the extended norm matrix, $G$, of the SO(6) spin chain
\begin{align}
&G\equiv \partial_J \phi_I,
\end{align}
where $I,J = 1,\ldots, M + N_+ + N_-$ and $\phi$ is the norm function obtained by taking the logarithm of the right hand side of the Bethe equations \eqref{eq:SO6BAE}, \textit{cf.} \eqref{eq:PhiSU3}. The determinant of the norm matrix again factorizes and in terms of these building blocks we can then give generalization of the one-point function formula to the full SO(6) sector
\begin{align}
C^{SO(6)}_k =
\sqrt{
\frac{Q_1(0)Q_1(\frac{i}{2})Q_1(\frac{ik}{2})Q_1(\frac{ik}{2})}{\bar{Q}_+(0)\bar{Q}_+(\frac{i}{2})\bar{Q}_-(0)\bar{Q}_-(\frac{i}{2})}
} \cdot \mathbb{T}_{k-1}(0) \cdot \sqrt{\frac{\det G_+}{\det G_-}},\label{SO6formula}
\end{align}
where
\begin{align}\label{TSO(6)}
\mathbb{T}_n(x) = \sum_{a=-\frac{n}{2}}^{\frac{n}{2}}(x+ia)^L \frac{Q_+(x+ia)Q_-(x+ia)}{Q_1(x+i(a+\frac{1}{2})) Q_1(x+i(a-\frac{1}{2}))}.
\end{align}
This formula can be proven for states with $M=2$ and it has been checked numerically for Bethe states with length up to 13. In particular, it contains the SU(2) and SU(3) cases. A direct proof is still missing and it would in particular be interesting to clarify the nature of the function $\mathbb{T}$. 

\section{Defect CFT at loop level}

So far, we only considered one-point functions at tree level. A natural question is to ask how it extends to quantum level. A framework to do quantum computations was formulated recently  \cite{Buhl-Mortensen:2016pxs,Buhl-Mortensen:2016jqo}. In this final section we will briefly review some of the applications and results of loop computations in the D3-D5 defect CFT.

\subsection{Quantum Field Theory}
In order to calculate quantum corrections in the defect CFT, the action of $\mathcal{N}=4$ SYM theory \eqref{SYMaction} has to be expanded around the classical solution \eqref{eq:vev}:
\begin{equation}
 \scal_i=\scalc_i+\scalq_i\eqndot
\end{equation}
It is easy to see that this expansion gives masses to the fields in the action.  Since the vacuum expectation values differ among the various flavours and are given by non-diagonal matrices in color space, this leads to a mass matrix that mixes the both the flavor and the color components of the fields. This mixing problem was solved in \cite{Buhl-Mortensen:2016pxs,Buhl-Mortensen:2016jqo}. The key was to use SU(2) representation theory.

Moreover, the vacuum expectation values are proportional to the inverse distance to the defect, $1/x_3$, such that the mass eigenvalues depend on $1/x_3$ as well. Via a Weyl transformation, this $x_3$-dependence can be absorbed to obtain standard propagators in an effective (auxiliary) $AdS_4$ space \cite{Nagasaki:2011ue,Buhl-Mortensen:2016pxs,Buhl-Mortensen:2016jqo}.

\begin{figure}[t]
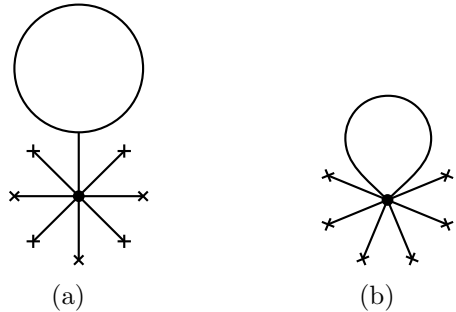

\centering
\subfigure[]{
  \includegraphics{fig2c.pdf}%
}\hspace{2cm}
\subfigure[]{
  \includegraphics{fig2b.pdf}%
}
\caption{Two diagrams have to be considered for the one-loop correction to a one-point function: the lollipop diagram (a) and the tadpole diagram (b).
 \label{fig: one-loop one-point functions}}
\end{figure}

\subsection{One-point functions}

Let us consider the one-loop correction to one-point functions in the SU(2) sector.

\paragraph{Computation}
At one-loop order, two different diagrams have to be considered for the one-loop correction to a one-point function of a single-trace operator built from scalars, see figure \ref{fig: one-loop one-point functions}.
The first of these diagrams, called the lollipop diagram was shown in \cite{Buhl-Mortensen:2016pxs,Buhl-Mortensen:2016jqo} to vanish provided that the employed renormalisation scheme preserves supersymmetry:
\begin{equation}
\label{eq: correction to scalar vev main text}
 \langle \scal_i\rangle_{\text{1-loop}}(x)= 0
 \eqndot
\end{equation}
This leaves us with the contribution of the diagram in figure \ref{subfig: tadpole}, called the tadpole diagram.
The tadpole diagram arises from expanding the composite operator to quadratic order in the quantum fields and contracting these two quantum fields. We also have to correctly normalize the operator using the renormalisation constant \eqref{eq:Z-our-scheme} in the renormalisation scheme that leaves the one-loop two-point function normalised. In the end, we find
\begin{align}
  &\langle\mathcal{Z}\mathcal{O}\rangle_{\text{1-loop,tad}}(x)=g^2\frac{1}{(x_3)^2}\sum_{j}\delta_{s_j=s_{j+1}}\Psi^{s_1\dots s_j\, s_{j+1} \dots i_L}\tr(\scalc_{s_1}\dots\scalc_{s_{j-1}} \scalc_{s_{j+2}}\dots\scalc_{s_L})(x)\nonumber\\
 &\quad+g^2
 \left(
 \frac{1}{2} - \log2+\gammaE - \log (x_3)+\digamma(\tfrac{k+1}{2})
 \right)
\Delta^{(1)}\langle\mathcal{O}\rangle_{\text{tree}}(x)
\label{eq: field theory result for the one-point function}
 \end{align}
for a one-loop eigenstate with one-loop anomalous dimension $\Delta^{(1)}$.
The term proportional to $\log (x_3)$ accounts for the correction to the scaling dimension.

\paragraph{Integrability and asymptotic Bethe Ansatz} 

When computing higher loop corrections to two-point functions, we find that the spin chain Hamiltonian gets corrected
\begin{align}
H = g^2\sum_n g^{2n} H_n .
\end{align}
The interaction range of the Hamiltonian increases with the loop order. Remarkably, the theory remains integrable at loop level even though the interactions are no longer nearest neighbor. Nonetheless, the spectrum can still be described by a Bethe Ansatz. This is done by introducing the coupling constant dependence via the Zhukovsky variable $x$~\cite{Beisert:2005fw}:
\begin{align}
&x + \frac{1}{x} = \frac{u}{g},
&&x = \frac{u}{g} - \frac{g}{u} + O(g^2)\eqncom
\end{align}
where the effective planar coupling constant $g^2$ is related to the 't Hooft coupling $\lambda=Ng_\YM^2$ as $g^2=\frac{\lambda}{16\pi^2}$. The cut of the function $x(u)$ is taken to be the straight line $[-2g,2g]$. The recipe to take  loop corrections into account is to replace the rapidity $u$ in the Bethe equations and the dispersion relation by $x(u)$ and to introduce an additional scattering phase.

More precisely, the all-loop asymptotic Bethe equations which determine the conformal operators of ${\cal N}=4$ SYM theory and their anomalous dimensions
are given by \cite{Beisert:2005fw}:
\begin{align}\label{eq:BAEquantum}
1 = &\left(\frac{x(u_k-\frac{i}{2})}{x(u_k+\frac{i}{2})}\right)^L \prod_{j\neq k} \frac{u_k-u_j+i}{u_k-u_j-i} \, \nonumber 
\exp (2i \theta(u_k,u_j))  \eqncom
\end{align}
where $\exp(2i\theta(u_k,u_j))$ is the so-called dressing phase \cite{Beisert:2006ez}. The anomalous dimension is then given by
\begin{align}
\Delta -L = \frac{i}{x(u+\ihalf)} - \frac{i}{x(u-\ihalf)} .
\end{align}
The asymptotic Bethe Ansatz holds up to the loop order $L$.

Following the recipe of the spectral problem, we are led to a natural generalization of \eqref{eq:SU2quotient}  by replacing the 
classical Bethe function $\Phi$ by the quantum Bethe function $\tilde{\Phi}$. Furthermore, 
the corresponding generalization of the the transfer matrix is the following one
\begin{align}\label{eq:Transfer quantum}
\tilde{T}_n(u) = g^L \sum_{a=-\frac{n}{2}}^{\frac{n}{2}} x(u+ia)^L\frac{Q(u+\frac{n+1}{2}i)Q(u-\frac{n+1}{2}i)}{Q(u+(a-\frac{1}{2})i)Q(u+(a+\frac{1}{2})i)}\eqndot
\end{align}
This gives an expression for  \eqref{eq:SU2genK} at the quantum level. 

Thus we would naturally expect the following Ansatz to work at loop level for one-point functions \cite{Buhl-Mortensen:2017ind}
\begin{align}\label{eq:Ansatz}
C_k =  i^L\tilde{T}_{k-1}(0)\sqrt{\frac{Q(\ihalf)Q(0)}{Q^2(\frac{ik}{2})} }\sqrt{\frac{\det  \tilde{G}_+}{\det  \tilde{G}_-}} 
\,\mathbb{F}_k\eqncom
\end{align}
where the Bethe roots are assumed to satisfy the all-loop asymptotic Bethe equations \eqref{eq:BAEquantum}. Moreover, the introduction of a flux factor $\mathbb{F}_k$ was needed in \eqref{eq:Ansatz}, and it was found to be of the form
\begin{align}
\mathbb{F}_k = 1+g^2 \Big[ \Psi(\textstyle{\frac{k+1}{2}}) + \gammaE - \log 2 \Big] \Delta^{(1)}+O(g^4)\,.
\end{align}
The above Ansatz can then be checked against explicit computations and perfect agreement has been found \cite{Buhl-Mortensen:2017ind}.

\subsection{Two-point functions}

Having set up the quantum field theoretical framework to compute one-point functions at loop level, a natural generalization is to consider two-point functions. This was undertaken in \cite{deLeeuw:2017dkd}. Several scalar two-point functions have been computed to leading order in the coupling constant. They are naturally expressed in terms of hypergeometric functions and indeed take the form of two-point functions in defect conformal field theories \eqref{eq:TwoPoint}.

A first step towards formulating an integrable approach to the computation of two-point functions was made in \cite{Widen:2017uwh}. In particular, it was shown that when one of the operators has length two, the Wick contractions can be interpreted as operators insertions in a spin chain. This reduces the computation of two-point functions to correlation functions on the Heisenberg spin chain and relates them to the one-point functions. 

\subsection{Wilson loops}

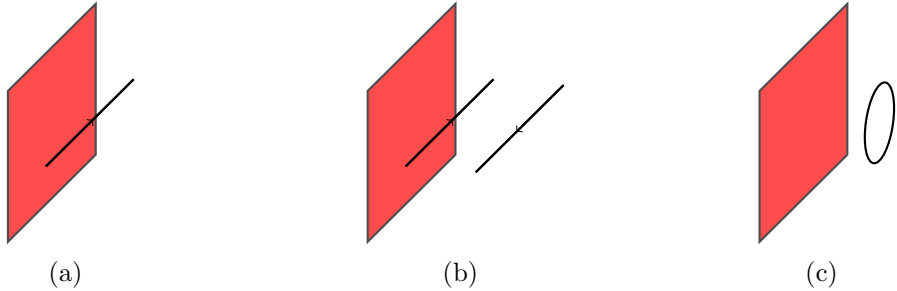
\begin{figure}[h]
\centering
 \subfigure[]{
\centering
\begin{tikzpicture}
	[	axis/.style={->,black,thick},
		axisline/.style={black,thick},
		cube/.style={opacity=.7, thick,fill=red}]
	\draw[cube] (0,-1,-1.5) -- (0,1,-1.5) -- (0,1,1.5) -- (0,-1,1.5) -- cycle;
\draw[axisline] (0.5,0,-1.5) -- (0.5,0,1.5) ;
\draw[middlearrow={<}] (0.5,0,-1.5) -- (0.5,0,1.5) ;
\draw[axisline] (0.5,0,1.5) -- (0.5,0,-1.5) ;
\end{tikzpicture}
  \label{subfig: tree}
} 
\hspace{2.5cm}
 \subfigure[]{
\centering
\begin{tikzpicture}
	[	axis/.style={->,black,thick},
		axisline/.style={black,thick},
		line/.style={black,thick},
		cube/.style={opacity=.7, thick,fill=red}]
	\draw[cube] (0,-1,-1.5) -- (0,1,-1.5) -- (0,1,1.5) -- (0,-1,1.5) -- cycle;
\draw[axisline] (0.5,0,-1.5) -- (0.5,0,1.5) ;
\draw[axisline] (0.5,0,1.5) -- (0.5,0,-1.5) ;
\draw[middlearrow={<}] (0.5,0,-1.5) -- (0.5,0,1.5) ;
\draw[axisline] (2,.5,0) -- (2,.5,3) ;
\draw[axisline] (2,.5,3) -- (2,.5,0) ;
\draw[middlearrow={<}] (2,.5,3) -- (2,.5,0)  ;
\end{tikzpicture}
  \label{subfig: lollipop}
} 
\hspace{2cm}
 \subfigure[]{
\centering
\begin{tikzpicture}
	[	axis/.style={->,black,thick},
		axisline/.style={black,thick},
		line/.style={black,thick},
		cube/.style={opacity=.7, thick,fill=red}]
	\draw[cube] (0,-1,-1.5) -- (0,1,-1.5) -- (0,1,1.5) -- (0,-1,1.5) -- cycle;
   \begin{scope}[canvas is yz plane at x=1]
     \draw[line] (0,0) circle (.5);
   \end{scope}
\end{tikzpicture}
  \label{subfig: tadpole}
}
\caption{The different Wilson line configurations that have been studied in the D3-D5 defect version of $\mathcal{N}=4$ SYM theory: 
(a) a single infinite Wilson line of length $T$,
(b) two antiparallel Wilson lines of length $T$,
(c) A circular Wilson loop parallel to the defect.
}
\label{fig: Wilson loop}
\end{figure}

Apart from one- and two-point functions, non-local operators in defect $\mathcal{N}=4$ SYM have been studied at tree-level and one-loop level as well. In particular, various configurations of Wilson loops have been studied. The presence of the defect clearly affects the vacuum expectation value of a Wilson line, which can, for instance, be used to compute the particle-interface potential. In Figure \ref{fig: Wilson loop}, the studied configurations are shown. Below we will briefly discuss each of the three configurations and refer the reader to the original papers for the details of the computations.

\paragraph{Single Wilson line}

The first set-up that was considered was a single, infinite Wilson line of length $T$ that runs parallel to the defect \cite{deLeeuw:2016vgp}. At tree-level the expectation value of the Wilson line can straightforwardly be computed
\begin{align}\label{eqtreeWilsonLine}
  \langle W\rangle_{\text{tree}} \sim N-k + e^{\frac{k-1}{2} \frac{\sin\chi}{x_3}T},
\end{align}
which holds in the large $T$ limit. The angle $\chi$ parameterizes the supersymmetric extension of the Wilson loop by including a coupling to the scalar fields.

The expectation value of the Wilson loop is related to the particle-interface potential as
\begin{equation}
 \langle W(x_3)\rangle \cong \exp(-T\,V(x_3))\eqncom 
\end{equation}
for $T\to\infty$.
At tree level, we therefore have from \eqref{eqtreeWilsonLine}
\begin{equation}
 V_{\text{tree}}(x_3)=-\frac{k-1}{2x_3}\sin\chi\eqncom
\end{equation}
which is a standard $1/x_3$ potential.

At one-loop level, one again finds contributions of lollipop and tadpole type. Evaluating the relevant Feynman diagrams then gives an explicit loop correction to the particle-interface potential, which simplifies in the large $k$ limit to
\begin{equation}
 V=V_{\text{tree}}\left[1+\frac{\lambda}{4\pi^2k^2}\frac{\sin\chi}{\cos^3\chi}\left(\frac{\pi}{2}-\chi-\frac{1}{2}\sin2\chi\right) +\mathcal{O}\left(\frac{\lambda^2}{k^4}\right)\right]\eqndot \label{V}
\end{equation}
We reproduce the string results from \cite{Nagasaki:2011ue}.

\paragraph{Antiparallel Wilson lines} A more general set-up was discussed in \cite{Preti:2017fhw}, where two antiparallel Wilson lines separated by a distance $d$ were placed at distance $L$ from the defect. From the perspective of the defect field theory, the configuration basically degenerates in the planar limit into two copies of the single Wilson line considered above. However, at strong coupling the system exhibits a Gross-Ooguri transition between a connected and a disconnected string phase.

\paragraph{Circular Wilson loop} Finally, in \cite{Aguilera-Damia:2016bqv} a circular Wilson loop parallel to the defect was considered. The authors considered a circle with radius $R$ at distance $L$ from the defect. This set-up was considered both from the string theory and the field theory side. The expectation value of this Wilson loop was computed to one-loop level in the field theory and in the large $k$ limit a simple expression can be derived analytically
\begin{align}
\log\langle W \rangle \sim \frac{k\pi R}{L}\left[1+\frac{\lambda}{4\pi^2k^2}\frac{\sin^2\chi + (\frac{L}{R})^2 }{\sin\chi\cos^3\chi}\left(\frac{\pi}{2}-\chi-\frac{1}{2}\sin2\chi\right) +\mathcal{O}\left(\frac{\lambda^2}{k^4}\right)\right]\eqndot 
\end{align}
Notice that in the $R\rightarrow\infty$ limit, the one-loop correction reduces to the Wilson line result \eqref{V}. For small values of the coupling $\chi$ the above expression could be compared to a string theory computation and agreement was found in the range of validity of the perturbative expansions \cite{Aguilera-Damia:2016bqv}.

\section{Conclusions and discussion}\label{sec:Disc}

In this review we discussed the computation of correlation functions in conformal field theories with a co-dimension one defect. We mainly focussed on the computation of one-point functions in the dCFT which is holographically dual to a D3-D5 brane set-up in $\ads$. The corresponding D3-D5 dCFT consists of a copy of $\mathcal{N}=4$ SYM theory on both sides of the defect with different ranks of the gauge group. The mismatch in degrees of freedom is offset by assigning vacuum expectation values to the scalar fields on one side of the defect. This breaks the symmetry and gives a non-zero vev to scalar operators. The vacuum expectation value is parameterized by a $k$-dimensional representation of $\alg{su}(2)$.

In the planar limit, conformal operators in $\mathcal{N}=4$ SYM theory can be mapped to the states of an integrable spin chain. The problem of computing the one-point function of an operator at tree level is then mapped to an overlap between the corresponding spin chain state and a Matrix Product State. Similar quantities appear in the context of quantum quenches in condensed matter physics. Building on these results we were able to derive a closed formula of determinant type for all one-point functions in the scalar sector of  $\mathcal{N}=4$ SYM with the D3-D5 defect.

Via the gauge-string duality our results could be directly compared against string theory computations and exact agreement has been found. This provides a new test of AdS/CFT where part of the supersymmetry and conformal symmetry is broken. In the SU(2) sector, the results were even extended to one-loop level and perfect agreement with the corresponding string theory computation was found.

While a lot of progress has been made, the field is still developing. Even for the D3-D5 defect, a lot of open questions remain. First, some results remain to be proven rigorously. For the SU(2) sector the formula still needs to be proven for $k=3$ and a proof for the full scalar sector formula \eqref{SO6formula} is also missing. Second, the realization of the quantum integrability of the full theory is an open problem. We were able to postulate a one-loop formula in the SU(2) sector, but it is unclear how it extends to the full quantum theory. At higher loop orders, the effects like wrapping also come into play and are important to be understood. These are open problems from the field theory side. On the string theory side of the duality, there are very few results. In particular, one-point functions of non-protected operators and two-point functions of any type have not been computed so far. As we have shown, there are now some field theory results for these quantities which could be compared with string theory computations.

There is much to more learn in the other defect set-ups as well. For the SO(5) D3-D7 defect, integrability was proven in \cite{deLeeuw:2018mkd} and some first results for one-point functions were found in \cite{deLeeuw:2016ofj}. However, a complete understanding and a closed formula in the scalar sector remains to be found. In this case, the quantum theoretical framework also remains to be formulated. This set-up breaks all supersymmetry and it would be interesting to see if this affects the computations at quantum level. Because of this, it would be particularly important to check the holographic duality for this system.

For the $\alg{su}(2)\times\alg{su}(2)$ D3-D7 defect, the quantum field theory was developed recently in \cite{Grau:2018keb}. However, the only one-point functions that have been computed so far are of chiral primaries. Actually, one can check that the matrix product state of this $\alg{su}(2)\oplus\alg{su}(2)$ defect CFT is not annihilated by the odd charges of the N = 4 SYM spin chain \cite{KdLV}. This violates the integrability criterion \eqref{eq:IntDef} and seems to suggest that no closed formula exists even in the SU(2) sector of $\mathcal{N}=4$ SYM theory. Indeed, so far it has only been possible to derive results for tree-level one-point functions of non-protected operators on a case by case basis \cite{KdLV}. 

A different interesting avenue is to consider theories that are related to $\mathcal{N}=4$ SYM theory by integrability preserving twists. An example of this is beta- or gamma-deformed $\mathcal{N}=4$ SYM. There one can also determine a MPS and compute one-point functions \cite{Widen:2018nnu}. Unfortunately also here integrability of the defect appears to be lost and closed formulas might be out of reach, but for some special states one-point functions can still be computed. It would be also interesting to understand the dual picture of this deformed dCFT.

Finally, there are a lot of interesting directions in related fields as well. We already mentioned the close link to quantum quenches and our understanding of integrability in nested systems is still incomplete and under development \cite{Piroli:2018ksf,Piroli:2018don}. A different interesting avenue is that of the boundary bootstrap program \cite{Liendo:2012hy,Liendo:2016ymz,Andrei:2018die}. This approach uses the remaining conformal symmetry to restrict and determine the conformal data of defect CFTs. In our system we were able to find closed formulas for one-point functions, which could, for instance, be used as additional input data for the bootstrap program.

\paragraph{Acknowledgements} First and foremost, I would like to thank all my collaborators I. Buhl-Mortensen, A. Ipsen, C. Kristjansen, G. Linardopoulos, S. Mori, K. Vardinghus, M. Wilhelm and K. Zarembo. I would also like to thank the students and fellow lecturers of the Young Researchers Integrability School and Workshop in Ascona for discussions and comments on the lecture notes. I would also like to thank C. Kristjansen, G. Linardopoulos and A. Pribytok for comments on the manuscript. I was supported by SFI and the Royal Society for funding under grant UF160578.

\begin{bibtex}[\jobname]
@article{Grau:2018keb,
      author         = "Grau, Aleix Gimenez and Kristjansen, Charlotte and Volk,
                        Matthias and Wilhelm, Matthias",
      title          = "{A Quantum Check of Non-Supersymmetric AdS/dCFT}",
      year           = "2018",
      eprint         = "1810.11463",
      archivePrefix  = "arXiv",
      primaryClass   = "hep-th",
      SLACcitation   = "
}

@article{KdLV,
      author         = "de Leeuw, Marius and Kristjansen, Charlotte and
                        Vardinghus, Kasper E.",
      title          = "{A non-integrable quench from AdS/dCFT}",
      year           = "2019",
      eprint         = "1906.10714",
      archivePrefix  = "arXiv",
      primaryClass   = "hep-th",
      SLACcitation   = "
}

@article{Liendo:2016ymz,
      author         = "Liendo, Pedro and Meneghelli, Carlo",
      title          = "{Bootstrap equations for $ \mathcal{N} $ = 4 SYM with
                        defects}",
      journal        = "JHEP",
      volume         = "01",
      year           = "2017",
      pages          = "122",
      doi            = "10.1007/JHEP01(2017)122",
      eprint         = "1608.05126",
      archivePrefix  = "arXiv",
      primaryClass   = "hep-th",
      SLACcitation   = "
}

@article{Andrei:2018die,
      author         = "Andrei, Natan and others",
      title          = "{Boundary and Defect CFT: Open Problems and
                        Applications}",
      year           = "2018",
      eprint         = "1810.05697",
      archivePrefix  = "arXiv",
      primaryClass   = "hep-th",
      SLACcitation   = "
}

@article{Buhl-Mortensen:2017ind,
      author         = "Buhl-Mortensen, Isak and de Leeuw, Marius and Ipsen,
                        Asger C. and Kristjansen, Charlotte and Wilhelm, Matthias",
      title          = "{Asymptotic One-Point Functions in Gauge-String Duality
                        with Defects}",
      journal        = "Phys. Rev. Lett.",
      volume         = "119",
      year           = "2017",
      number         = "26",
      pages          = "261604",
      doi            = "10.1103/PhysRevLett.119.261604",
      eprint         = "1704.07386",
      archivePrefix  = "arXiv",
      primaryClass   = "hep-th",
      SLACcitation   = "
}
@article{Levkovich-Maslyuk:2016kfv,
      author         = "Levkovich-Maslyuk, Fedor",
      title          = "{The Bethe ansatz}",
      journal        = "J. Phys.",
      volume         = "A49",
      year           = "2016",
      number         = "32",
      pages          = "323004",
      doi            = "10.1088/1751-8113/49/32/323004",
      eprint         = "1606.02950",
      archivePrefix  = "arXiv",
      primaryClass   = "hep-th",
      SLACcitation   = "
}

@article{deLeeuw:2018mkd,
      author         = "De Leeuw, Marius and Kristjansen, Charlotte and
                        Linardopoulos, Georgios",
      title          = "{Scalar one-point functions and matrix product states of
                        AdS/dCFT}",
      journal        = "Phys. Lett.",
      volume         = "B781",
      year           = "2018",
      pages          = "238-243",
      doi            = "10.1016/j.physletb.2018.03.083",
      eprint         = "1802.01598",
      archivePrefix  = "arXiv",
      primaryClass   = "hep-th",
      reportNumber   = "TCDMATH-18-04",
      SLACcitation   = "
}

@ARTICLE{Brockmann2,
       author = {{Brockmann}, M. and {De Nardis}, J. and {Wouters}, B. and {Caux}, J. -S.},
        title = "{N{\'e}el-XXZ state overlaps: odd particle numbers and Lieb-Liniger
        scaling limit}",
      journal = {Journal of Physics A Mathematical General},
     keywords = {Condensed Matter - Statistical Mechanics},
         year = 2014,
        month = Aug,
       volume = {47},
          eid = {345003},
        pages = {345003},
          doi = {10.1088/1751-8113/47/34/345003},
archivePrefix = {arXiv},
       eprint = {1403.7469},
 primaryClass = {cond-mat.stat-mech},
       adsurl = {https://ui.adsabs.harvard.edu/#abs/2014JPhA...47H5003B},
      adsnote = {Provided by the SAO/NASA Astrophysics Data System}
}

@ARTICLE{Brockmann1,
       author = {{Brockmann}, M. and {De Nardis}, J. and {Wouters}, B. and {Caux}, J. -S.},
        title = "{A Gaudin-like determinant for overlaps of N{\'e}el and XXZ Bethe states}",
      journal = {Journal of Physics A Mathematical General},
     keywords = {Condensed Matter - Statistical Mechanics, Mathematical Physics},
         year = 2014,
        month = Apr,
       volume = {47},
          eid = {145003},
        pages = {145003},
          doi = {10.1088/1751-8113/47/14/145003},
archivePrefix = {arXiv},
       eprint = {1401.2877},
 primaryClass = {cond-mat.stat-mech},
       adsurl = {https://ui.adsabs.harvard.edu/#abs/2014JPhA...47n5003B},
      adsnote = {Provided by the SAO/NASA Astrophysics Data System}
}

@article{Preti:2017fhw,
      author         = "Preti, Michelangelo and Trancanelli, Diego and Vescovi,
                        Edoardo",
      title          = "{Quark-antiquark potential in defect conformal field
                        theory}",
      journal        = "JHEP",
      volume         = "10",
      year           = "2017",
      pages          = "079",
      doi            = "10.1007/JHEP10(2017)079",
      eprint         = "1708.04884",
      archivePrefix  = "arXiv",
      primaryClass   = "hep-th",
      SLACcitation   = "
}

@article{Beisert:2010jr,
      author         = "Beisert, Niklas and others",
      title          = "{Review of AdS/CFT Integrability: An Overview}",
      journal        = "Lett. Math. Phys.",
      volume         = "99",
      year           = "2012",
      pages          = "3-32",
      doi            = "10.1007/s11005-011-0529-2",
      eprint         = "1012.3982",
      archivePrefix  = "arXiv",
      primaryClass   = "hep-th",
      reportNumber   = "AEI-2010-175, CERN-PH-TH-2010-306, HU-EP-10-87,
                        HU-MATH-2010-22, KCL-MTH-10-10, UMTG-270, UUITP-41-10",
      SLACcitation   = "
}

@inproceedings{DHoker:2002nbb,
      author         = "D'Hoker, Eric and Freedman, Daniel Z.",
      title          = "{Supersymmetric gauge theories and the AdS / CFT
                        correspondence}",
      booktitle      = "{Strings, Branes and Extra Dimensions: TASI 2001:
                        Proceedings}",
      year           = "2002",
      pages          = "3-158",
      eprint         = "hep-th/0201253",
      archivePrefix  = "arXiv",
      primaryClass   = "hep-th",
      reportNumber   = "UCLA-02-TEP-3, MIT-CTP-3242",
      SLACcitation   = "
}

@inproceedings{deLeeuw:2017cop,
      author         = "de Leeuw, Marius and Ipsen, Asger C. and Kristjansen,
                        Charlotte and Wilhelm, Matthias",
      title          = "{Introduction to Integrability and One-point Functions in
                        $\mathcal{N}=4$ SYM and its Defect Cousin}",
      booktitle      = "{Les Houches Summer School: Integrability: From
                        Statistical Systems to Gauge Theory Les Houches, France,
                        June 6-July 1, 2016}",
      url            = "https://inspirehep.net/record/1614793/files/arXiv:1708.02525.pdf",
      year           = "2017",
      eprint         = "1708.02525",
      archivePrefix  = "arXiv",
      primaryClass   = "hep-th",
      SLACcitation   = "
}

@article{tHooft:1973alw,
      author         = "'t Hooft, Gerard",
      title          = "{A Planar Diagram Theory for Strong Interactions}",
      journal        = "Nucl. Phys.",
      volume         = "B72",
      year           = "1974",
      pages          = "461",
      doi            = "10.1016/0550-3213(74)90154-0",
      reportNumber   = "CERN-TH-1786",
      SLACcitation   = "
}

@article{Fioravanti:2016bmi,
      author         = "Fioravanti, Davide and Nepomechie, Rafael I.",
      title          = "{An inhomogeneous Lax representation for the Hirota
                        equation}",
      year           = "2016",
      eprint         = "1609.06761",
      archivePrefix  = "arXiv",
      primaryClass   = "math-ph",
      reportNumber   = "UMTG-289",
      SLACcitation   = "
}

@article{Becken2000449,
title = "The analytic continuation of the Gaussian hypergeometric function 2F1(a,b;c;z) for arbitrary parameters ",
journal = "Journal of Computational and Applied Mathematics ",
volume = "126",
number = "1–2",
pages = "449 - 478",
year = "2000",
note = "",
issn = "0377-0427",
doi = "http://dx.doi.org/10.1016/S0377-0427(00)00267-3",
url = "http://www.sciencedirect.com/science/article/pii/S0377042700002673",
author = "W. Becken and P. Schmelcher",
abstract = "Our objective is to provide a complete table of analytic condinuation formulas for the Gaussian hypergeometric function 2F1(a,b;c;z) which allow its fast and accurate computation for arbitrary values of z and of the parameters a,b,c. To this end we distinguish 12 basis sets of the two-dimensional space of the solutions of the hypergeometric differential equation. Representing 2F1 in each of them yields 12 analytic continuation formulas. Each two of them are series in one of the arguments z, z/(z−1), (1−z), (1−1/z), 1/z, 1/(1−z), respectively, such that any given argument z, with the exception of two single points in the complex plane, lies in the convergence domain of at least one of them. We present rapidly converging series representations of 2F1 for all possible constellations of parameters. For minimizing the effort for the derivation of these series we have extensively used the symmetry group of the hypergeometric equation, which is shown to be isomorphic to the cubic group Oh. "
}

@ARTICLE{PozsgayXXZ,
       author = {{Pozsgay}, B.},
        title = "{Overlaps with arbitrary two-site states in the XXZ spin chain}",
      journal = {Journal of Statistical Mechanics: Theory and Experiment},
     keywords = {Condensed Matter - Statistical Mechanics, Nonlinear Sciences - Exactly Solvable and Integrable Systems},
         year = "2018",
        month = "May",
       volume = {5},
        pages = {053103},
          doi = {10.1088/1742-5468/aabbe1},
archivePrefix = {arXiv},
       eprint = {1801.03838},
 primaryClass = {cond-mat.stat-mech},
       adsurl = {https://ui.adsabs.harvard.edu/\#abs/2018JSMTE..05.3103P},
      adsnote = {Provided by the SAO/NASA Astrophysics Data System}
}

@article{Liendo:2012hy,
      author         = "Liendo, Pedro and Rastelli, Leonardo and van Rees, Balt
                        C.",
      title          = "{The Bootstrap Program for Boundary CFT$_\mathrm{d}$}",
      journal        = "JHEP",
      volume         = "07",
      year           = "2013",
      pages          = "113",
      doi            = "10.1007/JHEP07(2013)113",
      eprint         = "1210.4258",
      archivePrefix  = "arXiv",
      primaryClass   = "hep-th",
      reportNumber   = "YITP-SB-12-37",
      SLACcitation   = "
}

@article{deLeeuw:2015hxa,
      author         = "de Leeuw, Marius and Kristjansen, Charlotte and Zarembo,
                        Konstantin",
      title          = "{One-point Functions in Defect CFT and Integrability}",
      journal        = "JHEP",
      volume         = "08",
      year           = "2015",
      pages          = "098",
      doi            = "10.1007/JHEP08(2015)098",
      eprint         = "1506.06958",
      archivePrefix  = "arXiv",
      primaryClass   = "hep-th",
      reportNumber   = "NORDITA-2015-72, UUITP-12-15",
      SLACcitation   = "
}
@article{Nagasaki:2011ue,
      author         = "Nagasaki, Koichi and Tanida, Hiroaki and Yamaguchi,
                        Satoshi",
      title          = "{Holographic Interface-Particle Potential}",
      journal        = "JHEP",
      volume         = "01",
      year           = "2012",
      pages          = "139",
      doi            = "10.1007/JHEP01(2012)139",
      eprint         = "1109.1927",
      archivePrefix  = "arXiv",
      primaryClass   = "hep-th",
      reportNumber   = "OU-HET-723",
      SLACcitation   = "
}

@article{Alday:2009zm,
      author         = "Alday, Luis F. and Henn, Johannes M. and Plefka, Jan and
                        Schuster, Theodor",
      title          = "{Scattering into the fifth dimension of $\mathcal{N}=4$ super
                        Yang-Mills}",
      journal        = "JHEP",
      volume         = "01",
      year           = "2010",
      pages          = "077",
      doi            = "10.1007/JHEP01(2010)077",
      eprint         = "0908.0684",
      archivePrefix  = "arXiv",
      primaryClass   = "hep-th",
      reportNumber   = "HU-EP-09-35",
      SLACcitation   = "
}

@article{Allen:1985wd,
      author         = "Allen, Bruce and Jacobson, Theodore",
      title          = "{Vector Two Point Functions in Maximally Symmetric
                        Spaces}",
      journal        = "Commun. Math. Phys.",
      volume         = "103",
      year           = "1986",
      pages          = "669",
      doi            = "10.1007/BF01211169",
      reportNumber   = "UCSB-TH-4-1985",
      SLACcitation   = "
}

@article{Liu:1998ty,
      author         = "Liu, Hong and Tseytlin, Arkady A.",
      title          = "{On four point functions in the CFT / AdS
                        correspondence}",
      journal        = "Phys. Rev.",
      volume         = "D59",
      year           = "1999",
      pages          = "086002",
      doi            = "10.1103/PhysRevD.59.086002",
      eprint         = "hep-th/9807097",
      archivePrefix  = "arXiv",
      primaryClass   = "hep-th",
      reportNumber   = "IMPERIAL-TP-97-98-060",
      SLACcitation   = "
}

@article{Kent:2014nya,
      author         = "Kent, Carl and Winstanley, Elizabeth",
      title          = "{Hadamard renormalized scalar field theory on anti-de
                        Sitter spacetime}",
      journal        = "Phys. Rev.",
      volume         = "D91",
      year           = "2015",
      number         = "4",
      pages          = "044044",
      doi            = "10.1103/PhysRevD.91.044044",
      eprint         = "1408.6738",
      archivePrefix  = "arXiv",
      primaryClass   = "gr-qc",
      SLACcitation   = "
}

@article{Henningson:1998cd,
      author         = "Henningson, Mans and Sfetsos, Konstadinos",
      title          = "{Spinors and the AdS/CFT correspondence}",
      journal        = "Phys. Lett.",
      volume         = "B431",
      year           = "1998",
      pages          = "63-68",
      doi            = "10.1016/S0370-2693(98)00559-0",
      eprint         = "hep-th/9803251",
      archivePrefix  = "arXiv",
      primaryClass   = "hep-th",
      reportNumber   = "CERN-TH-98-078",
      SLACcitation   = "
}

@article{Mueck:1998iz,
      author         = "Mueck, Wolfgang and Viswanathan, K. S.",
      title          = "{Conformal field theory correlators from classical field
                        theory on anti-de Sitter space. 2. Vector and spinor
                        fields}",
      journal        = "Phys. Rev.",
      volume         = "D58",
      year           = "1998",
      pages          = "106006",
      doi            = "10.1103/PhysRevD.58.106006",
      eprint         = "hep-th/9805145",
      archivePrefix  = "arXiv",
      primaryClass   = "hep-th",
      SLACcitation   = "
}

@article{Ambrus:2015mfa,
      author         = "Ambrus, Victor E. and Winstanley, Elizabeth",
      title          = "{Renormalised fermion vacuum expectation values on
                        anti-de Sitter space-time}",
      journal        = "Phys. Lett.",
      volume         = "B749",
      year           = "2015",
      pages          = "597-602",
      doi            = "10.1016/j.physletb.2015.08.045",
      eprint         = "1505.04962",
      archivePrefix  = "arXiv",
      primaryClass   = "hep-th",
      SLACcitation   = "
}

@article{Camporesi:1992wn,
      author         = "Camporesi, R. and Higuchi, A.",
      title          = "{Stress energy tensors in anti-de Sitter space-time}",
      journal        = "Phys. Rev.",
      volume         = "D45",
      year           = "1992",
      pages          = "3591-3603",
      doi            = "10.1103/PhysRevD.45.3591",
      SLACcitation   = "
}

@article{Camporesi91,
  title = {$\ensuremath{\zeta}$-function regularization of one-loop effective potentials in anti-de Sitter spacetime},
  author = {Camporesi, Roberto},
  journal = {Phys. Rev. D},
  volume = {43},
  issue = {12},
  pages = {3958--3965},
  numpages = {0},
  year = {1991},
  month = {Jun},
  publisher = {American Physical Society},
  doi = {10.1103/PhysRevD.43.3958},
  url = {http://link.aps.org/doi/10.1103/PhysRevD.43.3958}
}

@article{Caldarelli99,
      author         = "Caldarelli, Marco M.",
      title          = "{Quantum scalar fields on anti-de Sitter space-time}",
      journal        = "Nucl. Phys.",
      volume         = "B549",
      year           = "1999",
      pages          = "499-515",
      doi            = "10.1016/S0550-3213(99)00137-6",
      eprint         = "hep-th/9809144",
      archivePrefix  = "arXiv",
      primaryClass   = "hep-th",
      reportNumber   = "UTF-422",
      SLACcitation   = "
}

@phdthesis{Hoppe82,
  author = {Jens Hoppe},
  title = {Quantum theory of a massless relativistic surface and a two-dimensional bound state problem},
  school = {MIT},
  year = 1982,
  url = {http://hdl.handle.net/1721.1/15717}
}

@article{deWit88,
  title = "On the quantum mechanics of supermembranes ",
  journal = "Nuclear Physics B ",
  volume = "305",
  number = "4",
  pages = "545 - 581",
  year = "1988",
  note = "",
  issn = "0550-3213",
  doi = "http://dx.doi.org/10.1016/0550-3213(88)90116-2",
  url = "http://www.sciencedirect.com/science/article/pii/0550321388901162",
  author = "B. de Wit and J. Hoppe and H. Nicolai"
}

@article{Iso01,
  title = "Noncommutative gauge theory on fuzzy sphere from matrix model ",
  journal = "Nuclear Physics B ",
  volume = "604",
  number = "1-2",
  pages = "121 - 147",
  year = "2001",
  note = "",
  issn = "0550-3213",
  doi = "http://dx.doi.org/10.1016/S0550-3213(01)00173-0",
  url = "http://www.sciencedirect.com/science/article/pii/S0550321301001730",
  author = "Satoshi Iso and Yusuke Kimura and Kanji Tanaka and Kazunori Wakatsuki"
}

@article{Kimura01,
  author = {Kimura, Yusuke}, 
  title = {Noncommutative Gauge Theories on Fuzzy Sphere and Fuzzy Torus from Matrix Model},
  volume = {106}, 
  number = {2}, 
  pages = {445-469}, 
  year = {2001}, 
  doi = {10.1143/PTP.106.445}, 
  URL = {http://ptp.oxfordjournals.org/content/106/2/445.abstract}, 
  journal = {Progress of Theoretical Physics} 
}

@article{Bertlmann08,
  author={Reinhold A Bertlmann and Philipp Krammer},
  title={Bloch vectors for qudits},
  journal={Journal of Physics A: Mathematical and Theoretical},
  volume={41},
  number={23},
  pages={235303},
  url={http://stacks.iop.org/1751-8121/41/i=23/a=235303},
  year={2008}
}

@article{Nagasaki:2012re,
      author         = "Nagasaki, Koichi and Yamaguchi, Satoshi",
      title          = "{Expectation values of chiral primary operators in
                        holographic interface CFT}",
      journal        = "Phys. Rev.",
      volume         = "D86",
      year           = "2012",
      pages          = "086004",
      doi            = "10.1103/PhysRevD.86.086004",
      eprint         = "1205.1674",
      archivePrefix  = "arXiv",
      primaryClass   = "hep-th",
      reportNumber   = "OU-HET-749",
      SLACcitation   = "
}

@article{Mueck00,
  author={Wolfgang M{\" u}ck},
  title={Spinor parallel propagator and Green function in maximally symmetric spaces},
  journal={Journal of Physics A: Mathematical and General},
  volume={33},
  number={15},
  pages={3021},
  url={http://stacks.iop.org/0305-4470/33/i=15/a=308},
  year={2000}
}

@article{Kawamoto:2015qla,
      author         = "Kawamoto, Shoichi and Kuroki, Tsunehide",
      title          = "{Existence of new nonlocal field theory on noncommutative
                        space and spiral flow in renormalization group analysis of
                        matrix models}",
      journal        = "JHEP",
      volume         = "06",
      year           = "2015",
      pages          = "062",
      doi            = "10.1007/JHEP06(2015)062",
      eprint         = "1503.08411",
      archivePrefix  = "arXiv",
      primaryClass   = "hep-th",
      SLACcitation   = "
}

@article{Cardy:1984bb,
      author         = "Cardy, John L.",
      title          = "{Conformal Invariance and Surface Critical Behavior}",
      journal        = "Nucl. Phys.",
      volume         = "B240",
      year           = "1984",
      pages          = "514-532",
      doi            = "10.1016/0550-3213(84)90241-4",
      SLACcitation   = "
}

@article{Beisert:2010jr,
      author         = "Beisert, Niklas and others",
      title          = "{Review of AdS/CFT Integrability: An Overview}",
      journal        = "Lett. Math. Phys.",
      volume         = "99",
      year           = "2012",
      pages          = "3-32",
      doi            = "10.1007/s11005-011-0529-2",
      eprint         = "1012.3982",
      archivePrefix  = "arXiv",
      primaryClass   = "hep-th",
      reportNumber   = "AEI-2010-175, CERN-PH-TH-2010-306, HU-EP-10-87,
                        HU-MATH-2010-22, KCL-MTH-10-10, UMTG-270, UUITP-41-10",
      SLACcitation   = "
}

@article{deLeeuw:2016umh,
      author         = "de Leeuw, Marius and Kristjansen, Charlotte and Mori,
                        Stefano",
      title          = "{AdS/dCFT one-point functions of the SU(3) sector}",
      journal        = "Phys. Lett.",
      volume         = "B763",
      year           = "2016",
      pages          = "197-202",
      doi            = "10.1016/j.physletb.2016.10.044",
      eprint         = "1607.03123",
      archivePrefix  = "arXiv",
      primaryClass   = "hep-th",
      SLACcitation   = "
}

@article{Buhl-Mortensen:2016pxs,
      author         = "Buhl-Mortensen, Isak and de Leeuw, Marius and Ipsen,
                        Asger C. and Kristjansen, Charlotte and Wilhelm, Matthias",
      title          = "{One-loop one-point functions in gauge-gravity dualities
                        with defects}",
      journal        = "Phys. Rev. Lett.",
      volume         = "117",
      year           = "2016",
      number         = "23",
      pages          = "231603",
      doi            = "10.1103/PhysRevLett.117.231603",
      eprint         = "1606.01886",
      archivePrefix  = "arXiv",
      primaryClass   = "hep-th",
      SLACcitation   = "
}

@article{Buhl-Mortensen:2015gfd,
      author         = "Buhl-Mortensen, Isak and de Leeuw, Marius and
                        Kristjansen, Charlotte and Zarembo, Konstantin",
      title          = "{One-point Functions in AdS/dCFT from Matrix Product
                        States}",
      journal        = "JHEP",
      volume         = "02",
      year           = "2016",
      pages          = "052",
      doi            = "10.1007/JHEP02(2016)052",
      eprint         = "1512.02532",
      archivePrefix  = "arXiv",
      primaryClass   = "hep-th",
      reportNumber   = "NORDITA-2015-132, UUITP-26-15",
      SLACcitation   = "
}

@article{Kristjansen:2012tn,
      author         = "Kristjansen, Charlotte and Semenoff, Gordon W. and Young,
                        Donovan",
      title          = "{Chiral primary one-point functions in the D3-D7 defect
                        conformal field theory}",
      journal        = "JHEP",
      volume         = "01",
      year           = "2013",
      pages          = "117",
      doi            = "10.1007/JHEP01(2013)117",
      eprint         = "1210.7015",
      archivePrefix  = "arXiv",
      primaryClass   = "hep-th",
      reportNumber   = "NORDITA-2012-81",
      SLACcitation   = "
}

@article{Maldacena:1997re,
      author         = "Maldacena, Juan Martin",
      title          = "{The Large N limit of superconformal field theories and
                        supergravity}",
      journal        = "Int. J. Theor. Phys.",
      volume         = "38",
      year           = "1999",
      pages          = "1113-1133",
      doi            = "10.1023/A:1026654312961",
      note           = "[Adv. Theor. Math. Phys.2,231(1998)]",
      eprint         = "hep-th/9711200",
      archivePrefix  = "arXiv",
      primaryClass   = "hep-th",
      reportNumber   = "HUTP-97-A097, HUTP-98-A097",
      SLACcitation   = "
}

@article{DeWolfe:2001pq,
      author         = "DeWolfe, Oliver and Freedman, Daniel Z. and Ooguri,
                        Hirosi",
      title          = "{Holography and defect conformal field theories}",
      journal        = "Phys. Rev.",
      volume         = "D66",
      year           = "2002",
      pages          = "025009",
      doi            = "10.1103/PhysRevD.66.025009",
      eprint         = "hep-th/0111135",
      archivePrefix  = "arXiv",
      primaryClass   = "hep-th",
      reportNumber   = "CALT-68-2359, CITUSC-01-041, NSF-ITP-01-172,
                        MIT-CTP-3212",
      SLACcitation   = "
}

@article{Karch:2000gx,
      author         = "Karch, Andreas and Randall, Lisa",
      title          = "{Open and closed string interpretation of SUSY CFT's on
                        branes with boundaries}",
      journal        = "JHEP",
      volume         = "06",
      year           = "2001",
      pages          = "063",
      doi            = "10.1088/1126-6708/2001/06/063",
      eprint         = "hep-th/0105132",
      archivePrefix  = "arXiv",
      primaryClass   = "hep-th",
      reportNumber   = "MIT-CTP-3146",
      SLACcitation   = "
}

@article{Nahm:1979yw,
      author         = "Nahm, W.",
      title          = "{A Simple Formalism for the BPS Monopole}",
      journal        = "Phys. Lett.",
      volume         = "B90",
      year           = "1980",
      pages          = "413-414",
      doi            = "10.1016/0370-2693(80)90961-2",
      reportNumber   = "CERN-TH-2796",
      SLACcitation   = "
}

@article{Diaconescu:1996rk,
      author         = "Diaconescu, Duiliu-Emanuel",
      title          = "{D-branes, monopoles and Nahm equations}",
      journal        = "Nucl. Phys.",
      volume         = "B503",
      year           = "1997",
      pages          = "220-238",
      doi            = "10.1016/S0550-3213(97)00438-0",
      eprint         = "hep-th/9608163",
      archivePrefix  = "arXiv",
      primaryClass   = "hep-th",
      SLACcitation   = "
}

@article{Constable:1999ac,
      author         = "Constable, Neil R. and Myers, Robert C. and Tafjord,
                        Oyvind",
      title          = "{The Noncommutative bion core}",
      journal        = "Phys. Rev.",
      volume         = "D61",
      year           = "2000",
      pages          = "106009",
      doi            = "10.1103/PhysRevD.61.106009",
      eprint         = "hep-th/9911136",
      archivePrefix  = "arXiv",
      primaryClass   = "hep-th",
      reportNumber   = "MCGILL-99-34, NSF-ITP-99-138",
      SLACcitation   = "
}
@article{Zarembo:2010rr,
      author         = "Zarembo, K.",
      title          = "{Holographic three-point functions of semiclassical
                        states}",
      journal        = "JHEP",
      volume         = "09",
      year           = "2010",
      pages          = "030",
      doi            = "10.1007/JHEP09(2010)030",
      eprint         = "1008.1059",
      archivePrefix  = "arXiv",
      primaryClass   = "hep-th",
      reportNumber   = "ITEP-TH-29-10, UUITP-25-10",
      SLACcitation   = "
}
@article{Klose:2011rm,
      author         = "Klose, Thomas and McLoughlin, Tristan",
      title          = "{A light-cone approach to three-point functions in $AdS_5\times S^5$}",
      journal        = "JHEP",
      volume         = "04",
      year           = "2012",
      pages          = "080",
      doi            = "10.1007/JHEP04(2012)080",
      eprint         = "1106.0495",
      archivePrefix  = "arXiv",
      primaryClass   = "hep-th",
      reportNumber   = "UUITP-16-11, AEI-2011-031",
      SLACcitation   = "
}
@article{Kristjansen:2002bb,
      author         = "Kristjansen, C. and Plefka, J. and Semenoff, G. W. and
                        Staudacher, M.",
      title          = "{A New double scaling limit of N=4 superYang-Mills theory
                        and PP wave strings}",
      journal        = "Nucl. Phys.",
      volume         = "B643",
      year           = "2002",
      pages          = "3-30",
      doi            = "10.1016/S0550-3213(02)00749-6",
      eprint         = "hep-th/0205033",
      archivePrefix  = "arXiv",
      primaryClass   = "hep-th",
      reportNumber   = "AEI-2002-036",
      SLACcitation   = "
}
@article{Lee:1998bxa,
      author         = "Lee, Sangmin and Minwalla, Shiraz and Rangamani, Mukund
                        and Seiberg, Nathan",
      title          = "{Three point functions of chiral operators in D = 4, N=4
                        SYM at large N}",
      journal        = "Adv. Theor. Math. Phys.",
      volume         = "2",
      year           = "1998",
      pages          = "697-718",
      eprint         = "hep-th/9806074",
      archivePrefix  = "arXiv",
      primaryClass   = "hep-th",
      reportNumber   = "PUPT-1796, IASSNS-HEP-98-51",
      SLACcitation   = "
}
@article{Vieira:2013wya,
      author         = "Vieira, Pedro and Wang, Tianheng",
      title          = "{Tailoring Non-Compact Spin Chains}",
      journal        = "JHEP",
      volume         = "10",
      year           = "2014",
      pages          = "35",
      doi            = "10.1007/JHEP10(2014)035",
      eprint         = "1311.6404",
      archivePrefix  = "arXiv",
      primaryClass   = "hep-th",
      SLACcitation   = "
}
@article{Escobedo:2010xs,
      author         = "Escobedo, Jorge and Gromov, Nikolay and Sever, Amit and
                        Vieira, Pedro",
      title          = "{Tailoring Three-Point Functions and Integrability}",
      journal        = "JHEP",
      volume         = "09",
      year           = "2011",
      pages          = "028",
      doi            = "10.1007/JHEP09(2011)028",
      eprint         = "1012.2475",
      archivePrefix  = "arXiv",
      primaryClass   = "hep-th",
      SLACcitation   = "
}

@article{Minahan:2002ve,
      author         = "Minahan, J. A. and Zarembo, K.",
      title          = "{The Bethe ansatz for N=4 superYang-Mills}",
      journal        = "JHEP",
      volume         = "03",
      year           = "2003",
      pages          = "013",
      doi            = "10.1088/1126-6708/2003/03/013",
      eprint         = "hep-th/0212208",
      archivePrefix  = "arXiv",
      primaryClass   = "hep-th",
      reportNumber   = "UUITP-17-02, ITEP-TH-73-02",
      SLACcitation   = "
}

@Article{Bethe1931,
author="Bethe, H.",
title="Zur Theorie der Metalle",
journal="Zeitschrift f{\"u}r Physik",
year="1931",
month="Mar",
day="01",
volume="71",
number="3",
pages="205--226",
}

@inproceedings{Faddeev:1996iy,
      author         = "Faddeev, L. D.",
      title          = "{How algebraic Bethe ansatz works for integrable model}",
      booktitle      = "{Relativistic gravitation and gravitational radiation.
                        Proceedings, School of Physics, Les Houches, France,
                        September 26-October 6, 1995}",
      year           = "1996",
      pages          = "pp. 149-219",
      eprint         = "hep-th/9605187",
      archivePrefix  = "arXiv",
      primaryClass   = "hep-th",
      SLACcitation   = "
}

@Article{Slavnov1989,
author="Slavnov, N. A.",
title="Calculation of scalar products of wave functions and form factors in the framework of the alcebraic Bethe ansatz",
journal="Theoretical and Mathematical Physics",
year="1989",
month="May",
day="01",
volume="79",
number="2",
pages="502--508",
}

@article{Piroli:2017sei,
      author         = "Piroli, Lorenzo and Pozsgay, Balázs and Vernier, Eric",
      title          = "{What is an integrable quench?}",
      journal        = "Nucl. Phys.",
      volume         = "B925",
      year           = "2017",
      pages          = "362-402",
      doi            = "10.1016/j.nuclphysb.2017.10.012",
      SLACcitation   = "
}

@article{Bazhanov:2010ts,
      author         = "Bazhanov, Vladimir V. and \L{}ukowski, Tomasz and
                        Meneghelli, Carlo and Staudacher, Matthias",
      title          = "{A Shortcut to the Q-Operator}",
      journal        = "J. Stat. Mech.",
      volume         = "1011",
      year           = "2010",
      pages          = "P11002",
      doi            = "10.1088/1742-5468/2010/11/P11002",
      eprint         = "1005.3261",
      archivePrefix  = "arXiv",
      primaryClass   = "hep-th",
      reportNumber   = "AEI-2010-023, HU-EP-10-24, HU-MATHEMATIK-2010-7",
      SLACcitation   = "
}

@article{Brockmann,
  author={M Brockmann},
  title={Overlaps of $q$-raised N{\' e}el states with XXZ Bethe states and their relation to the Lieb-Liniger Bose gas},
  journal={Journal of Statistical Mechanics: Theory and Experiment},
  volume={2014},
  number={5},
  pages={P05006},
  year={2014},
  }

@article{Pozsgay,
  author={Pozsgay, B.},
  title={Overlaps between eigenstates of the XXZ spin-1/2 chain and a class of simple product states},
  journal={Journal of Statistical Mechanics: Theory and Experiment},
  volume={2014},
  number={6},
  pages={P06011},
  year={2014},
}

@article{Foda:2015nfk,
      author         = "Foda, O. and Zarembo, K.",
      title          = "{Overlaps of partial N{\' e}el states and Bethe states}",
      journal        = "J. Stat. Mech.",
      volume         = "1602",
      year           = "2016",
      number         = "2",
      pages          = "023107",
      doi            = "10.1088/1742-5468/2016/02/023107",
      eprint         = "1512.02533",
      archivePrefix  = "arXiv",
      primaryClass   = "hep-th",
      reportNumber   = "NORDITA-2015-133, UUITP-27-15",
      SLACcitation   = "
}

@article{deLeeuw:2016vgp,
      author         = "de Leeuw, Marius and Ipsen, Asger C. and Kristjansen,
                        Charlotte and Wilhelm, Matthias",
      title          = "{One-loop Wilson loops and the particle-interface
                        potential in AdS/dCFT}",
      journal        = "Phys. Lett.",
      volume         = "B768",
      year           = "2017",
      pages          = "192-197",
      doi            = "10.1016/j.physletb.2017.02.047",
      eprint         = "1608.04754",
      archivePrefix  = "arXiv",
      primaryClass   = "hep-th",
      SLACcitation   = "
}

@article{Erdmenger:2002ex,
      author         = "Erdmenger, Johanna and Guralnik, Zachary and Kirsch,
                        Ingo",
      title          = "{Four-dimensional superconformal theories with
                        interacting boundaries or defects}",
      journal        = "Phys. Rev.",
      volume         = "D66",
      year           = "2002",
      pages          = "025020",
      doi            = "10.1103/PhysRevD.66.025020",
      eprint         = "hep-th/0203020",
      archivePrefix  = "arXiv",
      primaryClass   = "hep-th",
      reportNumber   = "HU-EP-02-07",
      SLACcitation   = "
}

@article{Buhl-Mortensen:2016jqo,
      author         = "Buhl-Mortensen, Isak and de Leeuw, Marius and Ipsen,
                        Asger C. and Kristjansen, Charlotte and Wilhelm, Matthias",
      title          = "{A Quantum Check of AdS/dCFT}",
      journal        = "JHEP",
      volume         = "01",
      year           = "2017",
      pages          = "098",
      doi            = "10.1007/JHEP01(2017)098",
      eprint         = "1611.04603",
      archivePrefix  = "arXiv",
      primaryClass   = "hep-th",
      SLACcitation   = "
}

@article{Liendo:2012hy,
      author         = "Liendo, Pedro and Rastelli, Leonardo and van Rees, Balt
                        C.",
      title          = "{The Bootstrap Program for Boundary CFT$_\mathrm{d}$}",
      journal        = "JHEP",
      volume         = "07",
      year           = "2013",
      pages          = "113",
      doi            = "10.1007/JHEP07(2013)113",
      eprint         = "1210.4258",
      archivePrefix  = "arXiv",
      primaryClass   = "hep-th",
      reportNumber   = "YITP-SB-12-37",
      SLACcitation   = "
}

@article{deLeeuw:2015hxa,
      author         = "de Leeuw, Marius and Kristjansen, Charlotte and Zarembo,
                        Konstantin",
      title          = "{One-point Functions in Defect CFT and Integrability}",
      journal        = "JHEP",
      volume         = "08",
      year           = "2015",
      pages          = "098",
      doi            = "10.1007/JHEP08(2015)098",
      eprint         = "1506.06958",
      archivePrefix  = "arXiv",
      primaryClass   = "hep-th",
      reportNumber   = "NORDITA-2015-72, UUITP-12-15",
      SLACcitation   = "
}
@article{Nagasaki:2011ue,
      author         = "Nagasaki, Koichi and Tanida, Hiroaki and Yamaguchi,
                        Satoshi",
      title          = "{Holographic Interface-Particle Potential}",
      journal        = "JHEP",
      volume         = "01",
      year           = "2012",
      pages          = "139",
      doi            = "10.1007/JHEP01(2012)139",
      eprint         = "1109.1927",
      archivePrefix  = "arXiv",
      primaryClass   = "hep-th",
      reportNumber   = "OU-HET-723",
      SLACcitation   = "
}

@article{Aguilera-Damia:2016bqv,
      author         = "Aguilera-Damia, Jeremias and Correa, Diego H. and
                        Giraldo-Rivera, Victor I.",
      title          = "{Circular Wilson loops in defect Conformal Field Theory}",
      journal        = "JHEP",
      volume         = "03",
      year           = "2017",
      pages          = "023",
      doi            = "10.1007/JHEP03(2017)023",
      eprint         = "1612.07991",
      archivePrefix  = "arXiv",
      primaryClass   = "hep-th",
      SLACcitation   = "
}
@article{Tseytlin:2010jv,
      author         = "Tseytlin, A. A.",
      title          = "{Review of AdS/CFT Integrability, Chapter II.1: Classical
                        $AdS_5\times S^5$ string solutions}",
      journal        = "Lett. Math. Phys.",
      volume         = "99",
      year           = "2012",
      pages          = "103-125",
      doi            = "10.1007/s11005-011-0466-0",
      eprint         = "1012.3986",
      archivePrefix  = "arXiv",
      primaryClass   = "hep-th",
      reportNumber   = "IMPERIAL-TP-AT-2010-05",
      SLACcitation   = "
}

@article{Nagasaki:2012re,
      author         = "Nagasaki, Koichi and Yamaguchi, Satoshi",
      title          = "{Expectation values of chiral primary operators in
                        holographic interface CFT}",
      journal        = "Phys. Rev.",
      volume         = "D86",
      year           = "2012",
      pages          = "086004",
      doi            = "10.1103/PhysRevD.86.086004",
      eprint         = "1205.1674",
      archivePrefix  = "arXiv",
      primaryClass   = "hep-th",
      reportNumber   = "OU-HET-749",
      SLACcitation   = "
}
@article{Widen:2017uwh,
      author         = "Widen, Erik",
      title          = "{Two-point functions of SU(2)-subsector and length-two
                        operators in dCFT}",
      journal        = "Phys. Lett.",
      volume         = "B773",
      year           = "2017",
      pages          = "435-439",
      doi            = "10.1016/j.physletb.2017.08.059",
      eprint         = "1705.08679",
      archivePrefix  = "arXiv",
      primaryClass   = "hep-th",
      SLACcitation   = "
}

@article{Cardy:1984bb,
      author         = "Cardy, John L.",
      title          = "{Conformal Invariance and Surface Critical Behavior}",
      journal        = "Nucl. Phys.",
      volume         = "B240",
      year           = "1984",
      pages          = "514-532",
      doi            = "10.1016/0550-3213(84)90241-4",
      SLACcitation   = "
}

@article{Beisert:2010jr,
      author         = "Beisert, Niklas and others",
      title          = "{Review of AdS/CFT Integrability: An Overview}",
      journal        = "Lett. Math. Phys.",
      volume         = "99",
      year           = "2012",
      pages          = "3-32",
      doi            = "10.1007/s11005-011-0529-2",
      eprint         = "1012.3982",
      archivePrefix  = "arXiv",
      primaryClass   = "hep-th",
      reportNumber   = "AEI-2010-175, CERN-PH-TH-2010-306, HU-EP-10-87,
                        HU-MATH-2010-22, KCL-MTH-10-10, UMTG-270, UUITP-41-10",
      SLACcitation   = "
}
@article{deLeeuw:2016umh,
      author         = "de Leeuw, Marius and Kristjansen, Charlotte and Mori,
                        Stefano",
      title          = "{AdS/dCFT one-point functions of the SU(3) sector}",
      journal        = "Phys. Lett.",
      volume         = "B763",
      year           = "2016",
      pages          = "197-202",
      doi            = "10.1016/j.physletb.2016.10.044",
      eprint         = "1607.03123",
      archivePrefix  = "arXiv",
      primaryClass   = "hep-th",
      SLACcitation   = "
}

@article{Buhl-Mortensen:2016pxs,
      author         = "Buhl-Mortensen, Isak and de Leeuw, Marius and Ipsen,
                        Asger C. and Kristjansen, Charlotte and Wilhelm, Matthias",
      title          = "{One-loop one-point functions in gauge-gravity dualities
                        with defects}",
      journal        = "Phys. Rev. Lett.",
      volume         = "117",
      year           = "2016",
      number         = "23",
      pages          = "231603",
      doi            = "10.1103/PhysRevLett.117.231603",
      eprint         = "1606.01886",
      archivePrefix  = "arXiv",
      primaryClass   = "hep-th",
      SLACcitation   = "
}

@article{Buhl-Mortensen:2015gfd,
      author         = "Buhl-Mortensen, Isak and de Leeuw, Marius and
                        Kristjansen, Charlotte and Zarembo, Konstantin",
      title          = "{One-point Functions in AdS/dCFT from Matrix Product
                        States}",
      journal        = "JHEP",
      volume         = "02",
      year           = "2016",
      pages          = "052",
      doi            = "10.1007/JHEP02(2016)052",
      eprint         = "1512.02532",
      archivePrefix  = "arXiv",
      primaryClass   = "hep-th",
      reportNumber   = "NORDITA-2015-132, UUITP-26-15",
      SLACcitation   = "
}

@article{Kristjansen:2012tn,
      author         = "Kristjansen, Charlotte and Semenoff, Gordon W. and Young,
                        Donovan",
      title          = "{Chiral primary one-point functions in the D3-D7 defect
                        conformal field theory}",
      journal        = "JHEP",
      volume         = "01",
      year           = "2013",
      pages          = "117",
      doi            = "10.1007/JHEP01(2013)117",
      eprint         = "1210.7015",
      archivePrefix  = "arXiv",
      primaryClass   = "hep-th",
      reportNumber   = "NORDITA-2012-81",
      SLACcitation   = "
}

@article{Maldacena:1997re,
      author         = "Maldacena, Juan Martin",
      title          = "{The Large N limit of superconformal field theories and
                        supergravity}",
      journal        = "Int. J. Theor. Phys.",
      volume         = "38",
      year           = "1999",
      pages          = "1113-1133",
      doi            = "10.1023/A:1026654312961",
      note           = "[Adv. Theor. Math. Phys.2,231(1998)]",
      eprint         = "hep-th/9711200",
      archivePrefix  = "arXiv",
      primaryClass   = "hep-th",
      reportNumber   = "HUTP-97-A097, HUTP-98-A097",
      SLACcitation   = "
}

@article{DeWolfe:2001pq,
      author         = "DeWolfe, Oliver and Freedman, Daniel Z. and Ooguri,
                        Hirosi",
      title          = "{Holography and defect conformal field theories}",
      journal        = "Phys. Rev.",
      volume         = "D66",
      year           = "2002",
      pages          = "025009",
      doi            = "10.1103/PhysRevD.66.025009",
      eprint         = "hep-th/0111135",
      archivePrefix  = "arXiv",
      primaryClass   = "hep-th",
      reportNumber   = "CALT-68-2359, CITUSC-01-041, NSF-ITP-01-172,
                        MIT-CTP-3212",
      SLACcitation   = "
}

@article{Karch:2000gx,
      author         = "Karch, Andreas and Randall, Lisa",
      title          = "{Open and closed string interpretation of SUSY CFT's on
                        branes with boundaries}",
      journal        = "JHEP",
      volume         = "06",
      year           = "2001",
      pages          = "063",
      doi            = "10.1088/1126-6708/2001/06/063",
      eprint         = "hep-th/0105132",
      archivePrefix  = "arXiv",
      primaryClass   = "hep-th",
      reportNumber   = "MIT-CTP-3146",
      SLACcitation   = "
}

@article{Nahm:1979yw,
      author         = "Nahm, W.",
      title          = "{A Simple Formalism for the BPS Monopole}",
      journal        = "Phys. Lett.",
      volume         = "B90",
      year           = "1980",
      pages          = "413-414",
      doi            = "10.1016/0370-2693(80)90961-2",
      reportNumber   = "CERN-TH-2796",
      SLACcitation   = "
}

@article{Diaconescu:1996rk,
      author         = "Diaconescu, Duiliu-Emanuel",
      title          = "{D-branes, monopoles and Nahm equations}",
      journal        = "Nucl. Phys.",
      volume         = "B503",
      year           = "1997",
      pages          = "220-238",
      doi            = "10.1016/S0550-3213(97)00438-0",
      eprint         = "hep-th/9608163",
      archivePrefix  = "arXiv",
      primaryClass   = "hep-th",
      SLACcitation   = "
}

@article{Constable:1999ac,
      author         = "Constable, Neil R. and Myers, Robert C. and Tafjord,
                        Oyvind",
      title          = "{The Noncommutative bion core}",
      journal        = "Phys. Rev.",
      volume         = "D61",
      year           = "2000",
      pages          = "106009",
      doi            = "10.1103/PhysRevD.61.106009",
      eprint         = "hep-th/9911136",
      archivePrefix  = "arXiv",
      primaryClass   = "hep-th",
      reportNumber   = "MCGILL-99-34, NSF-ITP-99-138",
      SLACcitation   = "
}
@article{Buhl-Mortensen:2017ind,
      author         = "Buhl-Mortensen, Isak and de Leeuw, Marius and Ipsen,
                        Asger C. and Kristjansen, Charlotte and Wilhelm, Matthias",
      title          = "{Asymptotic one-point functions in AdS/dCFT}",
      year           = "2017",
      eprint         = "1704.07386",
      archivePrefix  = "arXiv",
      primaryClass   = "hep-th",
      SLACcitation   = "
}
@article{Beisert:2003tq,
      author         = "Beisert, N. and Kristjansen, C. and Staudacher, M.",
      title          = "{The Dilatation operator of conformal $\mathcal{N}=4$ superYang-Mills
                        theory}",
      journal        = "Nucl. Phys.",
      volume         = "B664",
      year           = "2003",
      pages          = "131-184",
      doi            = "10.1016/S0550-3213(03)00406-1",
      eprint         = "hep-th/0303060",
      archivePrefix  = "arXiv",
      primaryClass   = "hep-th",
      reportNumber   = "AEI-2003-028",
      SLACcitation   = "
}
@article{Gaiotto:2008sa,
      author         = "Gaiotto, Davide and Witten, Edward",
      title          = "{Supersymmetric Boundary Conditions in $\mathcal{N}=4$ Super
                        Yang-Mills Theory}",
      journal        = "J. Statist. Phys.",
      volume         = "135",
      year           = "2009",
      pages          = "789-855",
      doi            = "10.1007/s10955-009-9687-3",
      eprint         = "0804.2902",
      archivePrefix  = "arXiv",
      primaryClass   = "hep-th",
      SLACcitation   = "
}

@article{deLeeuw:2016ofj,
      author         = "de Leeuw, Marius and Kristjansen, Charlotte and
                        Linardopoulos, Georgios",
      title          = "{One-point functions of non-protected operators in the
                        SO(5) symmetric D3-D7 dCFT}",
      journal        = "J. Phys.",
      volume         = "A50",
      year           = "2017",
      number         = "25",
      pages          = "254001",
      doi            = "10.1088/1751-8121/aa714b",
      eprint         = "1612.06236",
      archivePrefix  = "arXiv",
      primaryClass   = "hep-th",
      SLACcitation   = "
}

@article{Heslop:2001gp,
      author         = "Heslop, P. J. and Howe, Paul S.",
      title          = "{OPEs and three-point correlators of protected operators
                        in $\mathcal{N}=4$ SYM}",
      journal        = "Nucl. Phys.",
      volume         = "B626",
      year           = "2002",
      pages          = "265-286",
      doi            = "10.1016/S0550-3213(02)00023-8",
      eprint         = "hep-th/0107212",
      archivePrefix  = "arXiv",
      primaryClass   = "hep-th",
      SLACcitation   = "
}
@article{Baggio:2012rr,
      author         = "Baggio, Marco and de Boer, Jan and Papadodimas, Kyriakos",
      title          = "{A non-renormalization theorem for chiral primary 3-point
                        functions}",
      journal        = "JHEP",
      volume         = "07",
      year           = "2012",
      pages          = "137",
      doi            = "10.1007/JHEP07(2012)137",
      eprint         = "1203.1036",
      archivePrefix  = "arXiv",
      primaryClass   = "hep-th",
      SLACcitation   = "
}
@article{Minahan:2006sk,
      author         = "Minahan, J. A.",
      title          = "{A brief introduction to the Bethe ansatz in $\mathcal{N}=4$
                        super-Yang-Mills}",
      journal        = "J. Phys.",
      volume         = "A39",
      year           = "2006",
      pages          = "12657-12677",
      doi            = "10.1088/0305-4470/39/41/S02",
      SLACcitation   = "
}
@article{Serban:2010sr,
      author         = "Serban, Didina",
      title          = "{Integrability and the AdS/CFT correspondence}",
      journal        = "J. Phys.",
      volume         = "A44",
      year           = "2011",
      pages          = "124001",
      doi            = "10.1088/1751-8113/44/12/124001",
      eprint         = "1003.4214",
      archivePrefix  = "arXiv",
      primaryClass   = "hep-th",
      reportNumber   = "IPHT-T10-XXX",
      SLACcitation   = "
}
@article{Beisert:2004ry,
      author         = "Beisert, Niklas",
      title          = "{The Dilatation operator of $\mathcal{N}=4$ super Yang-Mills theory
                        and integrability}",
      journal        = "Phys. Rept.",
      volume         = "405",
      year           = "2004",
      pages          = "1-202",
      doi            = "10.1016/j.physrep.2004.09.007",
      eprint         = "hep-th/0407277",
      archivePrefix  = "arXiv",
      primaryClass   = "hep-th",
      reportNumber   = "AEI-2004-057",
      SLACcitation   = "
}
@article{Carlson:2011hy,
      author         = "Carlson, Warren and de Mello Koch, Robert and Lin, Hai",
      title          = "{Nonplanar Integrability}",
      journal        = "JHEP",
      volume         = "03",
      year           = "2011",
      pages          = "105",
      doi            = "10.1007/JHEP03(2011)105",
      eprint         = "1101.5404",
      archivePrefix  = "arXiv",
      primaryClass   = "hep-th",
      reportNumber   = "WITS-CTP-065",
      SLACcitation   = "
}
@article{Beisert:2005fw,
      author         = "Beisert, Niklas and Staudacher, Matthias",
      title          = "{Long-range psu$(2,2|4)$ Bethe Ans\"{a}tze for gauge theory and
                        strings}",
      journal        = "Nucl. Phys.",
      volume         = "B727",
      year           = "2005",
      pages          = "1-62",
      doi            = "10.1016/j.nuclphysb.2005.06.038",
      eprint         = "hep-th/0504190",
      archivePrefix  = "arXiv",
      primaryClass   = "hep-th",
      reportNumber   = "AEI-2005-092, PUTP-2159",
      SLACcitation   = "
}
@article{Vieira:2010kb,
      author         = "Vieira, Pedro and Volin, Dmytro",
      title          = "{Review of AdS/CFT Integrability, Chapter III.3: The
                        Dressing factor}",
      journal        = "Lett. Math. Phys.",
      volume         = "99",
      year           = "2012",
      pages          = "231-253",
      doi            = "10.1007/s11005-011-0482-0",
      eprint         = "1012.3992",
      archivePrefix  = "arXiv",
      primaryClass   = "hep-th",
      SLACcitation   = "
}
@article{Ambjorn:2005wa,
      author         = "Ambj{\o}rn, Jan and Janik, Romuald A. and Kristjansen,
                        Charlotte",
      title          = "{Wrapping interactions and a new source of corrections to
                        the spin-chain/string duality}",
      journal        = "Nucl. Phys.",
      volume         = "B736",
      year           = "2006",
      pages          = "288-301",
      doi            = "10.1016/j.nuclphysb.2005.12.007",
      eprint         = "hep-th/0510171",
      archivePrefix  = "arXiv",
      primaryClass   = "hep-th",
      reportNumber   = "NORDITA-2005-67",
      SLACcitation   = "
}
@article{Gromov:2012uv,
      author         = "Gromov, Nikolay and Vieira, Pedro",
      title          = "{Tailoring Three-Point Functions and Integrability IV.
                        Theta-morphism}",
      journal        = "JHEP",
      volume         = "04",
      year           = "2014",
      pages          = "068",
      doi            = "10.1007/JHEP04(2014)068",
      eprint         = "1205.5288",
      archivePrefix  = "arXiv",
      primaryClass   = "hep-th",
      SLACcitation   = "
}
@article{Beisert:2002bb,
      author         = "Beisert, N. and Kristjansen, C. and Plefka, J. and
                        Semenoff, G. W. and Staudacher, M.",
      title          = "{BMN correlators and operator mixing in $\mathcal{N}=4$
                        superYang-Mills theory}",
      journal        = "Nucl. Phys.",
      volume         = "B650",
      year           = "2003",
      pages          = "125-161",
      doi            = "10.1016/S0550-3213(02)01025-8",
      eprint         = "hep-th/0208178",
      archivePrefix  = "arXiv",
      primaryClass   = "hep-th",
      reportNumber   = "AEI-2002-061",
      SLACcitation   = "
}
@book{Ammon:2015wua,
      author         = "Ammon, Martin and Erdmenger, Johanna",
      title          = "{Gauge/gravity duality}",
      publisher      = "Cambridge Univ. Pr.",
      address        = "Cambridge, UK",
      year           = "2015",
      url            = "",
      ISBN           = "9781107010345",
      SLACcitation   = "
}
@book{Wess:1992cp,
      author         = "Wess, J. and Bagger, J.",
      title          = "{Supersymmetry and supergravity}",
      publisher =" Princeton Univ. Pr.",
      address= "Princeton, USA",
      year           = "1992",
      SLACcitation   = "
}

@article{Frassek:2015rka,
      author         = "Frassek, Rouven and Meidinger, David and Nandan,
                        Dhritiman and Wilhelm, Matthias",
      title          = "{On-shell diagrams, Gra{\ss}mannians and integrability for
                        form factors}",
      journal        = "JHEP",
      volume         = "01",
      year           = "2016",
      pages          = "182",
      doi            = "10.1007/JHEP01(2016)182",
      eprint         = "1506.08192",
      archivePrefix  = "arXiv",
      primaryClass   = "hep-th",
      reportNumber   = "HU-MATH-2015-10, HU-EP-15-30, DCPT-15-37",
      SLACcitation   = "
}
@article{Chicherin:2013ora,
      author         = "Chicherin, D. and Derkachov, S. and Kirschner, R.",
      title          = "{Yang-Baxter operators and scattering amplitudes in $\mathcal{N}=4$
                        super-Yang-Mills theory}",
      journal        = "Nucl. Phys.",
      volume         = "B881",
      year           = "2014",
      pages          = "467-501",
      doi            = "10.1016/j.nuclphysb.2014.02.016",
      eprint         = "1309.5748",
      archivePrefix  = "arXiv",
      primaryClass   = "hep-th",
      SLACcitation   = "
}
@article{Ferro:2012xw,
      author         = "Ferro, Livia and \L{}ukowski, Tomasz and Meneghelli, Carlo
                        and Plefka, Jan and Staudacher, Matthias",
      title          = "{Harmonic R-matrices for Scattering Amplitudes and
                        Spectral Regularization}",
      journal        = "Phys. Rev. Lett.",
      volume         = "110",
      year           = "2013",
      number         = "12",
      pages          = "121602",
      doi            = "10.1103/PhysRevLett.110.121602",
      eprint         = "1212.0850",
      archivePrefix  = "arXiv",
      primaryClass   = "hep-th",
      reportNumber   = "HU-EP-12-50, HU-MATHEMATIK:14-2012, DESY-12-228,
                        ZMP-HH-12-26, AEI-2012-198, --AEI-2012-198",
      SLACcitation   = "
}

@article{Ferro:2013dga,
      author         = "Ferro, Livia and \L{}ukowski, Tomasz and Meneghelli, Carlo
                        and Plefka, Jan and Staudacher, Matthias",
      title          = "{Spectral Parameters for Scattering Amplitudes in $\mathcal{N}=4$
                        Super Yang-Mills Theory}",
      journal        = "JHEP",
      volume         = "01",
      year           = "2014",
      pages          = "094",
      doi            = "10.1007/JHEP01(2014)094",
      eprint         = "1308.3494",
      archivePrefix  = "arXiv",
      primaryClass   = "hep-th",
      reportNumber   = "HU-MATHEMATIK-2013-12, HU-EP-13-33, AEI-2013-235,
                        DESY-13-488, --ZMP-HH-13-15",
      SLACcitation   = "
}
@article{Broedel:2014pia,
      author         = "Broedel, Johannes and de Leeuw, Marius and Rosso, Matteo",
      title          = "{A dictionary between R-operators, on-shell graphs and
                        Yangian algebras}",
      journal        = "JHEP",
      volume         = "06",
      year           = "2014",
      pages          = "170",
      doi            = "10.1007/JHEP06(2014)170",
      eprint         = "1403.3670",
      archivePrefix  = "arXiv",
      primaryClass   = "hep-th",
      SLACcitation   = "
}

@article{Muller:2013rta,
      author         = "M{\" u}ller, Dennis and M{\" u}nkler, Hagen and Plefka, Jan and
                        Pollok, Jonas and Zarembo, Konstantin",
      title          = "{Yangian Symmetry of smooth Wilson Loops in $\mathcal{N}
                        = $ 4 super Yang-Mills Theory}",
      journal        = "JHEP",
      volume         = "11",
      year           = "2013",
      pages          = "081",
      doi            = "10.1007/JHEP11(2013)081",
      eprint         = "1309.1676",
      archivePrefix  = "arXiv",
      primaryClass   = "hep-th",
      reportNumber   = "HU-EP-13-42, NORDITA-2013-64, UUITP-10-13",
      SLACcitation   = "
}
@article{Basso:2013vsa,
      author         = "Basso, Benjamin and Sever, Amit and Vieira, Pedro",
      title          = "{Spacetime and Flux Tube S-Matrices at Finite Coupling
                        for $\mathcal{N}=4$ Supersymmetric Yang-Mills Theory}",
      journal        = "Phys. Rev. Lett.",
      volume         = "111",
      year           = "2013",
      number         = "9",
      pages          = "091602",
      doi            = "10.1103/PhysRevLett.111.091602",
      eprint         = "1303.1396",
      archivePrefix  = "arXiv",
      primaryClass   = "hep-th",
      SLACcitation   = "
}
article{Basso:2013aha,
      author         = "Basso, Benjamin and Sever, Amit and Vieira, Pedro",
      title          = "{Space-time S-matrix and Flux tube S-matrix II.
                        Extracting and Matching Data}",
      journal        = "JHEP",
      volume         = "01",
      year           = "2014",
      pages          = "008",
      doi            = "10.1007/JHEP01(2014)008",
      eprint         = "1306.2058",
      archivePrefix  = "arXiv",
      primaryClass   = "hep-th",
      SLACcitation   = "
}
@article{Basso:2013aha,
      author         = "Basso, Benjamin and Sever, Amit and Vieira, Pedro",
      title          = "{Space-time S-matrix and Flux tube S-matrix II.
                        Extracting and Matching Data}",
      journal        = "JHEP",
      volume         = "01",
      year           = "2014",
      pages          = "008",
      doi            = "10.1007/JHEP01(2014)008",
      eprint         = "1306.2058",
      archivePrefix  = "arXiv",
      primaryClass   = "hep-th",
      SLACcitation   = "
}
@article{Basso:2017khq,
      author         = "Basso, Benjamin and Coronado, Frank and Komatsu, Shota
                        and Lam, Ho Tat and Vieira, Pedro and Zhong, De-liang",
      title          = "{Asymptotic Four Point Functions}",
      year           = "2017",
      eprint         = "1701.04462",
      archivePrefix  = "arXiv",
      primaryClass   = "hep-th",
      SLACcitation   = "
}
@article{Bargheer:2017eoz,
      author         = "Bargheer, Till",
      title          = "{Four-Point Functions with a Twist}",
      year           = "2017",
      eprint         = "1701.04424",
      archivePrefix  = "arXiv",
      primaryClass   = "hep-th",
      reportNumber   = "DESY-16-248",
      SLACcitation   = "
}
@article{Eden:2016xvg,
      author         = "Eden, Burkhard and Sfondrini, Alessandro",
      title          = "{Tessellating cushions: four-point functions in $\mathcal{N}=4$ SYM}",
      year           = "2016",
      eprint         = "1611.05436",
      archivePrefix  = "arXiv",
      primaryClass   = "hep-th",
      reportNumber   = "HU-EP-16-25",
      SLACcitation   = "
}
@article{Beisert:2010gn,
      author         = "Beisert, Niklas and Henn, Johannes and McLoughlin,
                        Tristan and Plefka, Jan",
      title          = "{One-Loop Superconformal and Yangian Symmetries of
                        Scattering Amplitudes in $\mathcal{N}=4$ Super Yang-Mills}",
      journal        = "JHEP",
      volume         = "04",
      year           = "2010",
      pages          = "085",
      doi            = "10.1007/JHEP04(2010)085",
      eprint         = "1002.1733",
      archivePrefix  = "arXiv",
      primaryClass   = "hep-th",
      reportNumber   = "AEI-2010-019, HU-EP-10-06",
      SLACcitation   = "
}
@article{Drummond:2009fd,
      author         = "Drummond, James M. and Henn, Johannes M. and Plefka, Jan",
      title          = "{Yangian symmetry of scattering amplitudes in $\mathcal{N}=4$ super
                        Yang-Mills theory}",
      booktitle      = "{Strangeness in quark matter. Proceedings, International
                        Conference, SQM 2008, Beijing, P.R. China, October 5-10,
                        2008}",
      journal        = "JHEP",
      volume         = "05",
      year           = "2009",
      pages          = "046",
      doi            = "10.1088/1126-6708/2009/05/046",
      eprint         = "0902.2987",
      archivePrefix  = "arXiv",
      primaryClass   = "hep-th",
      reportNumber   = "HU-EP-09-06, LAPTH-1308-09",
      SLACcitation   = "
}
@article{Basso:2015zoa,
      author         = "Basso, Benjamin and Komatsu, Shota and Vieira, Pedro",
      title          = "{Structure Constants and Integrable Bootstrap in Planar
                        $\mathcal{N}=4$ SYM Theory}",
      year           = "2015",
      eprint         = "1505.06745",
      archivePrefix  = "arXiv",
      primaryClass   = "hep-th",
      SLACcitation   = "
}
@article{Basso:2015eqa,
      author         = "Basso, Benjamin and Goncalves, Vasco and Komatsu, Shota
                        and Vieira, Pedro",
      title          = "{Gluing Hexagons at Three Loops}",
      journal        = "Nucl. Phys.",
      volume         = "B907",
      year           = "2016",
      pages          = "695-716",
      doi            = "10.1016/j.nuclphysb.2016.04.020",
      eprint         = "1510.01683",
      archivePrefix  = "arXiv",
      primaryClass   = "hep-th",
      SLACcitation   = "
}
@article{Basso:2017muf,
      author         = "Basso, Benjamin and Goncalves, Vasco and Komatsu, Shota",
      title          = "{Structure constants at wrapping order}",
      journal        = "JHEP",
      volume         = "05",
      year           = "2017",
      pages          = "124",
      doi            = "10.1007/JHEP05(2017)124",
      eprint         = "1702.02154",
      archivePrefix  = "arXiv",
      primaryClass   = "hep-th",
      SLACcitation   = "
}
@article{Escobedo:2010xs,
      author         = "Escobedo, Jorge and Gromov, Nikolay and Sever, Amit and
                        Vieira, Pedro",
      title          = "{Tailoring Three-Point Functions and Integrability}",
      journal        = "JHEP",
      volume         = "09",
      year           = "2011",
      pages          = "028",
      doi            = "10.1007/JHEP09(2011)028",
      eprint         = "1012.2475",
      archivePrefix  = "arXiv",
      primaryClass   = "hep-th",
      SLACcitation   = "
}
@article{Escobedo:2011xw,
      author         = "Escobedo, Jorge and Gromov, Nikolay and Sever, Amit and
                        Vieira, Pedro",
      title          = "{Tailoring Three-Point Functions and Integrability II.
                        Weak/strong coupling match}",
      journal        = "JHEP",
      volume         = "09",
      year           = "2011",
      pages          = "029",
      doi            = "10.1007/JHEP09(2011)029",
      eprint         = "1104.5501",
      archivePrefix  = "arXiv",
      primaryClass   = "hep-th",
      SLACcitation   = "
}
@article{Gao:2013dza,
      author         = "Gao, Zhiquan and Yang, Gang",
      title          = "{Y-system for form factors at strong coupling in $AdS_5$
                        and with multi-operator insertions in $AdS_3$}",
      journal        = "JHEP",
      volume         = "06",
      year           = "2013",
      pages          = "105",
      doi            = "10.1007/JHEP06(2013)105",
      eprint         = "1303.2668",
      archivePrefix  = "arXiv",
      primaryClass   = "hep-th",
      SLACcitation   = "
}

@article{Harmark:2017yrv,
      author         = "Harmark, Troels and Wilhelm, Matthias",
      title          = "{The Hagedorn temperature of AdS5/CFT4 via
                        integrability}",
      year           = "2017",
      eprint         = "1706.03074",
      archivePrefix  = "arXiv",
      primaryClass   = "hep-th",
      SLACcitation   = "
}

@article{Maldacena:2010kp,
      author         = "Maldacena, Juan and Zhiboedov, Alexander",
      title          = "{Form factors at strong coupling via a Y-system}",
      journal        = "JHEP",
      volume         = "11",
      year           = "2010",
      pages          = "104",
      doi            = "10.1007/JHEP11(2010)104",
      eprint         = "1009.1139",
      archivePrefix  = "arXiv",
      primaryClass   = "hep-th",
      SLACcitation   = "
}
@article{deLeeuw:2017dkd,
      author         = "de Leeuw, Marius and Ipsen, Asger C. and Kristjansen,
                        Charlotte and Vardinghus, Kasper E. and Wilhelm, Matthias",
      title          = "{Two-point functions in AdS/dCFT and the boundary
                        conformal bootstrap equations}",
      journal        = "JHEP",
      volume         = "08",
      year           = "2017",
      pages          = "020",
      doi            = "10.1007/JHEP08(2017)020",
      eprint         = "1705.03898",
      archivePrefix  = "arXiv",
      primaryClass   = "hep-th",
      SLACcitation   = "
}
@article{Gliozzi:2015qsa,
      author         = "Gliozzi, Ferdinando and Liendo, Pedro and Meineri, Marco
                        and Rago, Antonio",
      title          = "{Boundary and Interface CFTs from the Conformal
                        Bootstrap}",
      journal        = "JHEP",
      volume         = "05",
      year           = "2015",
      pages          = "036",
      doi            = "10.1007/JHEP05(2015)036",
      eprint         = "1502.07217",
      archivePrefix  = "arXiv",
      primaryClass   = "hep-th",
      reportNumber   = "HU-EP-15-08",
      SLACcitation   = "
}

@article{Billo:2016cpy,
      author         = "Bill{\' o}, Marco and Gon{\c c}alves, Vasco and Lauria, Edoardo
                        and Meineri, Marco",
      title          = "{Defects in conformal field theory}",
      journal        = "JHEP",
      volume         = "04",
      year           = "2016",
      pages          = "091",
      doi            = "10.1007/JHEP04(2016)091",
      eprint         = "1601.02883",
      archivePrefix  = "arXiv",
      primaryClass   = "hep-th",
      SLACcitation   = "
}
@article{Myers:2008me,
      author         = "Myers, Robert C. and Wapler, Matthias C.",
      title          = "{Transport Properties of Holographic Defects}",
      journal        = "JHEP",
      volume         = "12",
      year           = "2008",
      pages          = "115",
      doi            = "10.1088/1126-6708/2008/12/115",
      eprint         = "0811.0480",
      archivePrefix  = "arXiv",
      primaryClass   = "hep-th",
      SLACcitation   = "
}
@article{Klose:2010ki,
      author         = "Klose, Thomas",
      title          = "{Review of AdS/CFT Integrability, Chapter IV.3: N=6
                        Chern-Simons and Strings on $AdS_4\times CP^3$}",
      journal        = "Lett. Math. Phys.",
      volume         = "99",
      year           = "2012",
      pages          = "401-423",
      doi            = "10.1007/s11005-011-0520-y",
      eprint         = "1012.3999",
      archivePrefix  = "arXiv",
      primaryClass   = "hep-th",
      reportNumber   = "UUITP-37-10",
      SLACcitation   = "
}
@article{Gaiotto:2009tk,
      author         = "Gaiotto, Davide and Jafferis, Daniel Louis",
      title          = "{Notes on adding D6 branes wrapping $RP^3$ in $AdS_4 \times
                        CP^3$}",
      journal        = "JHEP",
      volume         = "11",
      year           = "2012",
      pages          = "015",
      doi            = "10.1007/JHEP11(2012)015",
      eprint         = "0903.2175",
      archivePrefix  = "arXiv",
      primaryClass   = "hep-th",
      SLACcitation   = "
}
@article{Hohenegger:2009as,
      author         = "Hohenegger, Stefan and Kirsch, Ingo",
      title          = "{A Note on the holography of Chern-Simons matter theories
                        with flavour}",
      journal        = "JHEP",
      volume         = "04",
      year           = "2009",
      pages          = "129",
      doi            = "10.1088/1126-6708/2009/04/129",
      eprint         = "0903.1730",
      archivePrefix  = "arXiv",
      primaryClass   = "hep-th",
      SLACcitation   = "
}
@article{Chandrasekhar:2009ey,
      author         = "Chandrasekhar, B. and Panda, Binata",
      title          = "{Brane Embeddings in AdS(4) x CP**3}",
      journal        = "Int. J. Mod. Phys.",
      volume         = "A26",
      year           = "2011",
      pages          = "2377-2404",
      doi            = "10.1142/S0217751X1105347X",
      eprint         = "0909.3061",
      archivePrefix  = "arXiv",
      primaryClass   = "hep-th",
      reportNumber   = "CERN-PH-TH-2009-159",
      SLACcitation   = "
}
@article{Ammon:2009wc,
      author         = "Ammon, Martin and Erdmenger, Johanna and Meyer, Rene and
                        O'Bannon, Andy and Wrase, Timm",
      title          = "{Adding Flavor to AdS$_4$/CFT$_3$}",
      journal        = "JHEP",
      volume         = "11",
      year           = "2009",
      pages          = "125",
      doi            = "10.1088/1126-6708/2009/11/125",
      eprint         = "0909.3845",
      archivePrefix  = "arXiv",
      primaryClass   = "hep-th",
      reportNumber   = "MPP-2009-52, ITP-UH-17-09",
      SLACcitation   = "
}
@article{tHooft:1973alw,
      author         = "'t Hooft, Gerard",
      title          = "{A Planar Diagram Theory for Strong Interactions}",
      journal        = "Nucl. Phys.",
      volume         = "B72",
      year           = "1974",
      pages          = "461",
      doi            = "10.1016/0550-3213(74)90154-0",
      reportNumber   = "CERN-TH-1786",
      SLACcitation   = "
}

@article{Erickson:2000af,
      author         = "Erickson, J. K. and Semenoff, G. W. and Zarembo, K.",
      title          = "{Wilson loops in $\mathcal{N}=4$ supersymmetric Yang-Mills theory}",
      journal        = "Nucl. Phys.",
      volume         = "B582",
      year           = "2000",
      pages          = "155-175",
      doi            = "10.1016/S0550-3213(00)00300-X",
      eprint         = "hep-th/0003055",
      archivePrefix  = "arXiv",
      primaryClass   = "hep-th",
      reportNumber   = "ITEP-TH-13-00",
      SLACcitation   = "
}
@article{Beisert:2003ys,
      author         = "Beisert, Niklas",
      title          = "{The su$(2|3)$ dynamic spin chain}",
      journal        = "Nucl. Phys.",
      volume         = "B682",
      year           = "2004",
      pages          = "487-520",
      doi            = "10.1016/j.nuclphysb.2003.12.032",
      eprint         = "hep-th/0310252",
      archivePrefix  = "arXiv",
      primaryClass   = "hep-th",
      reportNumber   = "AEI-2003-087",
      SLACcitation   = "
}
@article{Zwiebel:2008gr,
      author         = "Zwiebel, Benjamin I.",
      title          = "{Iterative Structure of the $\mathcal{N}=4$ SYM Spin Chain}",
      journal        = "JHEP",
      volume         = "07",
      year           = "2008",
      pages          = "114",
      doi            = "10.1088/1126-6708/2008/07/114",
      eprint         = "0806.1786",
      archivePrefix  = "arXiv",
      primaryClass   = "hep-th",
      reportNumber   = "DAMTP-2008-52",
      SLACcitation   = "
}
@article{Beisert:2007sk,
      author         = "Beisert, Niklas and Zwiebel, Benjamin I.",
      title          = "{On Symmetry Enhancement in the psu$(1,1|2)$ Sector of $\mathcal{N}=4$
                        SYM}",
      journal        = "JHEP",
      volume         = "10",
      year           = "2007",
      pages          = "031",
      doi            = "10.1088/1126-6708/2007/10/031",
      eprint         = "0707.1031",
      archivePrefix  = "arXiv",
      primaryClass   = "hep-th",
      reportNumber   = "AEI-2007-096, PUTP-2232",
      SLACcitation   = "
}

@book{Smirnov:2006ry,
      author         = "Smirnov, V. A.",
      title          = "{Feynman integral calculus}",
      publisher        = "Springer",
      address = "Berlin, Germany",
      year           = "2006",
      SLACcitation   = "
}

@article{Kazakov:1983ns,
      author         = "Kazakov, D. I.",
      title          = "{Calculation of Feynman diagrams by the 
                        'uniqueness' method}",
      journal        = "Theor. Math. Phys.",
      volume         = "58",
      year           = "1984",
      pages          = "223-230",
      doi            = "10.1007/BF01018044",
      note           = "[Teor. Mat. Fiz.58,343(1984)]",
      reportNumber   = "JINR-E2-83-323",
      SLACcitation   = "
}
@article{Gliozzi:1976qd,
      author         = "Gliozzi, F. and Scherk, Joel and Olive, David I.",
      title          = "{Supersymmetry, Supergravity Theories and the Dual Spinor
                        Model}",
      journal        = "Nucl. Phys.",
      volume         = "B122",
      year           = "1977",
      pages          = "253-290",
      doi            = "10.1016/0550-3213(77)90206-1",
      reportNumber   = "CERN-TH-2253",
      SLACcitation   = "
}
@article{Brink:1976bc,
      author         = "Brink, Lars and Schwarz, John H. and Scherk, Joel",
      title          = "{Supersymmetric Yang-Mills Theories}",
      journal        = "Nucl. Phys.",
      volume         = "B121",
      year           = "1977",
      pages          = "77-92",
      doi            = "10.1016/0550-3213(77)90328-5",
      reportNumber   = "CALT-68-574",
      SLACcitation   = "
}
@article{Minahan:2002ve,
      author         = "Minahan, J. A. and Zarembo, K.",
      title          = "{The Bethe ansatz for $\mathcal{N}=4$ superYang-Mills}",
      journal        = "JHEP",
      volume         = "03",
      year           = "2003",
      pages          = "013",
      doi            = "10.1088/1126-6708/2003/03/013",
      eprint         = "hep-th/0212208",
      archivePrefix  = "arXiv",
      primaryClass   = "hep-th",
      reportNumber   = "UUITP-17-02, ITEP-TH-73-02",
      SLACcitation   = "
}
@article{Gaudin:1976sv,
      author         = "Gaudin, M.",
      title          = "{Diagonalization of a Class of Spin Hamiltonians}",
      journal        = "J. Phys. France",
      volume = "37",
      year           = "1976",
      pages = "1086-1098",
      reportNumber   = "SACLAY-DPh-T/76-38",
      SLACcitation   = "
}
@article{Beisert:2003jj,
      author         = "Beisert, Niklas",
      title          = "{The complete one loop dilatation operator of $\mathcal{N}=4$
                        superYang-Mills theory}",
      journal        = "Nucl. Phys.",
      volume         = "B676",
      year           = "2004",
      pages          = "3-42",
      doi            = "10.1016/j.nuclphysb.2003.10.019",
      eprint         = "hep-th/0307015",
      archivePrefix  = "arXiv",
      primaryClass   = "hep-th",
      reportNumber   = "AEI-2003-056",
      SLACcitation   = "
}
@article{Sohnius:1981sn,
      author         = "Sohnius, Martin F. and West, Peter C.",
      title          = "{Conformal Invariance in $\mathcal{N}=4$ Supersymmetric Yang-Mills
                        Theory}",
      journal        = "Phys. Lett.",
      volume         = "B100",
      year           = "1981",
      pages          = "245",
      doi            = "10.1016/0370-2693(81)90326-9",
      reportNumber   = "ICTP/80-81/9",
      SLACcitation   = "
}
@article{Howe:1983sr,
      author         = "Howe, Paul S. and Stelle, K. S. and Townsend, P. K.",
      title          = "{Miraculous Ultraviolet Cancellations in Supersymmetry
                        Made Manifest}",
      journal        = "Nucl. Phys.",
      volume         = "B236",
      year           = "1984",
      pages          = "125-166",
      doi            = "10.1016/0550-3213(84)90528-5",
      reportNumber   = "ICTP-82-83-20",
      SLACcitation   = "
}
@article{Brink:1982wv,
      author         = "Brink, Lars and Lindgren, Olof and Nilsson, Bengt E. W.",
      title          = "{The Ultraviolet Finiteness of the $\mathcal{N}=4$ Yang-Mills
                        Theory}",
      journal        = "Phys. Lett.",
      volume         = "B123",
      year           = "1983",
      pages          = "323-328",
      doi            = "10.1016/0370-2693(83)91210-8",
      reportNumber   = "UTTG-1-82",
      SLACcitation   = "
}
@article{Tetelman,
      author         = "Tetelman, M.G.",
      title          = "{ Lorentz group for two-dimensional integrable lattice systems.}",
      journal        = "Sov. Phys. JETP",
      volume         = "55(2)",
      year           = "1982",
      pages          = "306-310",
      doi            = "",
      reportNumber   = "2",
      SLACcitation   = ""
}
@article{Sogo,
      author         = "Sogo, K. and Wadati, M.",
      title          = "{Boost Operator and Its Application to Quantum Gelfand-Levitan Equation for 
      Heisenberg-Ising Chain with Spin One-Half}",
      journal        = "Prog.Theor.Phys.",
      volume         = "69(2)",
      year           = "1983",
      pages          = "431-450",
      doi            = "10.1143/PTP.69.431",
      reportNumber   = "",
      SLACcitation   = "
}
@article{Korepin:1982gg,
      author         = "Korepin, V. E.",
      title          = "{Calculation of Norms of Bethe Wave Functions}",
      journal        = "Commun. Math. Phys.",
      volume         = "86",
      year           = "1982",
      pages          = "391-418",
      doi            = "10.1007/BF01212176",
      SLACcitation   = "
}
@article{Beisert:2007hz,
      author         = "Beisert, N. and McLoughlin, T. and Roiban, R.",
      title          = "{The Four-loop dressing phase of $\mathcal{N}=4$ SYM}",
      journal        = "Phys. Rev.",
      volume         = "D76",
      year           = "2007",
      pages          = "046002",
      doi            = "10.1103/PhysRevD.76.046002",
      eprint         = "0705.0321",
      archivePrefix  = "arXiv",
      primaryClass   = "hep-th",
      reportNumber   = "AEI-2007-026, PUTP-2233",
      SLACcitation   = "
}
@article{Staudacher:2004tk,
      author         = "Staudacher, Matthias",
      title          = "{The Factorized S-matrix of CFT/AdS}",
      journal        = "JHEP",
      volume         = "05",
      year           = "2005",
      pages          = "054",
      doi            = "10.1088/1126-6708/2005/05/054",
      eprint         = "hep-th/0412188",
      archivePrefix  = "arXiv",
      primaryClass   = "hep-th",
      reportNumber   = "AEI-2004-107, NSF-KITP-04-122",
      SLACcitation   = "
}

@article{Beisert:2006ez,
      author         = "Beisert, Niklas and Eden, Burkhard and Staudacher,
                        Matthias",
      title          = "{Transcendentality and Crossing}",
      journal        = "J. Stat. Mech.",
      volume         = "0701",
      year           = "2007",
      pages          = "P01021",
      doi            = "10.1088/1742-5468/2007/01/P01021",
      eprint         = "hep-th/0610251",
      archivePrefix  = "arXiv",
      primaryClass   = "hep-th",
      reportNumber   = "AEI-2006-079, ITP-UU-06-44, SPIN-06-34",
      SLACcitation   = "
}
@article{Constable:2001ag,
      author         = "Constable, Neil R. and Myers, Robert C. and Tafjord, Oyvind",
      title          = "{Non-abelian brane intersections}",
      journal        = "JHEP",
      volume         = "06",
      year           = "2001",
      pages          = "023",
      doi            = "10.1088/1126-6708/2001/06/023",
      eprint         = "hep-th/0102080",
      archivePrefix  = "arXiv",
      primaryClass   = "hep-th",
      reportNumber   = "MCGILL-01-01",
      SLACcitation   = "
}

@article{Castelino:1997rv,
      author         = "Castelino, Judith and Lee, Sangmin and Taylor, Washington",
      title          = "{Longitudinal 5-branes as 4-spheres in matrix theory}",
      journal        = "Nucl.Phys.",
      volume         = "B526",
      year           = "1998",
      pages          = "334",
      doi            = "10.1016/S0550-3213(98)00291-0",
      eprint         = "hep-th/9712105",
      archivePrefix  = "arXiv",
      primaryClass   = "hep-th",
      reportNumber   = "PUPT-1749",
      SLACcitation   = "
}

@article{Kristjansen:2012tn,
      author         = "Kristjansen, Charlotte and Semenoff, Gordon W. and Young,
                        Donovan",
      title          = "{Chiral primary one-point functions in the D3-D7 defect
                        conformal field theory}",
      journal        = "JHEP",
      volume         = "01",
      year           = "2013",
      pages          = "117",
      doi            = "10.1007/JHEP01(2013)117",
      eprint         = "1210.7015",
      archivePrefix  = "arXiv",
      primaryClass   = "hep-th",
      reportNumber   = "NORDITA-2012-81",
      SLACcitation   = "
}

@article{Piroli:2018ksf,
      author         = "Piroli, Lorenzo and Vernier, Eric and Calabrese, Pasquale
                        and Pozsgay, Balázs",
      title          = "{Integrable quenches in nested spin chains I: the exact
                        steady states}",
      year           = "2018",
      eprint         = "1811.00432",
      archivePrefix  = "arXiv",
      primaryClass   = "cond-mat.stat-mech",
      SLACcitation   = "
}
@article{Tsuchiya,
author = {Tsuchiya,Osamu },
title = {Determinant formula for the six-vertex model with reflecting end},
journal = {Journal of Mathematical Physics},
volume = {39},
number = {11},
pages = {5946-5951},
year = {1998},
doi = {10.1063/1.532606},
URL = { https://doi.org/10.1063/1.532606},
eprint = { https://doi.org/10.1063/1.532606}
}

@article{deLeeuw:2016ofj,
      author         = "de Leeuw, Marius and Kristjansen, Charlotte and
                        Linardopoulos, Georgios",
      title          = "{One-point functions of non-protected operators in the
                        SO(5) symmetric D3?D7 dCFT}",
      journal        = "J. Phys.",
      volume         = "A50",
      year           = "2017",
      number         = "25",
      pages          = "254001",
      doi            = "10.1088/1751-8121/aa714b",
      eprint         = "1612.06236",
      archivePrefix  = "arXiv",
      primaryClass   = "hep-th",
      SLACcitation   = "
}

@article{deLeeuw:2016ofj,
      author         = "de Leeuw, Marius and Kristjansen, Charlotte and
                        Linardopoulos, Georgios",
      title          = "{One-point functions of non-protected operators in the
                        SO(5) symmetric D3?D7 dCFT}",
      journal        = "J. Phys.",
      volume         = "A50",
      year           = "2017",
      number         = "25",
      pages          = "254001",
      doi            = "10.1088/1751-8121/aa714b",
      eprint         = "1612.06236",
      archivePrefix  = "arXiv",
      primaryClass   = "hep-th",
      SLACcitation   = "
}

@article{Sklyanin,
  author={E K Sklyanin},
  title={Boundary conditions for integrable quantum systems},
  journal={Journal of Physics A: Mathematical and General},
  volume={21},
  number={10},
  pages={2375},
  url={http://stacks.iop.org/0305-4470/21/i=10/a=015},
  year={1988},
  abstract={A new class of boundary conditions is described for quantum systems integrable by means of the quantum inverse scattering (R-matrix) method. The method proposed allows the author to treat open quantum chains with appropriate boundary terms in the Hamiltonian. The general considerations are applied to the XXZ and XYZ models, the nonlinear Schrodinger equation and Toda chain.}
}

@article{quench,
  author={Pasquale Calabrese and Fabian H L Essler and Giuseppe Mussardo},
  title={Introduction to 'Quantum Integrability in Out of Equilibrium Systems'},
  journal={Journal of Statistical Mechanics: Theory and Experiment},
  volume={2016},
  number={6},
  pages={064001},
  url={http://stacks.iop.org/1742-5468/2016/i=6/a=064001},
  year={2016},
  abstract={}
}

@article{Widen:2018nnu,
      author         = "Wid\'en, Erik",
      title          = "{One-point functions in $\beta$-deformed $ \mathcal{N}=4
                        $ SYM with defect}",
      journal        = "JHEP",
      volume         = "11",
      year           = "2018",
      pages          = "114",
      doi            = "10.1007/JHEP11(2018)114",
      eprint         = "1804.09514",
      archivePrefix  = "arXiv",
      primaryClass   = "hep-th",
      reportNumber   = "NORDITA-2018-034; UUITP-17/18, NORDITA-2018-034,
                        UUITP-17-18",
      SLACcitation   = "
}

@article{Semenoff:2018ffq,
      author         = "Semenoff, Gordon W.",
      title          = "{Lectures on the holographic duality of gauge fields and
                        strings}",
      booktitle      = "{Les Houches Summer School: Integrability: From
                        Statistical Systems to Gauge Theory Les Houches, France,
                        June 6-July 1, 2016}",
      year           = "2018",
      eprint         = "1808.04074",
      archivePrefix  = "arXiv",
      primaryClass   = "hep-th",
      SLACcitation   = "
}

@article{Mestyan:2017xyk,
      author         = "Mestyán, M. and Bertini, Bruno and Piroli, Lorenzo and
                        Calabrese, Pasquale",
      title          = "{Exact solution for the quench dynamics of a nested
                        integrable system}",
      journal        = "J. Stat. Mech.",
      volume         = "1708",
      year           = "2017",
      number         = "8",
      pages          = "083103",
      doi            = "10.1088/1742-5468/aa7df0",
      eprint         = "1705.00851",
      archivePrefix  = "arXiv",
      primaryClass   = "cond-mat.stat-mech",
      SLACcitation   = "
}
@article{Piroli:2018don,
      author         = "Piroli, Lorenzo and Vernier, Eric and Calabrese, Pasquale
                        and Pozsgay, Balázs",
      title          = "{Integrable quenches in nested spin chains II: the
                        Quantum Transfer Matrix approach}",
      year           = "2018",
      eprint         = "1812.05330",
      archivePrefix  = "arXiv",
      primaryClass   = "cond-mat.stat-mech",
      SLACcitation   = "
}

@article{Pozsgay:2018dzs,
      author         = "Pozsgay, B. and Piroli, L. and Vernier, E.",
      title          = "{Integrable Matrix Product States from boundary
                        integrability}",
      year           = "2018",
      eprint         = "1812.11094",
      archivePrefix  = "arXiv",
      primaryClass   = "cond-mat.stat-mech",
      SLACcitation   = "
}

@article{Ghoshal:1993tm,
      author         = "Ghoshal, Subir and Zamolodchikov, Alexander B.",
      title          = "{Boundary S matrix and boundary state in two-dimensional
                        integrable quantum field theory}",
      journal        = "Int. J. Mod. Phys.",
      volume         = "A9",
      year           = "1994",
      pages          = "3841-3886",
      eprint         = "hep-th/9306002",
      archivePrefix  = "arXiv",
      primaryClass   = "hep-th",
 }
 @Article{Izergin1989,
author="Izergin, A. G.
and Korepin, V. E.
and Reshetikhin, N. Yu.",
title="Field correlation functions in a one-dimensional Bose gas",
journal="Journal of Soviet Mathematics",
year="1989",
month="Jul",
day="01",
volume="46",
number="1",
pages="1581--1588",
abstract="Multifield correlation functions are calculated in the Bose gas model. It is shown that the multifield correlation functions like the correlator of currents previously calculated can be expressed in terms of the irreducible parts of the form factors and the solutions of nonlinear integral equations.",
issn="1573-8795",
doi="10.1007/BF01099188",
url="https://doi.org/10.1007/BF01099188"
}

@ARTICLE{PPV,
       author = {{Pozsgay}, B. and {Piroli}, L. and {Vernier}, E.},
        title = "{Integrable Matrix Product States from boundary integrability}",
      journal = {arXiv e-prints},
     keywords = {Condensed Matter - Statistical Mechanics, Nonlinear Sciences - Exactly Solvable and Integrable Systems},
         year = "2018",
        month = "Dec",
          eid = {arXiv:1812.11094},
        pages = {arXiv:1812.11094},
archivePrefix = {arXiv},
       eprint = {1812.11094},
 primaryClass = {cond-mat.stat-mech},
       adsurl = {https://ui.adsabs.harvard.edu/\#abs/2018arXiv181211094P},
      adsnote = {Provided by the SAO/NASA Astrophysics Data System}
}
@phdthesis{Escobedo:2012ama,
      author         = "Escobedo, Jorge",
      title          = "{Integrability in AdS/CFT: Exact Results for Correlation
                        Functions}",
      school         = "Waterloo U.",
      year           = "2012",
      url            = "http://uwspace.uwaterloo.ca/handle/10012/6788",
      SLACcitation   = "
}
\end{bibtex}

\bibliographystyle{nb}
\bibliography{\jobname}

\begin{thebibliography}{10}
\ifx\href\asklfhas\newcommand{\href}[2]{#2}\fi
\ifx\arxivref\asklfhas\newcommand{\arxivref}[2]{\href{http://arxiv.org/abs/#1}{#2}}\fi
\ifx\doiref\asklfhas\newcommand{\doiref}[2]{\href{http://dx.doi.org/#1}{#2}}\fi
\parskip 0pt
\normalsize

\bibitem{Maldacena:1997re}
J.M. Maldacena,
\textit{``{The Large N limit of superconformal field theories and
  supergravity}''},
\doiref{10.1023/A:1026654312961}{Int.~J.~Theor.~Phys. \textbf{38}, 1113
  (1999)},
\normalsize{\texttt{\arxivref{hep-th/9711200}{hep-th/9711200}}},
[Adv. Theor. Math. Phys.2,231(1998)].

\bibitem{Minahan:2002ve}
J.A. Minahan \& K.~Zarembo,
\textit{``{The Bethe ansatz for N=4 superYang-Mills}''},
\doiref{10.1088/1126-6708/2003/03/013}{JHEP \textbf{0303}, 013 (2003)},
\normalsize{\texttt{\arxivref{hep-th/0212208}{hep-th/0212208}}}.

\bibitem{Beisert:2010jr}
N.~Beisert et~al.,
\textit{``{Review of AdS/CFT Integrability: An Overview}''},
\doiref{10.1007/s11005-011-0529-2}{Lett.~Math.~Phys. \textbf{99}, 3 (2012)},
\normalsize{\texttt{\arxivref{1012.3982}{arXiv:1012.3982}}}.

\bibitem{DeWolfe:2001pq}
O.~DeWolfe, D.Z. Freedman \& H.~Ooguri,
\textit{``{Holography and defect conformal field theories}''},
\doiref{10.1103/PhysRevD.66.025009}{Phys.~Rev. \textbf{D66}, 025009 (2002)},
\normalsize{\texttt{\arxivref{hep-th/0111135}{hep-th/0111135}}}.

\bibitem{Karch:2000gx}
A.~Karch \& L.~Randall,
\textit{``{Open and closed string interpretation of SUSY CFT's on branes with
  boundaries}''},
\doiref{10.1088/1126-6708/2001/06/063}{JHEP \textbf{0106}, 063 (2001)},
\normalsize{\texttt{\arxivref{hep-th/0105132}{hep-th/0105132}}}.

\bibitem{Semenoff:2018ffq}
G.W. Semenoff,
\textit{``{Lectures on the holographic duality of gauge fields and strings}''},
\normalsize{\texttt{\arxivref{1808.04074}{arXiv:1808.04074}}},
in \textit{``{Les Houches Summer School: Integrability: From Statistical
  Systems to Gauge Theory Les Houches, France, June 6-July 1, 2016}''}.

\bibitem{deLeeuw:2018mkd}
M.~De~Leeuw, C.~Kristjansen \& G.~Linardopoulos,
\textit{``{Scalar one-point functions and matrix product states of
  AdS/dCFT}''},
\doiref{10.1016/j.physletb.2018.03.083}{Phys.~Lett. \textbf{B781}, 238 (2018)},
\normalsize{\texttt{\arxivref{1802.01598}{arXiv:1802.01598}}}.

\bibitem{deLeeuw:2015hxa}
M.~de~Leeuw, C.~Kristjansen \& K.~Zarembo,
\textit{``{One-point Functions in Defect CFT and Integrability}''},
\doiref{10.1007/JHEP08(2015)098}{JHEP \textbf{1508}, 098 (2015)},
\normalsize{\texttt{\arxivref{1506.06958}{arXiv:1506.06958}}}.

\bibitem{Buhl-Mortensen:2015gfd}
I.~Buhl-Mortensen, M.~de~Leeuw, C.~Kristjansen \& K.~Zarembo,
\textit{``{One-point Functions in AdS/dCFT from Matrix Product States}''},
\doiref{10.1007/JHEP02(2016)052}{JHEP \textbf{1602}, 052 (2016)},
\normalsize{\texttt{\arxivref{1512.02532}{arXiv:1512.02532}}}.

\bibitem{Mestyan:2017xyk}
M.~Mestyán, B.~Bertini, L.~Piroli \& P.~Calabrese,
\textit{``{Exact solution for the quench dynamics of a nested integrable
  system}''},
\doiref{10.1088/1742-5468/aa7df0}{J.~Stat.~Mech. \textbf{1708}, 083103 (2017)},
\normalsize{\texttt{\arxivref{1705.00851}{arXiv:1705.00851}}}.

\bibitem{Cardy:1984bb}
J.L. Cardy,
\textit{``{Conformal Invariance and Surface Critical Behavior}''},
\doiref{10.1016/0550-3213(84)90241-4}{Nucl.~Phys. \textbf{B240}, 514 (1984)}.

\bibitem{DHoker:2002nbb}
E.~D'Hoker \& D.Z. Freedman,
\textit{``{Supersymmetric gauge theories and the AdS / CFT correspondence}''},
in \textit{``{Strings, Branes and Extra Dimensions: TASI 2001: Proceedings}''},
p.~3-158.
\bibitem{deLeeuw:2017cop}
M.~de~Leeuw, A.C. Ipsen, C.~Kristjansen \& M.~Wilhelm,
\textit{``{Introduction to Integrability and One-point Functions in
  $\mathcal{N}=4$ SYM and its Defect Cousin}''},
in \textit{``{Les Houches Summer School: Integrability: From Statistical
  Systems to Gauge Theory Les Houches, France, June 6-July 1, 2016}''}.
\bibitem{tHooft:1973alw}
G.~'t~Hooft,
\textit{``{A Planar Diagram Theory for Strong Interactions}''},
\doiref{10.1016/0550-3213(74)90154-0}{Nucl.~Phys. \textbf{B72}, 461 (1974)}.

\bibitem{Sklyanin}
E.K. Sklyanin,
\textit{``Boundary conditions for integrable quantum systems''},
Journal~of~Physics~A:~Mathematical~and~General \textbf{21}, 2375 (1988),
\href{http://stacks.iop.org/0305-4470/21/i=10/a=015}{\texttt{http://stacks.iop.org/0305-4470/21/i=10/a=015}}.

\bibitem{Constable:1999ac}
N.R. Constable, R.C. Myers \& O.~Tafjord,
\textit{``{The Noncommutative bion core}''},
\doiref{10.1103/PhysRevD.61.106009}{Phys.~Rev. \textbf{D61}, 106009 (2000)},
\normalsize{\texttt{\arxivref{hep-th/9911136}{hep-th/9911136}}}.

\bibitem{Kristjansen:2012tn}
C.~Kristjansen, G.W. Semenoff \& D.~Young,
\textit{``{Chiral primary one-point functions in the D3-D7 defect conformal
  field theory}''},
\doiref{10.1007/JHEP01(2013)117}{JHEP \textbf{1301}, 117 (2013)},
\normalsize{\texttt{\arxivref{1210.7015}{arXiv:1210.7015}}}.

\bibitem{Constable:2001ag}
N.R. Constable, R.C. Myers \& O.~Tafjord,
\textit{``{Non-abelian brane intersections}''},
\doiref{10.1088/1126-6708/2001/06/023}{JHEP \textbf{0106}, 023 (2001)},
\normalsize{\texttt{\arxivref{hep-th/0102080}{hep-th/0102080}}}.

\bibitem{Castelino:1997rv}
J.~Castelino, S.~Lee \& W.~Taylor,
\textit{``{Longitudinal 5-branes as 4-spheres in matrix theory}''},
\doiref{10.1016/S0550-3213(98)00291-0}{Nucl.Phys. \textbf{B526}, 334 (1998)},
\normalsize{\texttt{\arxivref{hep-th/9712105}{hep-th/9712105}}}.

\bibitem{deLeeuw:2016ofj}
M.~de~Leeuw, C.~Kristjansen \& G.~Linardopoulos,
\textit{``{One-point functions of non-protected operators in the SO(5)
  symmetric D3-D7 dCFT}''},
\doiref{10.1088/1751-8121/aa714b}{J.~Phys. \textbf{A50}, 254001 (2017)},
\normalsize{\texttt{\arxivref{1612.06236}{arXiv:1612.06236}}}.

\bibitem{Bethe1931}
H.~Bethe,
\textit{``Zur Theorie der Metalle''},
Zeitschrift~f{\"u}r~Ph \textbf{71}, 205 (1931).

\bibitem{Faddeev:1996iy}
L.D. Faddeev,
\textit{``{How algebraic Bethe ansatz works for integrable model}''},
in \textit{``{Relativistic gravitation and gravitational radiation.
  Proceedings, School of Physics, Les Houches, France, September 26-October 6,
  1995}''},
p.~pp. 149-219.
\bibitem{Levkovich-Maslyuk:2016kfv}
F.~Levkovich-Maslyuk,
\textit{``{The Bethe ansatz}''},
\doiref{10.1088/1751-8113/49/32/323004}{J.~Phys. \textbf{A49}, 323004 (2016)},
\normalsize{\texttt{\arxivref{1606.02950}{arXiv:1606.02950}}}.

\bibitem{Izergin1989}
A.G. Izergin, V.E. Korepin \& N.Y. Reshetikhin,
\textit{``Field correlation functions in a one-dimensional Bose gas''},
\doiref{10.1007/BF01099188}{Journal~of~Soviet~Mathematics \textbf{46}, 1581
  (1989)},
\href{https://doi.org/10.1007/BF01099188}{\texttt{https://doi.org/10.1007/BF01099188}}.

\bibitem{Escobedo:2010xs}
J.~Escobedo, N.~Gromov, A.~Sever \& P.~Vieira,
\textit{``{Tailoring Three-Point Functions and Integrability}''},
\doiref{10.1007/JHEP09(2011)028}{JHEP \textbf{1109}, 028 (2011)},
\normalsize{\texttt{\arxivref{1012.2475}{arXiv:1012.2475}}}.

\bibitem{Slavnov1989}
N.A. Slavnov,
\textit{``Calculation of scalar products of wave functions and form factors in
  the framework of the alcebraic Bethe ansatz''},
Theoretical~and~Mathematical~Physics \textbf{79}, 502 (1989).

\bibitem{Gaudin:1976sv}
M.~Gaudin,
\textit{``{Diagonalization of a Class of Spin Hamiltonians}''},
J.~Phys.~France \textbf{37}, 1086 (1976).

\bibitem{Korepin:1982gg}
V.E. Korepin,
\textit{``{Calculation of Norms of Bethe Wave Functions}''},
\doiref{10.1007/BF01212176}{Commun.~Math.~Phys. \textbf{86}, 391 (1982)}.

\bibitem{quench}
P.~Calabrese, F.H.L. Essler \& G.~Mussardo,
\textit{``Introduction to 'Quantum Integrability in Out of Equilibrium
  Systems'''},
Journal~of~Statistical~Mechanics:~Theory~and~Experiment \textbf{2016}, 064001
  (2016),
\href{http://stacks.iop.org/1742-5468/2016/i=6/a=064001}{\texttt{http://stacks.iop.org/1742-5468/2016/i=6/a=064001}}.

\bibitem{Piroli:2017sei}
L.~Piroli, B.~Pozsgay \& E.~Vernier,
\textit{``{What is an integrable quench?}''},
\doiref{10.1016/j.nuclphysb.2017.10.012}{Nucl.~Phys. \textbf{B925}, 362
  (2017)}.

\bibitem{Ghoshal:1993tm}
S.~Ghoshal \& A.B. Zamolodchikov,
\textit{``{Boundary S matrix and boundary state in two-dimensional integrable
  quantum field theory}''},
Int.~J.~Mod.~Phys. \textbf{A9}, 3841 (1994),
\normalsize{\texttt{\arxivref{hep-th/9306002}{hep-th/9306002}}}.

\bibitem{Piroli:2018ksf}
L.~Piroli, E.~Vernier, P.~Calabrese \& B.~Pozsgay,
\textit{``{Integrable quenches in nested spin chains I: the exact steady
  states}''},
\normalsize{\texttt{\arxivref{1811.00432}{arXiv:1811.00432}}}.

\bibitem{Piroli:2018don}
L.~Piroli, E.~Vernier, P.~Calabrese \& B.~Pozsgay,
\textit{``{Integrable quenches in nested spin chains II: the Quantum Transfer
  Matrix approach}''},
\normalsize{\texttt{\arxivref{1812.05330}{arXiv:1812.05330}}}.

\bibitem{Pozsgay:2018dzs}
B.~Pozsgay, L.~Piroli \& E.~Vernier,
\textit{``{Integrable Matrix Product States from boundary integrability}''},
\normalsize{\texttt{\arxivref{1812.11094}{arXiv:1812.11094}}}.

\bibitem{PPV}
B.~{Pozsgay}, L.~{Piroli} \& E.~{Vernier},
\textit{``{Integrable Matrix Product States from boundary integrability}''},
arXiv~e-prints \textbf{A9}, arXiv:1812.11094 (2018),
\normalsize{\texttt{\arxivref{1812.11094}{arXiv:1812.11094}}}.

\bibitem{Tsuchiya}
O.~Tsuchiya,
\textit{``Determinant formula for the six-vertex model with reflecting end''},
\doiref{10.1063/1.532606}{Journal~of~Mathematical~Physics \textbf{39}, 5946
  (1998)},
\normalsize{\texttt{\arxivref{https://doi.org/10.1063/1.532606}{https://doi.org/10.1063/1.532606}}},
\href{https://doi.org/10.1063/1.532606}{\texttt{https://doi.org/10.1063/1.532606}}.

\bibitem{PozsgayXXZ}
B.~{Pozsgay},
\textit{``{Overlaps with arbitrary two-site states in the XXZ spin chain}''},
\doiref{10.1088/1742-5468/aabbe1}{Journal~of~Statistical~Mechanics:~Theory~and~Experiment
  \textbf{5}, 053103 (2018)},
\normalsize{\texttt{\arxivref{1801.03838}{arXiv:1801.03838}}}.

\bibitem{Foda:2015nfk}
O.~Foda \& K.~Zarembo,
\textit{``{Overlaps of partial N{\' e}el states and Bethe states}''},
\doiref{10.1088/1742-5468/2016/02/023107}{J.~Stat.~Mech. \textbf{1602}, 023107
  (2016)},
\normalsize{\texttt{\arxivref{1512.02533}{arXiv:1512.02533}}}.

\bibitem{Brockmann1}
M.~{Brockmann}, J.~{De Nardis}, B.~{Wouters} \& J.S. {Caux},
\textit{``{A Gaudin-like determinant for overlaps of N{\'e}el and XXZ Bethe
  states}''},
\doiref{10.1088/1751-8113/47/14/145003}{Journal~of~Physics~A~Mathematical~General
  \textbf{47}, 145003 (2014)},
\normalsize{\texttt{\arxivref{1401.2877}{arXiv:1401.2877}}}.

\bibitem{Brockmann2}
M.~{Brockmann}, J.~{De Nardis}, B.~{Wouters} \& J.S. {Caux},
\textit{``{N{\'e}el-XXZ state overlaps: odd particle numbers and Lieb-Liniger
  scaling limit}''},
\doiref{10.1088/1751-8113/47/34/345003}{Journal~of~Physics~A~Mathematical~General
  \textbf{47}, 345003 (2014)},
\normalsize{\texttt{\arxivref{1403.7469}{arXiv:1403.7469}}}.

\bibitem{Pozsgay}
B.~Pozsgay,
\textit{``Overlaps between eigenstates of the XXZ spin-1/2 chain and a class of
  simple product states''},
Journal~of~Statistical~Mechanics:~Theory~and~Experiment \textbf{2014}, P06011
  (2014).

\bibitem{Brockmann}
M.~Brockmann,
\textit{``Overlaps of $q$-raised N{\' e}el states with XXZ Bethe states and
  their relation to the Lieb-Liniger Bose gas''},
Journal~of~Statistical~Mechanics:~Theory~and~Experiment \textbf{2014}, P05006
  (2014).

\bibitem{Bazhanov:2010ts}
V.V. Bazhanov, T.~\L{}ukowski, C.~Meneghelli \& M.~Staudacher,
\textit{``{A Shortcut to the Q-Operator}''},
\doiref{10.1088/1742-5468/2010/11/P11002}{J.~Stat.~Mech. \textbf{1011}, P11002
  (2010)},
\normalsize{\texttt{\arxivref{1005.3261}{arXiv:1005.3261}}}.

\bibitem{deLeeuw:2016umh}
M.~de~Leeuw, C.~Kristjansen \& S.~Mori,
\textit{``{AdS/dCFT one-point functions of the SU(3) sector}''},
\doiref{10.1016/j.physletb.2016.10.044}{Phys.~Lett. \textbf{B763}, 197 (2016)},
\normalsize{\texttt{\arxivref{1607.03123}{arXiv:1607.03123}}}.

\bibitem{Escobedo:2012ama}
J.~Escobedo,
\textit{``{Integrability in AdS/CFT: Exact Results for Correlation
  Functions}''},
\href{http://uwspace.uwaterloo.ca/handle/10012/6788}{\texttt{http://uwspace.uwaterloo.ca/handle/10012/6788}}.

\bibitem{Basso:2017khq}
B.~Basso, F.~Coronado, S.~Komatsu, H.T. Lam, P.~Vieira \& D.l. Zhong,
\textit{``{Asymptotic Four Point Functions}''},
\normalsize{\texttt{\arxivref{1701.04462}{arXiv:1701.04462}}}.

\bibitem{Buhl-Mortensen:2016pxs}
I.~Buhl-Mortensen, M.~de~Leeuw, A.C. Ipsen, C.~Kristjansen \& M.~Wilhelm,
\textit{``{One-loop one-point functions in gauge-gravity dualities with
  defects}''},
\doiref{10.1103/PhysRevLett.117.231603}{Phys.~Rev.~Lett. \textbf{117}, 231603
  (2016)},
\normalsize{\texttt{\arxivref{1606.01886}{arXiv:1606.01886}}}.

\bibitem{Buhl-Mortensen:2016jqo}
I.~Buhl-Mortensen, M.~de~Leeuw, A.C. Ipsen, C.~Kristjansen \& M.~Wilhelm,
\textit{``{A Quantum Check of AdS/dCFT}''},
\doiref{10.1007/JHEP01(2017)098}{JHEP \textbf{1701}, 098 (2017)},
\normalsize{\texttt{\arxivref{1611.04603}{arXiv:1611.04603}}}.

\bibitem{Nagasaki:2011ue}
K.~Nagasaki, H.~Tanida \& S.~Yamaguchi,
\textit{``{Holographic Interface-Particle Potential}''},
\doiref{10.1007/JHEP01(2012)139}{JHEP \textbf{1201}, 139 (2012)},
\normalsize{\texttt{\arxivref{1109.1927}{arXiv:1109.1927}}}.

\bibitem{Beisert:2005fw}
N.~Beisert \& M.~Staudacher,
\textit{``{Long-range psu$(2,2|4)$ Bethe Ans\"{a}tze for gauge theory and
  strings}''},
\doiref{10.1016/j.nuclphysb.2005.06.038}{Nucl.~Phys. \textbf{B727}, 1 (2005)},
\normalsize{\texttt{\arxivref{hep-th/0504190}{hep-th/0504190}}}.

\bibitem{Beisert:2006ez}
N.~Beisert, B.~Eden \& M.~Staudacher,
\textit{``{Transcendentality and Crossing}''},
\doiref{10.1088/1742-5468/2007/01/P01021}{J.~Stat.~Mech. \textbf{0701}, P01021
  (2007)},
\normalsize{\texttt{\arxivref{hep-th/0610251}{hep-th/0610251}}}.

\bibitem{Buhl-Mortensen:2017ind}
I.~Buhl-Mortensen, M.~de~Leeuw, A.C. Ipsen, C.~Kristjansen \& M.~Wilhelm,
\textit{``{Asymptotic One-Point Functions in Gauge-String Duality with
  Defects}''},
\doiref{10.1103/PhysRevLett.119.261604}{Phys.~Rev.~Lett. \textbf{119}, 261604
  (2017)},
\normalsize{\texttt{\arxivref{1704.07386}{arXiv:1704.07386}}}.

\bibitem{deLeeuw:2017dkd}
M.~de~Leeuw, A.C. Ipsen, C.~Kristjansen, K.E. Vardinghus \& M.~Wilhelm,
\textit{``{Two-point functions in AdS/dCFT and the boundary conformal bootstrap
  equations}''},
\doiref{10.1007/JHEP08(2017)020}{JHEP \textbf{1708}, 020 (2017)},
\normalsize{\texttt{\arxivref{1705.03898}{arXiv:1705.03898}}}.

\bibitem{Widen:2017uwh}
E.~Widen,
\textit{``{Two-point functions of SU(2)-subsector and length-two operators in
  dCFT}''},
\doiref{10.1016/j.physletb.2017.08.059}{Phys.~Lett. \textbf{B773}, 435 (2017)},
\normalsize{\texttt{\arxivref{1705.08679}{arXiv:1705.08679}}}.

\bibitem{deLeeuw:2016vgp}
M.~de~Leeuw, A.C. Ipsen, C.~Kristjansen \& M.~Wilhelm,
\textit{``{One-loop Wilson loops and the particle-interface potential in
  AdS/dCFT}''},
\doiref{10.1016/j.physletb.2017.02.047}{Phys.~Lett. \textbf{B768}, 192 (2017)},
\normalsize{\texttt{\arxivref{1608.04754}{arXiv:1608.04754}}}.

\bibitem{Preti:2017fhw}
M.~Preti, D.~Trancanelli \& E.~Vescovi,
\textit{``{Quark-antiquark potential in defect conformal field theory}''},
\doiref{10.1007/JHEP10(2017)079}{JHEP \textbf{1710}, 079 (2017)},
\normalsize{\texttt{\arxivref{1708.04884}{arXiv:1708.04884}}}.

\bibitem{Aguilera-Damia:2016bqv}
J.~Aguilera-Damia, D.H. Correa \& V.I. Giraldo-Rivera,
\textit{``{Circular Wilson loops in defect Conformal Field Theory}''},
\doiref{10.1007/JHEP03(2017)023}{JHEP \textbf{1703}, 023 (2017)},
\normalsize{\texttt{\arxivref{1612.07991}{arXiv:1612.07991}}}.

\bibitem{Grau:2018keb}
A.G. Grau, C.~Kristjansen, M.~Volk \& M.~Wilhelm,
\textit{``{A Quantum Check of Non-Supersymmetric AdS/dCFT}''},
\normalsize{\texttt{\arxivref{1810.11463}{arXiv:1810.11463}}}.

\bibitem{KdLV}
M.~de~Leeuw, C.~Kristjansen \& K.E. Vardinghus,
\textit{``{A non-integrable quench from AdS/dCFT}''},
\normalsize{\texttt{\arxivref{1906.10714}{arXiv:1906.10714}}}.

\bibitem{Widen:2018nnu}
E.~Wid\'en,
\textit{``{One-point functions in $\beta$-deformed $ \mathcal{N}=4 $ SYM with
  defect}''},
\doiref{10.1007/JHEP11(2018)114}{JHEP \textbf{1811}, 114 (2018)},
\normalsize{\texttt{\arxivref{1804.09514}{arXiv:1804.09514}}}.

\bibitem{Liendo:2012hy}
P.~Liendo, L.~Rastelli \& B.C. van~Rees,
\textit{``{The Bootstrap Program for Boundary CFT$_\mathrm{d}$}''},
\doiref{10.1007/JHEP07(2013)113}{JHEP \textbf{1307}, 113 (2013)},
\normalsize{\texttt{\arxivref{1210.4258}{arXiv:1210.4258}}}.

\bibitem{Liendo:2016ymz}
P.~Liendo \& C.~Meneghelli,
\textit{``{Bootstrap equations for $ \mathcal{N} $ = 4 SYM with defects}''},
\doiref{10.1007/JHEP01(2017)122}{JHEP \textbf{1701}, 122 (2017)},
\normalsize{\texttt{\arxivref{1608.05126}{arXiv:1608.05126}}}.

\bibitem{Andrei:2018die}
N.~Andrei et~al.,
\textit{``{Boundary and Defect CFT: Open Problems and Applications}''},
\normalsize{\texttt{\arxivref{1810.05697}{arXiv:1810.05697}}}.

\end{thebibliography}

\end{document}